\documentclass[10pt]{article}
\usepackage{amsmath,bm,graphicx,amssymb,amsthm}
\usepackage{amsmath}
\usepackage{amsfonts}
\usepackage{amssymb}
\usepackage{graphicx}

\newtheorem{theorem}{Theorem}

\newtheorem{corollary}[theorem]{Corollary}
\newtheorem{definition}[theorem]{Definition}

\newtheorem{lemma}[theorem]{Lemma}
\newtheorem{proposition}[theorem]{Proposition}

\newtheorem{remark}[theorem]{Remark}

\begin{document}

\begin{center}
{\LARGE Gravitational solitons and $C^0$ vacuum metrics in}

\medskip
{\LARGE five-dimensional Lovelock gravity}

\medskip

\medskip

\medskip

{\large C. Garraffo}$^{1}${\large , G. Giribet}$^{2,3}${\large ,
E. Gravanis}$^{4}${\large , S. Willison}$^{5}$

\medskip

$^{1}$ Instituto de Astronom\'{\i}a y F\'{\i}sica del Espacio, IAFE, CONICET, Argentina.

{\it Ciudad Universitaria, IAFE, C.C. 67, Suc. 28, 1428, Buenos
Aires, Argentina}.

\medskip

$^{2}$ Center for Cosmology and Particle Physics, New York
University, NYU,

{\it 4 Washington Place, NY10003, New York, US}.

\medskip

$^{3}$ Departamento de F\'{\i}sica, FCEN, Universidad de Buenos
Aires, Argentina,

{\it Ciudad Universitaria, Pabell\'on 1, 1428, Buenos Aires,
Argentina}.

\medskip

$^{4}$ Department of Physics, Kings College London, UK.

\medskip

$^{5}$ Centro de Estudios Cient\'{\i}ficos CECS,

{\it Casilla 1469, Valdivia, Chile.}

\medskip

\end{center}

\begin{abstract}

Junction conditions for vacuum solutions in five-dimensional
Einstein-Gauss-Bonnet gravity are studied. We focus on those cases
where two spherically symmetric regions of space-time are joined in
such a way that the induced stress tensor on the junction surface
vanishes. So a spherical vacuum shell, containing no matter, arises
as a boundary between two regions of the space-time. A general
analysis is given of solutions that can be constructed by this
method of geometric surgery. Such solutions are a generalized kind
of spherically symmetric empty space solutions, described by metric
functions of the class $C^0$. New global structures arise with
surprising features. In particular, we show that vacuum spherically
symmetric wormholes do exist in this theory. These can be regarded
as gravitational solitons, which connect two asymptotically (Anti)
de-Sitter spaces with different masses and/or different effective
cosmological constants. We prove the existence of both static and
dynamical solutions and discuss their (in)stability under
perturbations that preserve the symmetry. This leads us to discuss a
new type of instability that arises in five-dimensional Lovelock
theory of gravity for certain values of the coupling of the
Gauss-Bonnet term. The issues of existence and uniqueness of
solutions and determinism in the dynamical evolution are also
discussed.

\end{abstract}

\newpage

\tableofcontents

\newpage

\section{Introduction}

In this article we shall be concerned with the Einstein-Gauss-Bonnet
theory of gravity, whose action is given by the Einstein-Hilbert
term plus the Einstein cosmological constant term $\Lambda$, and
both supplemented with the Gauss-Bonnet term, quadratic in the
curvature. The action of the theory reads\footnote{Capital Roman
letters $A$, $B$ etc. have been employed for five-dimensional tensor
indices. ${\cal R}^{A}_{\ BCD}$ is the five-dimensional Riemann
tensor.}
\begin{equation}
S=\frac{1}{2\kappa ^{2}}\int d^{5}x\sqrt{-g}\, \Big( {\cal R}
-2\Lambda +\alpha \left({\cal R}^2 - 4 {\cal R}_{AB} {\cal R}^{AB} +
{\cal R}_{ABCD} {\cal R}^{ABCD} \right)  \Big) ,\label{The_Action}
\end{equation}
where $\kappa ^{2}=8\pi G$ and $\alpha $ represents the coupling
constant of the quadratic term. The quadratic term is often called the
Gauss-Bonnet term because it is the dimensional extension of the
Gauss-Bonnet topological invariant in four dimensions.

In five dimensions, the action (\ref{The_Action}) contains all of
the non-zero terms of the Lovelock series. It is thus the most
general metric torsion-free theory of gravity which leads to
conserved equations of motion which are second order in
derivatives \cite{Lovelock}. The perturbation theory about the
maximally symmetric vacuum is free of ghosts
\cite{Zwiebach,Zumino} which suggests that it could appear as a
higher order correction to Einstein's theory in the effective
action coming from some more fundamental quantum theory. In fact,
the Gauss-Bonnet term naturally arises as a higher order
correction to gravity within string theory. Although the
fourth-order derivative corrections are known to appear as the
next-to-leading-order correction in the Type II strings\cite{GW},
the quadratic corrections are present in both the heterotic and
bosonic string theory \cite{Zwiebach,GS,mas}. In those cases, the
coupling of the Gauss-Bonnet term is given by $\alpha^\prime$
multiplied by a function of the dilaton, and so corresponding to
powers of the string coupling. The five-dimensional Gauss-Bonnet
term also arises in the Calabi-Yau compactification of M-theory,
where the coupling of the second-order corrections turns out to be
given in terms of the K\"{a}hler moduli of the six-dimensional
compact manifold \cite{Ferrara:1996hh}.

The presence of the Gauss-Bonnet term introduces some exotic
features not found in General Relativity. One such feature is
related to the problem of causality; this was treated in Ref.
\cite{Teitelboim-87} in the Hamiltonian formalism (see also Ref.
\cite{Choquet-Bruhat-88} for an alternative treatment of the Cauchy
problem). Because of the non-linearity of the theory, the canonical
momenta are not linear in the extrinsic curvature; and there exist
quite generically points in the phase space where the Hamiltonian
turns out to be multiple-valued. In such a situation, there is a
breakdown in the deterministic evolution of the metric from the
initial data. This can also be seen explicitly using the junction
conditions \cite{Davis:2002gn,Gravanis:2002wy}. In fact, it can be
shown that there exist vacuum solutions where the extrinsic
curvature can jump spontaneously at some spacelike hypersurface in a
way that is not predicted by the initial data\footnote{The junction
condition in vacuum gives precisely that the jump in the canonical
momenta is zero. The existence of solutions with non-zero jump in
the extrinsic curvature at a spacelike shell is therefore equivalent
to the problem of a multiple-valued Hamiltonian. }. This breakdown
in predictability is induced by the presence of terms in the
junction conditions which, unlike the Israel conditions valid for
Einstein's theory, contain non-linear contributions coming from the
Gauss-Bonnet term.

On the other hand, the timelike version of such a jump in the
extrinsic curvature is also of great interest. This is realized by
the existence of a kind of gravitational solitons in the theory,
which resemble a kink solution. These solitons correspond to
spacetimes that contain timelike hypersurfaces where the metric is
$C^0$ continuous but where the extrinsic curvature jumps. Although
the Riemann curvature tensor contains delta-function singularities
on the hypersurface, these spacetimes can still be vacuum solutions
because of a nontrivial cancelation coming from additional terms in
the junction conditions. Some explicit examples have appeared in the
literature \cite{Meissner}, and a spherically symmetric realization
of such solutions were studied in detail in Ref.
\cite{Gravanis:2007ei} for the case of pure Gauss-Bonnet
gravitational theory. Here, the systematical analysis made in Ref.
\cite{Gravanis:2007ei} will be extended to the more
phenomenologically important case where Einstein-Hilbert term and
cosmological constant are included in the gravitational action. We
will show that vacuum shell solutions are indeed found in
Einstein-Gauss-Bonnet theory described by the action
(\ref{The_Action}).

So then we will consider the junction conditions for spherical thin
shells in Einstein-Gauss-Bonnet theory in the case that the induced
stress tensor on the shell vanishes. Then, we will show that
geometries associated with two different spherically symmetric
spaces can be joined without resorting to the introduction of matter
fields as a source. Depending on the orientation of the two spaces,
different global structures may arise. For instance, for one choice
of orientation we get vacuum wormholes in five-dimensions. These
wormholes are gravitational solitons that connect two regions with
different masses which can be asymptotically either flat, Anti de
Sitter (AdS) or de Sitter (dS) depending on the sign of the
effective cosmological constant in each region. Other choices of
orientation are possible, such as spherical bubbles, inside of which
the value of the effective cosmological drastically changes. All the
cases we will study in detail are such that the singular
hypersurfaces where the jump in extrinsic curvature is located
correspond to a sphere. We will call them``vacuum shells".

This paper is organised as follows. We begin section 2 by presenting
some preliminary material that will be used in the rest of the
paper. First, we review basic aspects of the spherically symmetric
solution of Einstein-Gauss-Bonnet theory: the well known
Boulware-Deser metric \cite{Boulware:1985wk,Wheeler}. Secondly, we
review the junction conditions for this theory. We discuss both the
cases where the junction hypersurface is of timelike and spacelike
signature, we describe the different orientations allowed. At the
end of the second section we derive the equation that contains all
the information about the junction of two spherically symmetric
vacuum solutions in the five-dimensional Einstein-Gauss-Bonnet
theory. In section 3 we focus our attention on the static case
corresponding to the timelike time-independent junctions; also we
study the instantaneous case corresponding to the spacelike
analogue. We explore the space of parameters of the theory for which
solutions describing wormhole-like and bubble-like geometries exist.
 We see that
such vacuum shells can also contain interior regions where naked
singularities arise. In section 4 we survey the catalogue of these
curious geometries, and we discuss the qualitative aspects of static
solutions, emphasizing the most relevant properties. In section 5 we
analyze the dynamical solutions. This includes a discussion of the
(in)stability of the static solutions under perturbations that
preserve the symmetry. Also some general results about the behavior
of time-dependent solutions are given in Propositions \ref{No
minimum} and \ref{Repulsion}. In section 6 we give an exhaustive
parametrization of the space of constant radius solutions. Section 7
is devoted to a discussion of the $C^0$ class metrics and the
topology of the solutions. We also discuss there the uniqueness and
staticity of the spherically symmetric solutions, concerning the
global validity of the Birkhoff-type theorems in Lovelock gravity.
Section 8 contains the conclusions.

With respect to the style of presentation, we have chosen to
organize our results in a series of remarks, propositions and
theorems in order to highlight key facts, but descriptions such as
`theorem' should not be taken in the most strict mathematical sense.

\section{The setup}

First, we will present some introductory material and notation and
conventions. The spherically symmetric solutions of
Einstein-Gauss-Bonnet gravitational theory will be reviewed. Then we
will discuss the junction conditions in this theory.

Then we will show how these junction conditions permit to join two
spherically symmetric spaces without resorting to the introduction
of matter source.

\subsection{The bulk metric}

Let us consider the Einstein-Gauss-Bonnet theory.
The field
equations associated with the action (\ref{The_Action}) coupled to
some matter action take the form
\begin{gather}\label{The_field_equations}
 G^A_{B} + \Lambda \delta^A_B + \alpha H^A_B
 = \kappa^2 T^A_B\,,
\end{gather}
where $T^A_B$ is the stress tensor, $G^A_B  \equiv -\frac{1}{4}\,
\delta^{A CD}
 _{B EF}\, {\cal R}^{EF}_{AB}
 = {\cal R}^A_{\ B}- \frac{1}{2} \delta^A_B\, {\cal R}$ is the
 Einstein tensor and
\begin{align*}
 H^A_B & \equiv -\frac{1}{8}\, \delta^{A C_1 \dots C_4}
 _{B D_1 \dots D_4}\, {\cal R}^{D_1 D_2}_{\ \ \ C_1
 C_2} {\cal R}^{D_3 D_4}_{\ \ \ C_3 C_4} \, ,
\end{align*}
and where the antisymmetrized Kronecker delta is defined as
$\delta^{A_1 \dots A_p}_{B_1 \dots B_p} \equiv p!
\delta^{A_1}_{[B_1} \cdots  \delta^{ A_p}_{B_p]}$.

We are mainly interested in the static spherically symmetric
solution (without matter) to Einstein-Gauss-Bonnet theory in five
dimensions. In this case, of space-times fibered over (constant
radius) 3-spheres, the solutions correspond to the analogues of the
Schwarzschild geometry, and its form was found by D. Boulware and S.
Deser in Ref. \cite{Boulware:1985wk}. More generally, the solutions
that correspond to fiber bundles over 3-surfaces of constant
negative (or vanishing) curvature were subsequently studied in Ref.
\cite{Cai:2001dz} (and also Ref. \cite{Aros:2000ij}
 in a special class of Lovelock theories in arbitrary
dimension \cite{BHscan}). Let us discuss these solutions here.
First, let us write the ansatz for the metric as follows
\begin{equation}
 ds^2 = -f(r) dt^2 + \frac{dr^2}{f(r)} + r^2 \, d\Omega^2_k \, , \label{ansatz}
\end{equation}
where $d\Omega^2_k$ is the metric of the constant curvature
three-manifold (of normalized curvature $k = +1$, $-1$ or $0$). From
$T_{0}^{0}= 0$ (the other field equations are equivalent to it) one
obtains
\begin{equation}
f^{\prime}\big\{r^{2}+4\alpha\,(k-f)\big\}=-2 r^{3}\frac{\Lambda}%
{3}+2r\,(k-f)\label{field eq}\, .
\end{equation}
This is integrated for $(k-f)$ to give
\begin{equation}\label{BD_metric}
f(r)=k+\frac{r^{2}}{4\alpha}\left(  1+\xi\sqrt{1+\frac
{4\Lambda\alpha}{3}+\frac{16 M \alpha}{r^{4}}}\right)
\end{equation}
where $\xi^{2}=1$. The case $\xi=+1$ corresponds to the ``exotic
branch'' of the Boulware-Deser metrics which for $\Lambda=0$ and
$M=0$ gives a ``microscopic'' anti-de Sitter or de Sitter metric,
with $f(r)=1+r^{2}/2\alpha$. It is usually argued that this exotic
branch turns out to be an unstable vacuum of the theory, containing
ghost excitations \cite{Boulware:1985wk,Zwiebach}. Unlike the case
$\xi =-1$, this branch does not have a well defined
$\alpha\rightarrow0$ limit. As in the case of Schwarzschild
solution, $M$ here is a constant of integration, and is also
associated with the mass of the solution. In fact, when there is an
asymptotic region at the infinity of the coordinate $r$, i.e.
$1+\frac {4\alpha\Lambda}{3}\geq0$, the total energy w.r.t. each
constant curvature background is calculated to be
\begin{equation}
\textrm{mass}=
%xi^2\, M\,\frac{6\pi^{2}}{\kappa^{2}}=
M\,\frac{6\pi^{2}}{\kappa^{2}}\,,\label{m}%
\end{equation}
so that, in general, we will call $M$ the mass parameter or simply
the mass of the metric\footnote{It should be kept in mind that the
masses $M$ in each branch $\xi$, by being the total energy w.r.t.
the $M=0$ spacetime in \emph{that branch}, can not be directly
compared.}.

The general features of the black holes
(\ref{ansatz})-(\ref{BD_metric}), such as horizons structure,
singularities, etc, were studied systematically in
Ref.~\cite{Torii-05}; for further details see the Appendix. Unlike
General Relativity, the Einstein-Gauss-Bonnet theory admits massive
solutions with no horizon but with a naked singularity at the
origin. From (\ref{BD_metric}) we see that this always happens for
the exotic branch $\xi=+1$, and might also happen for the branch
$\xi=-1$, provided $M<\alpha $. A related feature occurs for
electrically charged solutions \cite{Wiltshire, Wiltshire2}. Among
other interesting properties, it can be seen that charged black
holes in Einstein-Gauss-Bonnet theory have a single horizon if the
mass reaches a certain critical value. Another substantial
difference between the Schwarzschild solution and the Boulware-Deser
solution concerns thermodynamics. Unlike black holes in General
Relativity, the Einstein-Gauss-Bonnet black holes turn out to be
eternal. The thermal evaporation process leads to eternal remnants
due to a change of the sign in the specific heat for sufficiently
small black holes. This and the other unusual phenomena discussed
above are ultimately due to the ultraviolet corrections introduced
by the Gauss-Bonnet term.

The discussion about a spherically symmetric solution of a given
theory of gravity immediately raises the obvious question about its
uniqueness. Regarding this, there is a subtlety that deserves to be
pointed out. The uniqueness of the Boulware-Deser solution,
discussed previously in Refs. \cite{Wheeler, Charmousis} (see
\cite{Aliev:2007dp} for a uniqueness result in axi-dilaton gravity
with Gauss-Bonnet term), is only valid under certain assumptions.
This was formalized in a theorem proven by R. Zegers \cite{Z}, and
which also holds for generic Lovelock theory in any dimension. Let
us state the result as applies for Einstein-Gauss-Bonnet theory in
five dimensions:
\begin{theorem}[Ref.~\cite{Z}]
%\begin{theorem}[Zegers]
Any solution with spherical (or planar or hyperbolic) symmetry in
the second-order Einstein-Gauss-Bonnet theory of gravity has to be
locally static and given by the Boulware-Deser solution provided two
key conditions are satisfied: i) The coefficients of the Lovelock
expansion are generic enough, which means that the exceptional
combination $\alpha\Lambda =-3/4$ is excluded; ii) the solution is
$C^2$ smooth.
\end{theorem}
Condition {\it i)} is certainly a necessary assumption. Indeed, the
non-uniqueness in the case of $\alpha\Lambda =-3/4$, corresponding
to the (A)dS-invariant Chern-Simons theory, is a well-known result
and was explicitly shown in Refs. \cite{Charmousis,GOT07}. In this
paper, we will see that condition {\it ii)} is also necessary. In
fact, the vacuum shell solutions we will present are $C^0$
spacetimes which are only piecewise of the Boulware-Deser form.

In order to analyze $C^0$ solutions, we will need to use the
junction conditions in the theory, which will now be discussed.

\subsection{Junction Conditions}\label{the sting}

The next ingredient in our discussion is the junction conditions in
Einstein-Gauss-Bonnet theory. These are the analogues of the Israel
conditions~\cite{I} in General Relativity, and were worked out in
Refs.~\cite{Davis:2002gn,Gravanis:2002wy}. In particular, the
junction conditions will be employed to join two different
spherically symmetric spaces.

We will organize the discussion as follows: First, we will discuss
the timelike junction condition; namely, the case where the surgery
is performed on a timelike hypersurface, which we shall call a
timelike shell. After studying this we will briefly discuss its
spacelike analogue.

\subsubsection{Timelike shell}

Let $\Sigma$ be a timelike hypersurface separating two bulk regions
of spacetime, region ${\cal V}_L$ and region ${\cal V}_R$ (``left" and ``right").
Conveniently, we introduce the coordinates ($t_L,r_L$) and ($t_R,r_R$) and the
metrics
\begin{equation}
ds_{L}^{2}=-f_{L}\,dt_{L}^{2}+\frac{dr_{L}^{2}}{f_{L}}+r_{L}^{2}d\Omega
^{2}\, ,\label{bulk L}%
\end{equation}
\begin{equation}
ds_{R}^{2}=-f_{R}\,
dt_{R}^{2}+\frac{dr_{R}^{2}}{f_{R}}+r_{R}^{2}d\Omega ^{2}\label{bulk
R}\, ,
\end{equation}
in the respective regions. We shall be interested in the case where
the bulk regions are empty of matter so $f_L(r_L)$ and $f_R(r_R)$
are the Boulware-Deser metric functions given by equation
(\ref{BD_metric}). In general, the mass parameter $M_R$ will be
different from $M_L$. Moreover, we will also consider the
possibility of having $\xi _R$ different from $\xi _L$, so that the
two different branches of the Boulware-Deser solution can be
considered to the two spaces to be joined.

It is convenient to parameterize the shell's motion in the $r-t$
plane using the proper time $\tau$ on $\Sigma$. In region ${\cal
V}_L$ we have $r_{L}=a(\tau)$, $t_{L}=T_{L}(\tau)$ and in region
${\cal V}_R$ we have $r_{R}=a(\tau)$, $t_{R}=T_{R}(\tau)$. The
induced metric on $\Sigma$ induced from region ${\cal V}_L$ is the
same as
that induced from region ${\cal V}_R$, and is given by%
\begin{equation}
d\hat{s}^2=-d\tau^{2}+a(\tau)^{2}d\Omega^{2}\, .\label{4-geometry}%
\end{equation}
 This guarantees the existence of a coordinate system where the
metric is continuous ($C^0$).

Here, $d\Omega^{2}$ will be chosen to be the line element of a
3-manifold with (intrinsic) curvature $k=\pm1,0$ i.e. it is a unit
sphere, a hyperboloid or flat space respectively. Although our
interest will be mainly focused on the spherical shell similar
features to those we will discuss hold also for the cases $k=0$
and $k=-1$. The hypersurface $\Sigma$ is the shell's world-volume,
i.e. the four-dimensional history of the shell in spacetime. The
intrinsic geometry is well defined on $\Sigma$ and given by
(\ref{4-geometry}). However, since the metric is only $C^0$ and
not necessarily differentiable, the geometry of the embedding of
$\Sigma$ into ${\cal V}_{L}$ is independent of the embedding of
$\Sigma$ into ${\cal V}_R$. The geometric information about the
embedding is quantified by the extrinsic curvature as well as the
orientation of $\Sigma$ with respect to each bulk region.

To be precise, let us consider the following conventions for a
timelike shell outside of any event horizon:
\begin{itemize}
\item The hypersurface $\Sigma$ has a single unit normal vector
$\bm{n}$
 which points from left to
right.

\item The orientation factor $\eta$ of each bulk region is defined
as follows: $\eta = +1$ if the radial coordinate $r$ points from
left to right, while $\eta = -1$ if the radial coordinate $r$
points from right to left.

\end{itemize}

\begin{figure}[t]
\begin{center}
  \includegraphics[height=\textwidth,  angle=270]{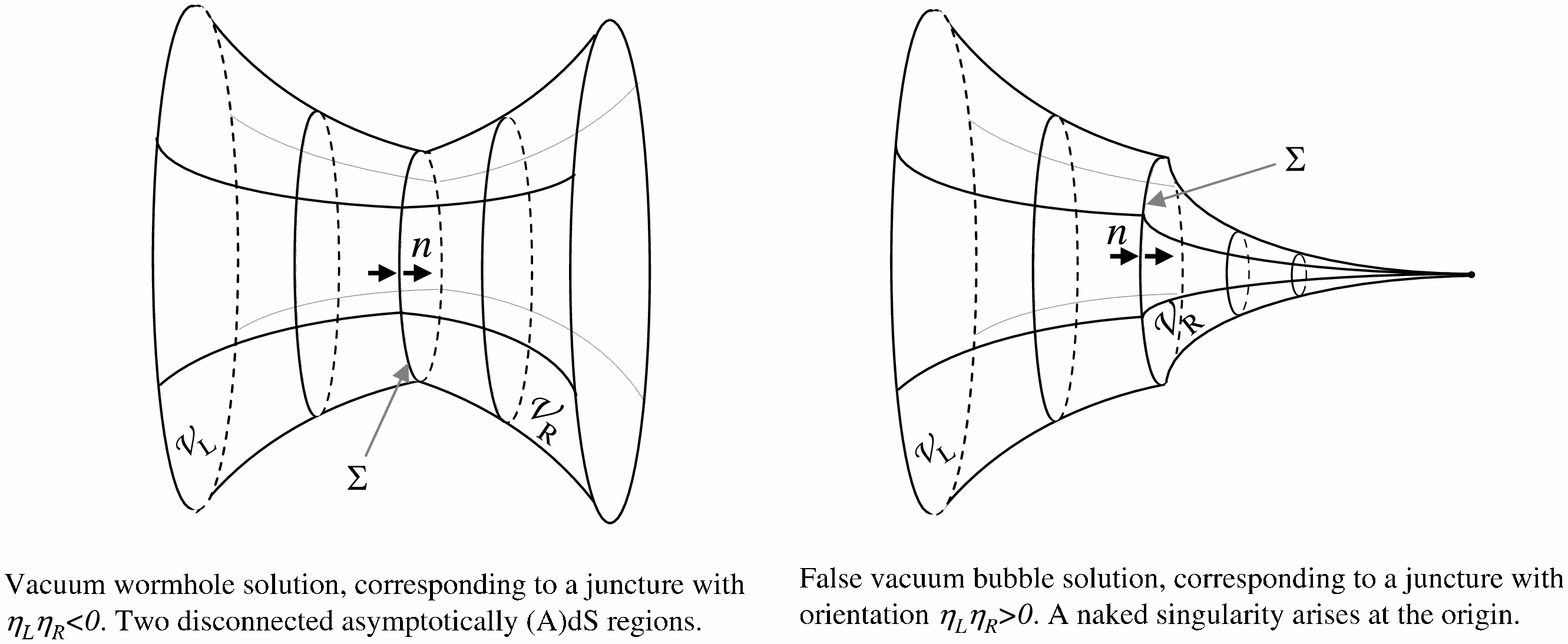}
  \\
  \caption{The figure on the left corresponds to a wormhole-like solution,
  defining the orientation $\eta _L \eta_R <0$.
  The throat connects two different asymptotically (Anti) de-Sitter spaces.
  The figure on the right corresponds to a vacuum shell with standard orientation
  ($\eta _L \eta_R >0$), connecting
  two Boulware-Deser solutions of different branches.
  }\label{Wormhole1}
  \end{center}
\end{figure}

This is depicted in Fig. \ref{Wormhole1}. Notice that the wormhole
depicted on the left of that figure is not the only possibility for
$\eta_L \eta_R <0$. While this geometry roughly speaking corresponds
to joining two ``exterior regions'' of a spherical solution, it is
also feasible to construct a space by joining two ``interior
regions'', instead. This corresponds to the case $\eta_L \eta_R <0$
as well.

\begin{definition}\label{the definition}
 The orientation defined by $\eta_L\eta_R>0$ will be
called the standard orientation. A shell with standard orientation
will be called a standard shell. The orientation defined by
$\eta_L\eta_R<0$ will be called the wormhole orientation. $[$This
makes actual sense when $\eta_R=+1$. When $\eta_R=-1$ the latter
case represents a closed universe, containing singularities.$]$
\end{definition}

The components of the normal vector with respect to the basis
$\bm{e}_A := (\partial_{ t_L},\partial_{r_L},
\bm{e}_\theta,\bm{e}_\chi,\bm{e}_\varphi)$ of ${\cal V_L}$ and the
basis $\bm{e}_{A^{\prime}} := (\partial_{t_R},\partial_{r_R},
\bm{e}_\theta,\bm{e}_\chi,\bm{e}_\varphi)$ of ${\cal V_R}$ are
respectively given by
\begin{equation*}
n^A = \eta_{L}\left( \frac{\dot a}{f_{L}} , \sqrt{f_{L} + \dot
a^{2}},0,0,0\right)\, , \qquad n^{A^{\prime}} = \eta_{R}\left(
\frac{\dot a}{f_{R}} , \sqrt{f_{R} + \dot a^{2}},0,0,0\right)\, .
\end{equation*}
 where dot denotes differentiation with respect
to $\tau$. This formula for the normal vector extends the definition
of the orientation factors to the situation where the shell is
inside the horizon when $r$ is a timelike coordinate.

We can introduce the basis $\bm{e}_a =
(\partial_\tau,\bm{e}_\theta,\bm{e}_\chi,\bm{e}_\varphi)$ intrinsic
to $\Sigma$. The extrinsic curvature is then defined as $K_{ab} :=
\bm{e}_{a}\cdot \nabla_{\bm{e}_b} \bm{n} = - \bm{n} \cdot
\nabla_{\bm{e}_b} \bm{e}_{a}$. In terms of a coordinate basis we
have $e^A_a =\frac{\partial X^A}{\partial \zeta^a}$ and the
extrinsic curvature takes the explicit form
\[
K_{ab} = - n_A \left( \frac{\partial^2 X^A}{\partial
\zeta^a\partial \zeta^b} + \Gamma^A_{BC}\frac{\partial
X^B}{\partial \zeta^a}\frac{\partial X^C}{\partial \zeta^b}\right)
,
\]
and in our case the components read
\begin{gather}\label{Ks_Timelike}
K_{\tau}^{\tau}   = \eta\frac{\ddot{a}+\frac{1}{2}f^{\prime}}%
{\sqrt{\dot{a}^{2}+f}}\, , \qquad
K_{\theta}^{\theta}   = K_{\chi}^{\chi}= K_\varphi^\varphi = \frac{\eta}{a}%
\sqrt{\dot{a}^{2}+f} \, .
\end{gather}
We denote the extrinsic curvature with respect to the embedding into
${\cal V}_L$ and ${\cal V}_R$ by $(K_L)_{ab}$ and $(K_R)_{ab}$
respectively. At a singular shell $(K_L)_{ab} \neq (K_R)_{ab}$, i.e.
the extrinsic curvature jumps from one side to the other. This is a
covariant way of expressing the fact that the metric is not $C^1$
(i.e. there does not exist any coordinate system where the metric is
$C^1$). In General Relativity this amounts to saying that (non-null)
vacuum shells do not exist since Israel conditions cannot be
satisfied without the introduction of a induced stress tensor on the
spherical shell. Things are different in the case of the gravity
theory defined by action (\ref{The_Action}). This is because the
Gauss-Bonnet term induces additional terms in the junction
conditions, which supplements the Israel equation. In section
\ref{Junction_section} we will show how both contributions can be
combined to yield vacuum spherically symmetric thin shells. First we
briefly discuss spacelike shells.

\subsubsection{Spacelike shell}

Solutions of a different sort are those constructed by joining two
spaces through a spacelike juncture. Let us suppose now that
$\Sigma$ is now a spacelike hypersurface. The motion of the shell in
the $r-t$ plane is parameterized by $(t,r)=(T(\tau),a(\tau))$, where
it is necessary to remember that $\tau$ is now a spacelike
coordinate on $\Sigma$. The induced metric on $\Sigma$ is then given
by
\begin{equation}
 d\hat{s}^2 = +d\tau^{2}+a(\tau)^{2}d\Omega^{2}\, .\label{4-geometry-spacelike}%
\end{equation}
The components of the normal vector with respect to the basis
$\bm{e}_A := (\partial_{ t_L},\partial_{r_L},
\bm{e}_\theta,\bm{e}_\chi,\bm{e}_\varphi)$ of ${\cal V}_L$ and the
basis $\bm{e}_{A^{\prime}} := (\partial_{t_R},\partial_{r_R},
\bm{e}_\theta,\bm{e}_\chi,\bm{e}_\varphi)$ of ${\cal V}_R$ are
respectively:
\begin{equation*}
n^A = \eta_{L}\left( \frac{\dot a}{f_{L}} , \sqrt{\dot a^{2}-f_{L}
},0,0,0\right)\, , \qquad n^{A^{\prime}} = \eta_{R}\left( \frac{\dot
a}{f_{R}} , \sqrt{\dot a^{2}-f_{R}},0,0,0\right)\, .
\end{equation*}
This defines the orientation factors in the case of a spacelike
shell. The components of the extrinsic curvature are:
\begin{gather}\label{Ks_Spacelike}
K_{\tau}^{\tau}   = \eta\frac{\ddot{a}-\frac{1}{2}f^{\prime}}%
{\sqrt{\dot{a}^{2}-f}}\, , \qquad
K_{\theta}^{\theta}   =K_{\phi}^{\phi}=K_{\chi}^{\chi}= \frac{\eta}{a}%
\sqrt{\dot{a}^{2}-f} \, .
\end{gather}

\subsection{The junction condition for a vacuum
shell}\label{Junction_section}

The Einstein-Gauss-Bonnet field equations are well-defined
distributionally at $\Sigma$ due to the property of quasi-linearity
in second derivatives (see e.g.
Refs~\cite{Deruelle-03,Gravanis:2007ei} ). Thus, one can define a
distributional stress tensor $T_{AB} = \delta(\Sigma)\, e_A^b e_B^b
S_{ab}$, where $S_{ab}$ is the intrinsic stress tensor induced on
the shell and $\delta(\Sigma)$ denotes a Dirac delta function with
support on the shell world-volume $\Sigma$.

Integrating the field equations from left to right in an
infinitesimally thin region across $\Sigma$ one obtains the junction
condition. This relates the discontinuous change of spacetime
geometry across $\Sigma$ with the stress tensor $S_{a}^{b}$. For the
Einstein-Gauss-Bonnet theory the general formulas can be found in
the Refs.~\cite{Davis:2002gn,Gravanis:2002wy,Gravanis:2007ei}.
\begin{equation}
\label{explicit junction ij}   ({\frak Q}_R)_{a}^{b}-({\frak
Q}_L)_{a}^{b} =- \kappa^{2} S_{a}^{b}\ ,
\end{equation}
where the symmetric tensor ${\frak Q}^a_{\ b}$ is given by
\begin{gather}\label{explicit_Q}
 {\frak Q}^a_{\ b} =  \mp \delta^{ac}_{bd} K^d_c +\alpha\, \delta^{acde}_{bfgh}
 \Big( \mp  K^f_c R^{gh}_{\ \ de}
 + \frac{2}{3} K^f_cK_d^gK_e^h \Big)\, .
\end{gather}
Above, the sign $\mp $ depends on the signature of the junction
hypersurface: it is minus for the timelike case and plus for the
spacelike case. In this expression, lower case Roman letters from
the beginning of the alphabet $a$, $b$ etc. represent
four-dimensional tensor indices on the tangent space of the
world-volume of the shell. The symbol $K^a_b$ refers to the
extrinsic curvature, while the symbol $R^{ab}_{\ \ cd}$ appearing
here corresponds to the four-dimensional intrinsic curvature (see
the appendix for details).

Once applied to the spherically symmetric (or $k=-1$, $0$) case
the tensor ${\frak Q}_{a}^{b}$ turns out to be diagonal with
components
\begin{align}
{\frak Q}^{\tau}_{\tau} & = -3 \sigma  \ a^{-3}   \bigg( \eta\
a^{2}\, \sqrt{\dot a^{2}+f} +\, 4 \alpha\, \eta\,\sqrt{\dot a^{2}+f}
\ \big(k+\frac{2}{3}\sigma \dot a^{2}-\frac
{1}{3}f\big) \bigg) \ ,\label{Qtt}\\
{\frak Q}_\theta^\theta & = {\frak Q}_\chi^\chi = {\frak
Q}_\varphi^\varphi\, .
\end{align}
The precise form of ${\frak Q}_\theta^\theta$ will not be needed
but is given in the appendix for completeness. The above formula
was written in a way that is valid for both timelike and spacelike
shells, where we have defined
\begin{equation*}
  \sigma = +1\quad \text{(timelike shell)}\, , \qquad \sigma =-1\, \quad
  \text{(spacelike shell)}\, .
\end{equation*}
Also, let us be reminded of the fact that $\eta_L$ and $\eta_R$
(with $\eta^2 =1$) are the orientation factors in each region, which
are independent one from each other. Above, the subscripts $L$, $R$
signify the quantity evaluated on $\Sigma$ induced by regions ${\cal
V}_L$ and ${\cal V}_R$ respectively (e.g. ${\frak Q}_L$ is a
function of $\eta_L$ and $f_L(a)$)\footnote{From now on we shall be
concerned with $f(a)$, i.e. the metric function evaluated at the
shell. In an abuse of notation we shall just write $f$ instead of
$f(a)$.}.

One may verify that the following equation is satisfied
\begin{equation}
\label{Q conservation}\frac{d}{d\tau}\big(a^{3} {\frak Q}^{
\tau}_{\tau}\big)=\dot a \, 3 a^{2} {\frak Q}_{\theta}^{\theta}\ ,
\end{equation}
which expresses the conservation of $S_{a}^{b}$. The reason why one
obtains exact conservation, i.e. no energy flow to the bulk, is that
the normal-tangential components of the energy tensor in the bulk is
the same in both sides of the junction
hypersurface~\cite{Davis:2002gn,Gravanis:2007ei}.

The main point here is that, unlike the Israel conditions in
Einstein gravity, non-trivial solutions to (\ref{explicit junction
ij}) are possible even when $S_{a}^{b}=0$. That is, the extrinsic
curvature can be discontinuous across $\Sigma$ with no matter on the
shell to serve as a source. The discontinuity is then self-supported
gravitationally and this is due to non-trivial cancelations between
the terms of the junction conditions. Similar configurations are
impossible in Einstein gravity. From now on we consider the vacuum
case
\begin{gather}
S_{a}^{b}=0\ .
\end{gather}

In the next section we will treat the static shell in detail. An
exhaustive study of the space of solutions describing both static
and dynamical shells is left until sections 5 and 6. Let us now
first  briefly introduce the basic features of the general solution
for a dynamical shell.

Equation (\ref{Q conservation}) tells us that when $\dot a \ne0$,
the components of the junction condition are not independent;
namely
\begin{equation*}
({\frak Q}_{R})_{\tau}^{\tau}-({\frak Q}_{L})_{\tau}^{\tau}=0
\quad\Rightarrow\quad ({\frak Q}_{R})_{\theta}^{\theta}-({\frak
Q}_{L})_{\theta}^{\theta}=0 \ .
\end{equation*}
Therefore, for time-dependent solutions it suffices to impose only
the first condition. This can be factorized as follows,
\begin{align}
&\left(\eta_R \sqrt{\dot{a}^{2}+ \sigma f_R } - \eta_L
\sqrt{\dot{a}^{2}+ \sigma f_L }\right)\times    \nonumber
\\&\qquad \times\left\{ a^{2} +
4\alpha(k+\sigma \dot{a}^{2}) -\sigma \frac{4\alpha}{3} \left( f_{R}
+ f_{L} + 2\sigma \dot
{a}^{2} + \eta_{R} \eta_{L} \sqrt{\sigma f_{R} +\dot{a}^{2}} \sqrt{\sigma f_{L} +\dot{a}%
^{2}}\right) \right\}  =0\, .    \label{explicit general junction}
\end{align}

Equation (\ref{explicit general junction}) contains all the
information about the spherically symmetric junctions in empty
space, which we generically call vacuum shells. Certainly, there
exist several cases to be explored. First of all, there are the
parameters $k,\ M$ and $\xi $, which characterize each of the two
Boulware-Deser metrics to be joined. On the other hand, there are
two possible orientations for each one of the spaces, and this is
given by the sign of the respective $\eta $. The solutions to
(\ref{explicit general junction}) include both wormhole-like and
bubble-like geometries, depending on whether the orientation is $
\eta_L \eta_R <0$ or $\eta_L \eta_R>0$ respectively.
 Furthermore, there is
the sign of $\sigma $, what tells us whether the signature of the
junction hypersurface is timelike ($\sigma =+1$) or spacelike
($\sigma =-1$). So, this permits a very interesting catalogue of
geometries which we survey in section \ref{Surveying static vacuum
solutions} and further explore in subsequent sections.

The vanishing of the first factor in (\ref{explicit general
junction}) would imply that the metric is smooth across $\Sigma$.
Rejecting this as the trivial solution, we demand that the second
factor vanishes. From the second factor, squaring appropriately, we
obtain
\begin{equation}
\label{solved-explicit} \dot a^{2}=\ \sigma \,
\frac{\Big(f_{R}+f_{L} -3(k+a^{2}/4\alpha) \Big)^{2}-f_{R}f_{L}} {3
\Big(f_{R}+f_{L}-2(k+a^{2}/4\alpha) \Big)}=:- V(a) \, ,
\end{equation}
along with two inequalities discussed below. The system, because of
the symmetry, has reduced to an essentially one-dimensional problem,
given by the ordinary differential equation (\ref{solved-explicit}).
It is seemingly equivalent to the problem of a particle moving in a
potential\footnote{Notice that the effective potential for the
spacelike shell is simply minus the potential for the timelike
shell.} $V(a)$. Nevertheless, it is worth pointing out that, unlike
the equation for a single particle, here we find that the energy $h$
is unavoidably fixed to zero instead of arising as a constant of
motion. An important difference arises in the case where there is a
minimum of $V(a)$ precisely at $V =0$. The constraint $h=0$,
provided the fact that the minimum of $V(a)$ is precisely at zero
energy, would lead to the conclusion that the shell can not move but
it would be stacked at the bottom of the potential. Actually, this
is the case if no external system acts as a perturbation. One such
perturbation can be thought of as being an incoming particle which,
after perturbing the shell, scatters back to infinity spending an
energy $\delta h$ through the process. This would provide energy for
the vacuum shell to move. One can also think about a slight change
in the parameters of the solution yielding a shifting $V(a)\to
V(a)-\delta h$, see ~\cite{VisserWiltshire}.

Now, let us notice that since we have squared the junction
condition, we must substitute (\ref{solved-explicit}) back into
(\ref{explicit general junction}) to check the consistency. When
doing so, the solutions of equation (\ref{solved-explicit}) are
solutions of the junction condition if and only if the following
restrictions are obeyed
\begin{equation}
-\eta _{R}\eta _{L}\ (2f_{R}+f_{L}-3(k+a^{2}/4\alpha ))\
(2f_{L}+f_{R}-3(k+a^{2}/4\alpha ))\geq 0\ ; \label{timed ineq}
\end{equation}
and
\begin{align}
\label{timelike_real_roots} &(f_{R} + f_{L} -2(k+a^{2}/4\alpha) ) >
0\  & \text{timelike shell}\; \\
\label{spacelike real roots} &(f_{R} + f_{L} -2(k+a^{2}/4\alpha) )
<0\  & \text{spacelike shell}\, .
\end{align}

Furthermore, we also have an inequality which is not an extra
condition but rather follows as a consequence of equation
(\ref{solved-explicit}). The fact that $\dot a^2$ is positive in
(\ref{solved-explicit}) implies that
\begin{gather}
\label{ineq 2} \Big(f_{R}+f_{L} -3(k+a^{2}/4\alpha)
\Big)^{2}-f_{R}f_{L} \geq0\ .
\end{gather}
for both timelike and spacelike. This inequality provides further
information about the space of solutions of
(\ref{solved-explicit}).

\begin{proposition}\label{general_solution}
For a dynamical vacuum shell with a timelike world-volume $\Sigma$,
the scale factor of the metric (\ref{4-geometry}) on $\Sigma$ is
governed by (\ref{solved-explicit}), under the inequalities
(\ref{timed ineq}) and (\ref{timelike_real_roots}).
\\
On the other hand, for a dynamical vacuum shell with a spacelike
world-volume $\Sigma$, the scale factor of the metric
(\ref{4-geometry-spacelike}) on $\Sigma$ is governed by
(\ref{solved-explicit}), under the inequalities (\ref{timed ineq})
and (\ref{spacelike real roots}).
\end{proposition}
\vspace{.2in}

Now, let us begin by studying the inequalities to give idea of what
kinds of solutions exist. With this in mind, let us translate the
restrictive inequalities (\ref{timed ineq}-\ref{spacelike real
roots}) into simpler terms. The metric function evaluated on the
hypersurface is
\begin{gather}
 f_L(a) = k+\frac{a^2}{4\alpha}\Big( 1 + \xi_L Y_L(a) \Big)
 \, ,\qquad Y_L(a)\equiv\sqrt{1+\frac{4\alpha\Lambda}{3}+\frac{16\alpha
M_L}{a^4}}\ ,
\end{gather}
and similarly for $f_R$. Recall that $\xi_L$ and $\xi_R$ are
independent of each other, with $\xi =+1$ being the exotic branch of
the Boulware-Deser solution. It is convenient to write the
inequalities in terms of the square roots $Y_L(a)$ and $Y_R(a)$;
namely
\begin{equation}
-\eta _{R}\eta _{L}\ (2 \xi_R Y_R +  \xi_L Y_L )\ (2 \xi_L Y_L+
\xi_R Y_R)\geq 0\ ; \label{timed ineq_Ys}
\end{equation}
and
\begin{align}
\label{timelike_real_roots_Ys} &\alpha (\xi_R Y_R + \xi_L Y_L )
>
0\  & \text{timelike shell}\; \\
\label{spacelike real roots_Ys} &\alpha (\xi_R Y_R + \xi_L Y_L ) <0\
& \text{spacelike shell}\, .
\end{align}
These inequalities contain relevant information about the global
structure of the solutions. Let us summarize
this information in the following table\\

\begin{center}

\begin{tabular}{c||c|c|c|}

 \textbf{Timelike}& Product of     & Product of     &   Inequalities \\
 \textbf{shells} & orientation   &  branch signs & imposed on \\
    $(\sigma=+1)$   & factors ($\eta_L \eta_R$) & ($\xi_L \xi_R$) & solutions \\
     \hline\hline&              &                &  \\
  Standard    & +1             & +1             &  No solution \\
  orientation &                &                &                  \\
 & +1             & -1            & $\frac{1}{2}Y_L \le Y_R(a) \le 2 Y_L(a)\ ;$
  \\          &                 &               &    $\xi_R (M_R-M_L)>0$       \\\hline
             &                 &                &           \\
  ``Wormhole" & -1             & +1            & $\alpha \xi_R >0$ ; $Y_L , Y_R >0$\\
   orientation&                &                &    \\
              & -1             & -1             & $Y_R \geq 2Y_L\ $  or $\ Y_R \leq \frac{1}{2}Y_L$ ;\\
              &                &                &          $\xi_R (M_R-M_L)>0$    \\
  \hline
\end{tabular}

\begin{tabular}{c||c|c|c|}

 \textbf{Spacelike}& Product of     & Product of     &   Inequalities \\
 \textbf{shells} & orientation   &  branch signs & imposed on \\
    $(\sigma =-1)$   & factors ($\eta_L \eta_R$) & ($\xi_L \xi_R$) & solutions \\
     \hline\hline&              &                &  \\
  Standard    & +1             & +1             &  No solution \\
  orientation &                &                &                  \\
  & +1             & -1            & $\frac{1}{2}Y_L \le Y_R(a) \le 2 Y_L(a)\ ;$
  \\          &                 &               &   $\xi_R (M_R-M_L)>0$        \\\hline
             &                 &                &           \\
  ``Wormhole" & -1             & +1            & $\alpha \xi_R < 0$ ; $Y_L , Y_R >0$\\
   orientation&                &                &    \\
              & -1             & -1             & $Y_R \geq 2Y_L\ $  or $\ Y_R \leq\frac{1}{2}Y_L$ ;\\
              &                &                &          $\xi_R (M_R-M_L)<0$    \\
  \hline
\end{tabular}

\end{center}

\[
\]
From the conditions obtained here we conclude the following:

\begin{remark}\label{vacuum_bubble_remark}
 Vacuum shells with the standard orientation always
 involve the gluing of a plus branch $(\xi = +1)$ metric with a
 minus branch $(\xi = -1)$ metric.
\end{remark}
Now the plus branch has a different effective cosmological constant
to the minus branch. In this sense, standard shells are a kind of
false vacuum bubble. This is discussed further in section
\ref{Bubble_section}.

\begin{remark}\label{minus_minus_remark}
 Vacuum shells which involve the gluing of two
 minus branch $(\xi = -1)$ metrics exist only
 when the Gauss-Bonnet coupling constant $\alpha$ satisfies $\alpha<0$.
 They always have the wormhole orientation.
\end{remark}

In the analysis above it has been explicitly assumed that $\dot{a}
\neq 0$. Nevertheless, the case $\dot{a}=0$ is also of considerable
interest. This describes static shells in the timelike case, and
also an analogous situation for the spacelike case which we call
instantaneous shells. In the next section, the case of constant $a$
shells is considered in detail. It can be checked that, as expected,
all the information about the constant $a$ solutions can be obtained
from the dynamical case by imposing both $V(a_0)=0$ and $
V'(a_0)=0$. Thus, proposition \ref{general_solution} gives the
general solution of all the vacuum shells, including the static
ones.

Closing the general discussion of the dynamical vacuum we note the
following. The potential $V(a)$ in (\ref{solved-explicit}) and the
restrictive inequalities (\ref{timed ineq}) and
(\ref{timelike_real_roots}), (\ref{spacelike real roots}) are
symmetric in the exchange
\begin{equation}
\xi_L, M_L \leftrightarrow \xi_R, M_R\ .
\end{equation}
That is, the same kinds of motion are possible for the two
situations obtained if we swap the values of the parameters $\xi, M$
in $\mathcal{V}_L$ and $\mathcal{V}_R$. In the constant $a$ case,
governed by $V(a_0)=0=V'(a_0)$, the symmetry means that the value of
$a_0$ is left unchanged under the swapping.

%\begin{definition} The space of solutions of equations $V(a_0)=0$
%and $ V'(a_0)=0$  for the vacuum shell with timelike world sheet
%will be called the moduli space.
%\end{definition}

\section{Static vacuum shells}\label{Static_Section}

Now, let us discuss the solutions at constant $a$. That is, the
static and instantaneous solutions, depending on whether the
juncture corresponds to the timelike or spacelike case
respectively.

The bulk metric in each of the two region is assumed to be of the
Boulware-Deser form (\ref{BD_metric}) with ($k=\pm1, 0$) and
considering $a=a_0$ fixed. Although the main focus will be on the
spherically symmetric case $k=+1$, the analysis can be
straightforwardly extended to the cases $k=-1$ and $k=0$. Then,
there are two possibilities to be distinguished; namely,

\begin{itemize}

\item Static shell: For the timelike case the shell is located at
fixed radius $r_L= r_R=a_0$. The proper time on the shell's
world-volume is $\tau = t_L\sqrt{f_L(a)}= t_R\sqrt{f_R(a)}$ so that
the induced metric on $\Sigma$ turns out to be $d\hat{s}^2 =
-d\tau^2 + a_0^2 d\Omega^2$. Then, the extrinsic curvature
components are $ K_{\tau}^{\tau}   = \eta\frac{f^{\prime}}
{2\sqrt{f}}$, $K_{\theta}^{\theta}
=K_{\chi}^{\chi}=K_{\varphi}^{\varphi}= \frac{\eta \sqrt{f}}{a}$ and
the intrinsic curvature components are $R_{\ \
\theta\varphi}^{\theta\varphi} =k/a_0^{2}$, etc.

\item Instantaneous shell: In the spacelike case there is an
exotic kind of shell, which exists when $f$ is negative. The metric
function is negative inside of an event horizon or outside of a
cosmological horizon, where $r$ actually plays the role of  a
timelike coordinate. Matching two metrics at time $r_\pm=a_0$
therefore describes an instantaneous transition from one smooth
metric to another. We can introduce $\tau = t_L\sqrt{-f_L(a)}=
t_R\sqrt{-f_R(a)}$ which is a spacelike intrinsic coordinate on the
shell, so that the induced metric on $\Sigma$ is $ds^2 = +d\tau^2 +
a_0^2 d\Omega^2$. The extrinsic curvature components are $
K_{\tau}^{\tau} = -\eta\frac{f^{\prime}} {2\sqrt{-f}}$,
$K_{\theta}^{\theta} =K_{\chi}^{\chi}=K_{\varphi}^{\varphi}=
\frac{\eta \sqrt{-f}}{a}$.

\end{itemize}

It is worth noticing that both the static and instantaneous shells
can be analyzed together, provided the presence of $\sigma$ in the
equations. Recall that the sign of $\sigma $ carries the information
about the signature of the junction hypersurface. Then, by
considering the quantities introduced above, and by substituting
this in the junction conditions with $S^a_{b} = 0$, we get
\begin{align}
\label{static j 00}&S^\tau_\tau =0\ \Rightarrow &\big( \eta_{R}
\sqrt{f_{R}}- \eta_{L} \sqrt{f_{L}} \big)
\Big(a_{0}^{2}+\frac{4\alpha}{3} \big\{3k- f_{R} - f_{L} -
\sigma\eta_{L} \eta_{R} \sqrt{f_{L} f_{R}} \big\} \Big)=0\ .
\\
\label{static j ij}
&S^\theta_\theta = 0\ \Rightarrow \ &
\Big(\frac{\eta_{R}}{\sqrt{f_{R}}}-\frac{\eta_{L}}%
{\sqrt{f_{L}}}\Big) \Big(k-\frac{\Lambda
a_{0}^{2}}{3}-\sigma\eta_{L} \eta_{R} \sqrt{f_{L} f_{R}} \Big)=0\ ,
\end{align}
where $\sigma =+1$ is the static shell and $\sigma =-1$ is the
instantaneous shell. The l.h.s. of (\ref{field eq}) conveniently
appears in the $\theta-\theta$ component of the junction condition
and we have used it to eliminate the derivative of $f$ from the
formula. This is why $\Lambda$ appears explicitly in equation
(\ref{static j ij}).

In both equations (\ref{static j 00}) and (\ref{static j ij}), the
first factor vanishes if and only if the metric is smooth. Again,
rejecting this as the trivial solution, we demand that the second
factor vanishes in both equations.
%That is
%\begin{equation}
%f_{+} + f_{-} +\sigma \eta_+\eta_- \sqrt{f_{+} f_{-}}=3k+\frac{3
%a_{0}^{2}}{4 \alpha}\ , \ \ \ \sigma\eta_+\eta_-\sqrt{f_{+}
%f_{-}}=k-\frac{\Lambda a_{0}^{2}}{3}\ ,
%\end{equation}
So, we have
\begin{proposition}\label{static instant}
A static vacuum shell is described by
\begin{align}
f_{L} + f_{R} & =2k+\frac{3 a_{0}^{2}}{4 \alpha} + \frac{\Lambda
a_{0}^{2}}{3}\ ,\\
 \eta_L\eta_R\sqrt{f_{L} f_{R}} & =k-\frac{\Lambda
a_{0}^{2}}{3}\ ,
\end{align}
under the condition $f_L, f_R>0$. On the other hand, an
instantaneous vacuum shell is described by
\begin{align}
f_{L} + f_{R} & =2k+\frac{3 a_{0}^{2}}{4 \alpha} + \frac{\Lambda
a_{0}^{2}}{3}\ ,\\
 -\eta_L\eta_R\sqrt{f_{L} f_{R}} & =k-\frac{\Lambda
a_{0}^{2}}{3}\ ,
\end{align}under the condition $f_L, f_R<0$.
\end{proposition}

We have included for completeness the instantaneous shells. Now, let
us consider some examples of the static case with more attention. As
mentioned, a more complete analysis of the space of solutions will
be given in sections 5 and 6.

\subsection{The moduli space of solutions}

Now, to continue the study of the different solutions we find it
convenient to introduce some notation. For the rest of this section
it is convenient to define the dimensionless parameters
\begin{equation}\label{faithless}
x\equiv\frac{4\alpha\Lambda}{3}\, , \qquad
y\equiv\frac{\Lambda}{3}a_{0}^{2} \, , \qquad \bar{M}
\equiv\frac{M}{\alpha}\,. \
\end{equation}
By $x$ and $y$ we measure the Gauss-Bonnet coupling and the vacuum
shell radius respectively in units of $\Lambda$. The parameter $y$
is useful for our purposes but it is meaningful only when $\Lambda
\ne 0$. In terms of these parameters, the Boulware-Deser solution
evaluated at $r=a_{0}$ has the form
\begin{gather}\label{BD}
f_{L,R}(a_{0}) \equiv1+ \frac{y}{x} \left(  k + \xi_{L,R}%
\sqrt{1+x + \frac{x^2}{y^{2}}\bar{M}_{L,R}}\  \right) \, .
\end{gather}

The general solution will be derived in the following way: We will
solve the junction conditions for $\bar{M}_L$ and $\bar{M}_R$ in
terms of $(x,y)$. The range of admissible values of ($x,y$) turns
out to be restricted by inequalities coming from demanding the
metric to be real-valued. So there is a continuous space of
solutions.
\begin{definition}
 The range of values of $(x,y)$ for which solutions exist will be
 called the moduli space.
\end{definition}
The parameters $x$ and $y$ are coordinates of this moduli space. The
complete description of the moduli space will be given in more
appropriate parameters introduced in section 6. For the moment, let
us consider $x$, $y$ and $\bar {M}$.

Since the moduli space is two dimensional, it can be plotted. So by
obtaining a formula for the masses and by plotting the moduli space,
we obtain all the solutions. Let us now do this explicitly for the
case of non-vanishing cosmological constant.

\subsection{Static spherical shells with $\Lambda \neq 0$}\label{x y
shells section}

Consider static spherically symmetric shells with $\Lambda \neq 0$.
For definiteness, let us focus on the case of timelike shells with
$k=1$. From Proposition \ref{static instant} we have the following
pair of equations
\begin{align}
f_{L}+f_{R} &  =\frac{y}{x}(3+x)+2\ ,\label{F-F+}\\
\sqrt{f_{L}f_{R}} &  =\eta_L\eta_R\, (1-y)\ ,\label{F+F-}%
\end{align}
where $f_L,f_R>0$. We can see immediately from (\ref{F+F-}) that
solutions with the wormhole orientation, i.e. $\eta_L\eta_R =-1$,
only exist for $y\equiv \Lambda a_0^2/3 >1$.
\begin{remark}
 Static vacuum shell wormholes exist only when $\Lambda >0$.
\end{remark}
Solving the equations above we see that $f_L$ and $f_R$ obey the
same quadratic equation where one $f$ has the $+$ root of the
solution and the other has the $-$ root. So we define a solution
$f_{(+)}$ which corresponds to the $+$ root of the solution and an
$f_{(-)}$ which corresponds to the $-$ root. So there are two
solutions to the problem:
\begin{equation}\label{swap}
f_{L}=f_{(-)}\ \ , \ \ f_R=f_{(+)} \qquad  \text{or}, \qquad
f_{L}=f_{(+)}\ \ , \ \ f_{R}=f_{(-)}\ .
\end{equation}

Substituting the explicit expression (\ref{BD}) for $f_{L,R}(a_0)$
we have:  In the first case of (\ref{swap}), $M_L=M_{(-)}$,
$\xi_L=\xi_{(-)}$ and $M_R=M_{(+)}$, $\xi_R=\xi_{(+)}$, and in the
second case $+\leftrightarrow-$, for constants $\xi_{(\pm)}$ and
$M_{(\pm)}$ satisfying
\begin{gather}
1+x -\sqrt{3} \sqrt{x(1+x)\Big(\frac{4}{y}+\frac{3}{x}-1\Big)} =2
\xi_{(-)}
\sqrt{1+x+\frac{x^2 \bar M_{(-)}}{y^{2}}}\, ,\label{General_Solution_mess_1}%
\\
1+x + \sqrt{3} \sqrt{x(1+x)\Big(\frac{4}{y}+\frac{3}{x}-1\Big)} =2
\xi_{(+)}
\sqrt{1+x+\frac{x^2 \bar M_{(+)}}{y^{2}}}\, .\label{General_Solution_mess_2}%
\end{gather}

For a solution to exist, the square root in the l.h.s. of the above
equations must be real, so that we demand
\begin{equation}\label{reality condition}
 x(1+x)\Big(\frac{4}{y}+\frac{3}{x}-1\Big) \ge 0 \qquad
 \text{(Existence of solutions)}\, .
\end{equation}
Since we have squared the equations we must substitute back to check
the consistency. So we get the following inequalities:\footnote{Note
that the timelike condition (\ref{causality condition}), when
combined with the reality condition (\ref{reality condition}) can be
equivalently stated
\begin{equation*}
 x y (1+x)>0 \qquad \text{(Timelike shell)}\, .
\end{equation*}
This is useful for plotting the graphs.}
\begin{equation}\label{causality condition}
 \frac{y}{x}(3+x)+2>0 \qquad
 \text{(Timelike shells)}\, ;
\end{equation}
\begin{equation}\label{orientation condition}
 y<1 \quad \text{(Standard orientation)}\,, \qquad  y>1 \qquad
 \text{(Wormhole orientation)}\, .
\end{equation}
The above inequalities are plotted in figures
\ref{SpacelikeAndTimelike_fig} and \ref{WormholeAndPlain_fig}. Also
we find the regions of the moduli space corresponding to the allowed
branch signs $(\xi_{(-)}, \xi_{(+)})$.
\begin{center}
\begin{tabular}{|c|c|c|}
  \hline
  $\xi_{(-)}$ & $\xi_{(+)}$ & Inequality \\\hline
  +1 &  &  $1+x >0\ \bigcap\ x\left(\frac{3}{y}+\frac{2}{x}-1\right)<0$\\
  $-1$ &  &  $1+x <0\ \bigcup\ x\left(\frac{3}{y}+\frac{2}{x}-1\right)>0$\\
   & +1 &  $1+x >0\ \bigcup\ x\left(\frac{3}{y}+\frac{2}{x}-1\right)<0$\\
   & $-1$ & $1+x <0\ \bigcap\ x\left(\frac{3}{y}+\frac{2}{x}-1\right)>0$ \\
  \hline
\end{tabular}\quad
(Branches).
\end{center}
The standard shells are always $(-,+)$. The regions $(+,+)$, $(-,+)$
and $(-,-)$ for the wormholes are shown in figure
\ref{BranchesTimelikeWormhol_fig}.

\begin{remark}
Provided $\Lambda >0$ and assuming the existence of two asymptotic
regions we find $\alpha >0$. Consequently, at least one of the two
spherically symmetric spaces connected through the throat turns out
to be asymptotically Anti-de Sitter.
\end{remark}

The next step is computing the masses. We can solve
(\ref{General_Solution_mess_1}) and (\ref{General_Solution_mess_2})
to give the parameter $M$ in each region, namely
\begin{align}
 \bar M_{(-)} = \frac{y(1+x)}{2x^2}\left\{ 6x +3y -xy - y\sqrt{3}\sqrt{x(1+x)\Big(\frac{4}{y}+\frac{3}{x}-1\Big)}
 \ \right\}\, ,\label{Mass_formula_1}
 \\
 \bar M_{(+)} = \frac{y(1+x)}{2x^2}\left\{ 6x +3y -xy + y\sqrt{3}\sqrt{x(1+x)
 \Big(\frac{4}{y}+\frac{3}{x}-1\Big)} \ \right\}\,
 .\label{Mass_formula_2}
\end{align}
As mentioned above, relations (\ref{swap}), the left-metric can be
either a metric with parameters $(\xi_{(-)},M_{(-)})$ or a metric
with  $(\xi_{(+)},M_{(+)})$, and the other way around for the
right-metric. For wormholes the two solutions (\ref{swap})
correspond to the same spacetime looked at from the opposite way
around. In the case of standard shells, they correspond to swapping
the mass and branch sign of the interior with those of the exterior
region.

The metrics with parameters $(\xi_{(-)},M_{(-)})$ and
$(\xi_{(+)},M_{(+)})$ as determined by the solutions we found above
have different properties. We will call these metrics minus- and
plus-metrics respectively.

% And these can be clearly written as
%\begin{equation*}
% \bar M_{R,L} =
% \frac{y^2(1+x)}{2x}\left( \frac{4}{y} + \frac{3}{x} - 1\right)
% + \frac{y(1+x)}{x} \pm \frac{y^2(1+x)}{2x^2}
% \sqrt{3}\sqrt{x(1+x)\Big(\frac{4}{y}+\frac{3}{x}-1\Big)}\, ,
%\end{equation*}
Also we note the following useful expression: we can eliminate $y$
to get an implicit equation for the masses and $x$. The solution
lies on sections of the curves
\begin{align}
 1+ \frac{9}{4} \frac{x(x+1)}{3-x} \frac{1}{\bar{M}_{(+)} + \bar{M}_{(-)}}
 \left( 1+(-1)^p \sqrt{1 + \frac{4}{9}\frac{(3-x)}{x(x+1)}(\bar{M}_{(+)}+\bar{M}_{(-)})}
 \right)\qquad \nonumber\\ = (-1)^q \sqrt{1+\frac{(3-x)}{(1+x)}\frac{(\bar{M}_{(+)}
 -\bar{M}_{(-)})^2}{(\bar{M}_{(+)}
 +\bar{M}_{(-)})^2}}
\end{align}
where the signs $(-1)^p$ and $(-1)^q$ are to be determined by
consistency.

\subsection{Instantaneous shells}
Before concluding this section, let us briefly comment on spacelike
junction conditions with $\dot{a}=0$. For instance, consider the
case $\Lambda \neq 0$. From Proposition \ref{static instant} we have
the following pair of equations:
\begin{align}
 f_{L}+f_{R} &  =\frac{y}{x}(3+x)+2\ ,\label{F-F+_spacelike}\\
 \sqrt{f_{L}f_{R}} &  = -\eta_L\eta_R\, (1-y)\ ,\label{F+F-_spacelike}%
\end{align}
The solution is exactly the same as the above except that the
inequalities (\ref{causality condition}) and (\ref{orientation
condition}) are reversed. That means
\begin{equation}\label{causality condition_spacelike}
 \frac{y}{x}(3+x)+2<0 \qquad
 \text{(Spacelike shells)}\, ;
\end{equation}
\begin{equation}\label{orientation condition_spacelike}
 y>1 \quad \text{(Standard orientation)}\,, \qquad  y<1 \qquad
 \text{(Wormhole orientation)}\, .
\end{equation}
The inequality (\ref{reality condition}) and mass formulae are the
same. The moduli space of these solutions is plotted in figure
\ref{SpacelikeRegions_fig}. They exist for $\alpha<0$.

\subsection{Static spherical shells with $\Lambda =0$}

Now, we will consider the case of static spherically symmetric
shells with $\Lambda =0$. This is an interesting special case. The
analysis simplifies considerably and, besides, there are some
qualitative differences between this and the case $\Lambda \neq 0$.
In this case, the equations reduce to
\begin{align}
f_{L} + f_{R} & =2+\frac{3 a_{0}^{2}}{4 \alpha}\ ,\\
 \eta_L\eta_R\sqrt{f_{L} f_{R}} & =1\ ,
\end{align}
We see from the second equation that $\eta_L \eta_R$ must be $+1$,
i.e. static wormholes do not exist for $\Lambda =0$. Then, the
solution is either $M_L = M_{(-)}, M_R = M_{(+)}$ or $M_L = M_{(+)},
M_R = M_{(-)}$ where
\begin{align}
 \frac{M_{(\pm)}}{\alpha} & = \frac{1}{2} \frac{ a_{0}^{2}}{4 \alpha}  \left(
 6 + \frac{3 a_{0}^{2}}{4 \alpha} \pm \sqrt{12 \frac{ a_{0}^{2}}{4 \alpha}
 + 9\left(\frac{ a_{0}^{2}}{4 \alpha}\right)^2}\right)\,  .
\end{align}
The consistency of the solution requires
\begin{equation}
 \qquad \alpha >0\, , \qquad (\xi_{(-)}, \xi_{(+)}) = (-1,+1)\, ,
\end{equation}
so that $M_{(-)}$ and $M_{(+)}$ correspond to minus branch and
exotic plus branch metrics respectively. There are solutions for all
positive values of $M_{(-)}$ (the plus branch mass parameter is also
positive but in that case the bulk spacetime asymptotically takes
the form of a negative mass AdS-Schwarzschild solution). When the
throat radius is small compared to the scale set by the Gauss-Bonnet
coupling constant, $a_0^2 << \alpha$, the masses are also small
compared to $\alpha$, namely $M_{(-)}/\alpha \sim M_{(+)}/\alpha
\sim 3a_0^2/ 4\alpha$. On the other hand, for large radius $a_0^2
>> \alpha$, the masses are large, $M_{(-)}/\alpha \sim
a_0^2/2\alpha$, $M_{(+)}/\alpha \sim 3a_0^4/ 16\alpha^2$. Figure
\ref{ZeroLmassPics} shows a plot of the masses as a function of
$\alpha$ and also an implicit plot of ${M}_{(+)}$ as a function of
$M_{(-)}$.

\section{Surveying static vacuum solutions}\label{Surveying static vacuum
solutions}

In the previous section we have shown the existence of static vacuum
shells in the spherically symmetric case and found some basic
qualitative features, as well as a formula for the mass parameters
in each region. A more exhaustive treatment of the static shells
will be left for section \ref{Moduli_section}. Before going any
further let us summarize the catalogue of vacuum solutions that
arise through the geometric surgery we described above. The first
cases of interest are those corresponding to the standard
orientation $\eta _L \eta _R >0$.

\subsection{Standard shells and false vacuum
bubbles}\label{Bubble_section}

The vacuum shells with the standard orientation are always
$(\xi_{(-)} ,\xi _{(+)})=(-1,+1)$ branch. So region ${\cal V}_L$ has
a different effective cosmological constant to region ${\cal V}_R$,
as can be seen from the expansion of the metric for large $r$. For
example, when the bare cosmological constant $\Lambda =0$ we have on
one side of the shell the effective cosmological constant
$\Lambda_{d}^{(+)} = -3 /2\alpha$ and on the other
$\Lambda_{d}^{(-)} = 0$. In the region with $\Lambda_{d}^{(+)}$ the
graviton is expected to have ghost instability. In this sense the
shell is like the false vacuum bubbles\footnote{Strictly speaking,
this label of false vacuum bubble would be correct if the minus
branch metric were lower total energy with respect to the plus
branch metric and if the classical transition were impossible.}
studied in Refs. \cite{Vacuum bubbles}, but for a false vacuum which
is of purely gravitational origin.

%A spherical bubble of false vacuum inside a
%spacetime of true vacuum can be studied in this way. Such solutions
%were considered in General Relativity in refs. \cite{Vacuum
%bubbles}. Here the false vacuum is a purely gravitational origin.
\begin{figure}[t]
\begin{center}
  \includegraphics[height=\textwidth,angle=270]{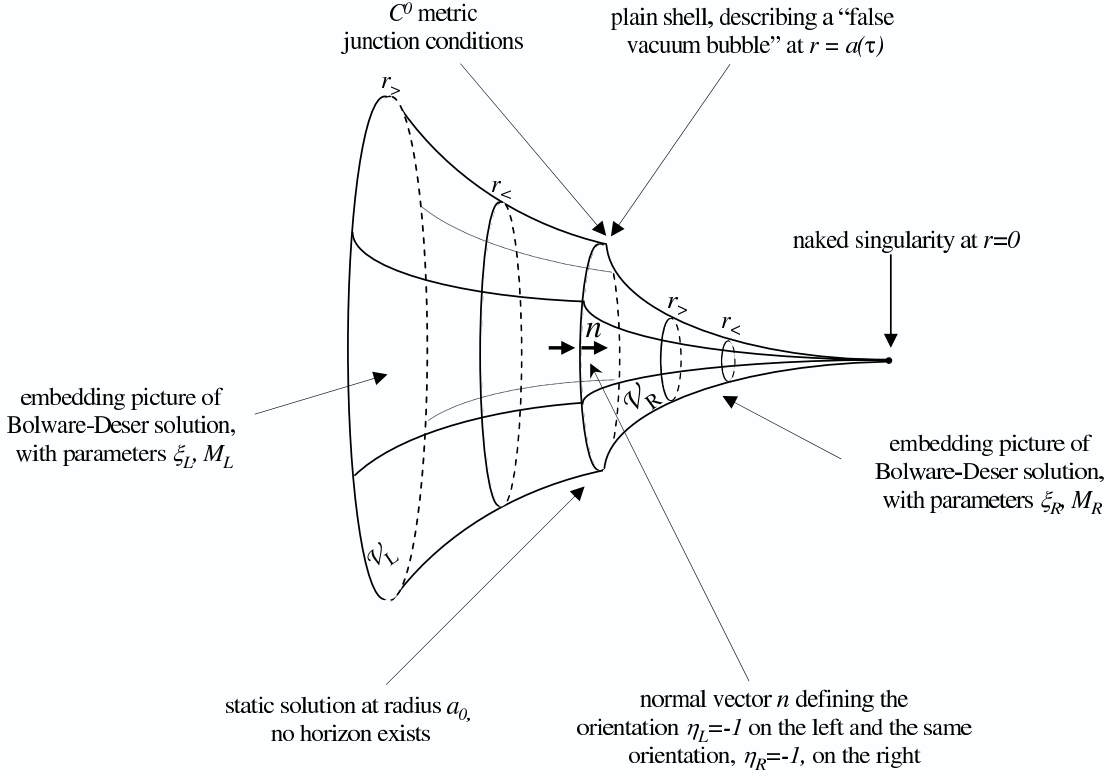}\\
  \caption{A spherically symmetric spacetime with metric of the class $C^0$.
  A vacuum shell with the standard orientation always
  connects two regions with different branch
  signs $\xi $ (and generically with different mass parameters $M$). Each region has a
  different effective cosmological constant.}\label{Wormhole3}
  \end{center}
\end{figure}

These kind of solutions might lead to curious implications. For
instance, let us consider the following construction: Suppose we
have a ``well behaved" minus branch ($\xi _L =-1$) solution with
positive mass $M_L$; where by ``well behaved'' we mean a solution in
which the singularity is hidden behind an event horizon and for
which we get a suitable GR limit for small $\alpha $. Now, let us
cut out the black hole at some radius $r=a(\tau) > r_H$ and then
replace it with the interior of a plus branch ($\xi _R =+1$)
solution, i.e. a naked singularity. By doing this we would be
constructing a vacuum solution whose geometry, from the point of
view of an external observer, would coincide with that of a black
hole but, instead, would not possess a horizon. A particle in free
fall would not find a horizon but rather a naked singularity as soon
as it passes through the $C^0$ junction hypersurface located at
$r=a>r_H$. The solutions with $\Lambda =0$ and $\alpha>0$ are a
clear example of this. As can be seen from figure
\ref{ZeroLmassPics} there are solutions for all positive $M_{(-)}$;
so that we can indeed cut out the event horizon and replace it with
a naked singularity!

Also, for $\Lambda \neq 0$ ``false vacuum bubble'' solutions gluing
a positive mass Boulware-Deser branch with a naked singularity do
exist. This is seen by looking at the moduli space described in
figure \ref{WormholeAndPlain_fig}. One might expect that such
cosmic-censorship-spoiling shells be unstable and in section
\ref{dynamical general section} we will confirm that they are
unstable with respect to small perturbations.

\subsection{Vacuum wormhole-like geometries}

So far, we have discussed different kinds of geometries constructed
by a cut and paste procedure of two spaces that were initially
provided with the Boulware-Deser metric on them. The strategy was to
make use of the junction conditions holding in Einstein-Gauss-Bonnet
theory and, in particular, we have shown that solutions with
non-trivial topology, which have no analogues in Einstein gravity,
do arise through this method. A remarkable example is the existence
of vacuum wormhole-like geometries\footnote{Smooth wormhole
solutions in Lovelock theory have been found previously with matter
source in refs. \cite{Ghoroku:1992tz} and without matter for special
choice of coupling constants in refs. \cite{wormjulin,GOT07}.},
corresponding to the case $\eta_L \eta_R <0$. These ``wormholes''
can be thought of as belonging to two different classes: The first
class describes actual wormholes, presenting two different
asymptotic regions which are connected through a throat located at
radius $r_L = r_R =a$; the radius of the throat being larger than
the radius where the event horizons (or naked singularities) would
be. The two asymptotic regions are $r_L\to \infty$ and $r_R \to
\infty$ as measured by the radial coordinate in the respective sides
of the junction. This type of geometry is an example of a vacuum
spherically symmetric wormhole solution in Lovelock theory and its
existence is a remarkable fact on its own. On the other hand, a
second class of wormhole-like geometry with no asymptotic regions
also exists. This second class is obtained also by considering the
orientation $\eta _L \eta _R <0$, this time cutting away the
exterior region of both geometries and gluing the two interior
regions together. We shall discuss this later; first let us discuss
the static wormhole solutions with two asymptotic regions (actual
wormholes).

\subsubsection{Geometries presenting two asymptotic regions}

Let us begin by emphasizing that such static wormhole solutions only
exist if at least one of the two bulk regions corresponds to $\xi
=+1$. That is, at least one of the two Boulware-Deser metrics has to
correspond to what we have called the exotic branch. This could have
deep implications in what regards semiclassical stability
\cite{Boulware:1985wk}. It is also remarkable that for these static
wormholes to exist it is necessary that $\Lambda >0$. Furthermore,
the existence of two asymptotic regions demands $\alpha >0$ (for
values $\alpha <0$ there are only solutions with``closed universe"
geometry to be discussed below). Moreover, since the static
wormholes only exist if at least one of the branches corresponds to
$\xi =+1$, then at least one of the regions connected through the
throat possesses a negative effective cosmological constant.

\begin{figure}[t]
\begin{center}
  \includegraphics[height=0.9\textwidth,angle=270]{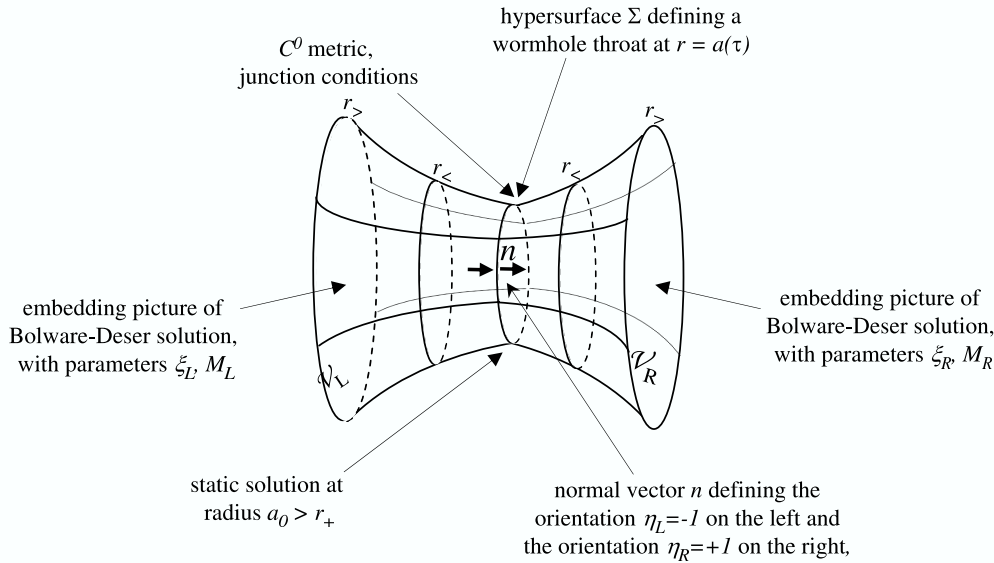}\\
  \caption{ Static vacuum wormhole solution.
  The junction conditions are satisfied on the timelike hypersurface
  $\Sigma$ for the case of the orientation $\eta _L \eta_R <0$.
  This solution presents two disconnected asymptotically de-Sitter
  regions.}\label{Wormhole2}
  \end{center}
\end{figure}

Another interesting feature concerns the stability under radial
perturbations. This is seen in Fig \ref{V''}. In particular, it can
be shown that stable static wormholes only exist for the case $\xi_L
= \xi_R = +1$; namely, the case where both Boulware-Deser metrics
correspond to the exotic branch. Nevertheless, no stable wormholes
exist for the case $M_L = M_R$, and thus, concisely, the static
symmetric wormholes are unstable under perturbations that preserve
the spherical symmetry.

An interesting possibility is that of having wormhole solutions
whose Boulware-Deser metrics would correspond to negative mass
parameters. For instance, one can construct a static wormhole with
one side being of the ``good branch'' $\xi _L =-1$ and having a
negative mass $M_L <0$. In that case, from the point of view of a
naive external observer, the vacuum solution would seem to
correspond to a naked singularity. However, now we know that the
inclusion of non-trivial junctures makes it possible to replace
such a singularity by an exterior region on the other side of a
non-smooth wormhole throat. This has a deep implication in what
concerns the ``cosmic censorship principle'' since for the
appropriate values of the coupling constants, and unlike what
usually happens in pure gravitational theories, the spherically
symmetric vacuum solutions presenting naked singularity cannot be
unambiguously classified (and consequently systematically
excluded) in terms of the mass parameter.

Another particular case that deserves to be mentioned as a special
one is that of having a massless solution in one of the sides of the
wormhole geometry. For instance, such a construction is achieved if
the massless side corresponds to the exotic branch $\xi =+1$ and the
massive side to the branch $\xi =-1$. In these cases, the wormhole
throat turns out to be a  kind of puncture of the (A)dS spacetime,
let us call it a ``hole in the vacuum''. Since (A)dS is
homogeneously isotropic, a spherically symmetric matching can be
done anywhere: remarkably, several of these ``holes" could be
located at different places in the spacetime and each ``hole'' would
not influence the others.
We shall discuss this kind of geometry in more detail in section
\ref{Octopus section}. The massless side may then correspond to a
microscopic de-Sitter geometry and, presumably, its cosmological
horizon, yielding thermal radiation, could be seen from the massive
sides. This is an intriguing possibility that deserves to be further
explored.

\subsubsection{Closed universe type geometries}

Now, let us comment on the second class of wormholes; namely those
with no asymptotic regions. As mentioned, these geometries are
constructed by gluing the interior of the throat of both regions,
instead of the exterior. One can perform the matching by keeping the
region that is inside the throat but still outside the horizons.
Consequently, one gets a geometry that resembles a ``static closed
universe'' with horizons. This exotic geometry has no asymptotic
regions at all, and, because of this, this second type of geometry
does not represent what one would usually call a wormhole.
Nevertheless, we shall abuse the notation and call ``wormhole'' any
timelike junction with the orientation $\eta_L \eta _R <0$.

\begin{figure}[t]
\begin{center}
  \includegraphics[height=.85\textwidth,angle=270]{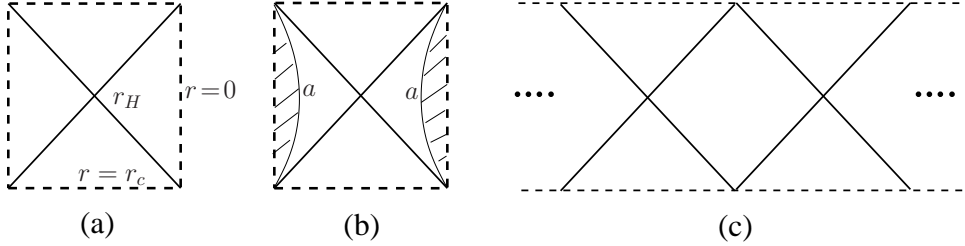}
  \caption{(a) The causal diagram of the smooth spherically symmetric solution for
  $\alpha <0$, $1+ \frac{4\alpha \Lambda}{3} <0$.  (b) The vacuum shell is introduced at radius
  $r=a$ cutting out the $r=0$ singularity. (c) Causal diagram of the resulting spacetime
  (a $C^0$, spherically symmetric
  vacuum solution).  }
  \end{center}\label{Penrose_closed_universe}
\end{figure}

Static solutions of this kind without naked singularities (i.e. with
horizon) exist for negative values of the coupling $\alpha $
%\begin{figure}[t]
%\begin{center}
%  \includegraphics[height=.80\textwidth,angle=270]{Wormhole4.ps}
%  \caption{Another kind of wormhole-like geometry is obtained by
%  joining two different spherically symmetric solutions but keeping
%  the region between the event horizon and the throat. This resembles a
%  sort of static closed universe.}\label{Wormhole4}
%  \end{center}
%\end{figure}
and $x\equiv 4\alpha\Lambda/3<-1$. In this range of the coupling
constants the Boulware-Deser metric develops a branch singularity at
fixed radius $r_c^4 = \frac{16M\alpha }{|x|-1}$, where the curvature
diverges. This branch singularity represents the maximum
three-sphere radius: the metric becomes non-real for $r>r_c$. In
addition there is a curvature singularity at $r=0$. In this region
of the space of parameters we would say that the Boulware-Deser
geometry is somehow pathological. However, if junction conditions
are appropriately applied, then a well-behaved $C^0$ vacuum geometry
can be constructed by simply taking a pair of such pathological
spaces, cutting out the naked singularities and joining them
together. To see that this is possible, it is sufficient to consider
the symmetrical case. It can be checked from equations
(\ref{Mass_formula_1}) and (\ref{Mass_formula_2}) that two bulk
regions with equal masses $M_L/\alpha = M_R/\alpha =
\frac{4(1+x)}{(x-3)}$ can be matched at a throat radius $a^2 =
\frac{16|\alpha|}{3-x}$. Consulting figure
\ref{BranchesTimelikeWormhol_fig} (these solutions are located on
the upper bounding curve of the left part of the moduli space) we
see that wormhole solutions exist when the bulk regions have branch
signs $(\xi _L, \xi _R)=(-1,-1)$. The bulk metric has a horizon
$r_H$, which separates $r=0$ (a timelike naked singularity) from
$r=r_c$, which is a spacelike singularity. The static shell is
located at $a<r_H$. So by cutting out the regions $r<a$ and joining
with the wormhole orientation the naked singularities can be
removed. The causal diagram of the original pathological spacetimes
and the extended causal diagram of the $C^0$ closed universe, which
results from the matching, with horizons are shown in figure
\ref{Penrose_closed_universe}.

\section{Dynamical vacuum shells}

In general, vacuum shells will be dynamical objects. We discuss the
dynamics here and also discuss the issue of radial stability of the
static solutions.

\subsection{General solution}\label{dynamical general section}

Let us briefly recapitulate upon the equation
(\ref{solved-explicit}), which governs the dynamics of the shells.
We can treat both the timelike and spacelike together since, as we
noticed in section 2, the analysis is completely analogous. A
dynamical vacuum shell is governed by a differential equation of the
form
\begin{equation}
 \dot a^{2}+ V(a) = 0 \, ;  \label{tonga2}
\end{equation}
see (\ref{solved-explicit}) above. It is useful to express $V(a)$ in
terms of the non-negative quantity $ Y= \sqrt{1+ \frac{4\alpha
\Lambda}{3}+ \frac{16M \alpha}{a^4}}$, and the effective potential
then reads
\begin{equation}
 V(a)=  \sigma\left(k+\frac{a^2}{4\alpha}\right)
 - \frac{\sigma a^2}{4\alpha}
 \left(  \frac{3(\xi_R  Y_R + \xi_L Y_L )^2 + (\xi_R  Y_R - \xi_L Y_L )^2 }
 {12(\xi_R Y_R + \xi_L Y_L)}\right) \, . \label{tonga}
\end{equation}
In addition to the differential equation, the solution must obey the
inequalities (\ref{timed ineq_Ys})-(\ref{spacelike real roots_Ys}).
It is convenient to rewrite them as follows
\begin{equation}\label{Ineq_prod}
 -\eta_R \eta_L \Big( 9(\xi_R  Y_R + \xi_L Y_L )^2 - (\xi_R  Y_R - \xi_L Y_L )^2
 \Big) \geq 0\, ,
\end{equation}
\begin{equation}\label{Ineq_Sum}
 \sigma \alpha (\xi_R Y_R + \xi_L Y_L) >0\, .
\end{equation}

Note that the effective potential (\ref{tonga}), $ V(a) =
\frac{\sigma a^2}{4\alpha} + \sigma k + \Delta V(a)$, consists of a
quadratic piece, a constant determined by the three-dimensional
curvature $k$ of the shell, and another piece which, by inequality
(\ref{Ineq_Sum}) obeys $\Delta V <0$.

To analyze the motion of a shell we shall need to know the
derivatives of the potential. This is worked out in appendix
\ref{Potential_Appendix}. Differentiating the potential we get the
following expression for the acceleration of a moving shell,
\begin{equation}\label{force}
\ddot a=-\sigma\,\frac{a}{4\alpha} \Big[1-\frac{1 + 4\alpha\Lambda/3
}{\xi_R Y_R + \xi_L Y_L}\Big]\ .
\end{equation}
Considering the sign of this acceleration and making use of
inequality (\ref{Ineq_Sum}) we can make some general observations:
\begin{remark}\label{Force_Remark}
For a timelike shell ($\sigma=+1$):\\
 When $1 + \frac{4\alpha\Lambda}{3} \geq 0$ and $\alpha <0$
 a vacuum shell always experiences a repulsive force away from  $r = 0$;
 \\
 When $1 + \frac{4\alpha\Lambda}{3} \leq 0$ and $\alpha > 0$
 a vacuum shell always experiences an
attractive force towards  $r =
 0$.
\end{remark}

In the situations not covered by Remark \ref{Force_Remark} the
potential may have an extremum. From (\ref{force}) we deduce that
there is an extremum at $r= a_e$ iff \begin{equation}
 \xi_R Y_R(a_e)
+ \xi_L Y_L(a_e) = 1 +
 \frac{4\alpha\Lambda}{3}\,.\label{extremum xis
and w}
\end{equation}
Recalling inequality (\ref{Ineq_Sum}), we conclude that an extremum
can exist only if
\begin{equation}\label{sigma alpha w}
\sigma\alpha\left(1+\frac{4\alpha\Lambda}{3}\right)>0\, .
\end{equation}

The extremum will be a minimum or maximum depending on the sign of
the second derivative of the potential evaluated there,
\begin{gather}\label{extremum V'' in Ys}
  V^{\prime\prime}(a_e) =  \frac{\sigma}{\alpha}\left(
 \frac{  1+4\alpha \Lambda / 3  }{\xi_R \xi_L Y_R(a_e) Y_L(a_e) }  -1\right)\,
 .
\end{gather}
There is a general result for vacuum shells separating different
branch metrics. From (\ref{sigma alpha w}) we see that $V''(a_e)$ in
(\ref{extremum V'' in Ys}) must be negative for
$1+\frac{4\alpha\Lambda}{3} \ge 0$.
\begin{proposition}\label{No minimum}
 In the range\footnote{In the case
of the wormhole orientation, by using the inequality
(\ref{Ineq_prod}), the result can be extended to apply to the range
$1+\frac{4\alpha\Lambda}{3}>-\frac{1}{2}$.}
$1+\frac{4\alpha\Lambda}{3} \geq 0$
 for the product of Gauss-Bonnet coupling and cosmological constant:
 Let $\Sigma$ be a vacuum shell such that $\xi_L\xi_R=-1$.
 Then the potential never has a minimum. If $\Sigma$ is a timelike shell
 it will either be in an (unstable) static state, or, if it is moving,
 will either expand or
 collapse, it can not be bound.
\end{proposition}

We have already remarked in section 2 that any shell with standard
orientation \emph{must} match two bulk metrics of opposite branch
sign ($\xi_L\xi_R =-1$), except in the trivial case of a smooth
matching. So we obtain the following general result about
instability of standard shells:
\begin{corollary}
 Let $1+\frac{4\alpha\Lambda}{3} \ge 0$. A timelike shell with
 standard orientation is either in an (unstable) static motion, or, if it is moving,
 will either expand or
 collapse, it can not be bound.
\end{corollary}

We have already seen in section \ref{Static_Section} that static
shells with standard orientation are always in a state of unstable
equilibrium in the (physical) regime $1+\frac{4\alpha\Lambda}{3}
\geq 0$. The proposition above strengthens this result to include
dynamical shells. A dynamical shell with standard orientation can
not be oscillatory. It must either disappear into a singularity or
fly out towards spatial infinity.

There is not such a strong result for shells with the wormhole
orientation. Indeed in section \ref{Static_Section} we found stable
static wormholes for $1+\frac{4\alpha\Lambda}{3} \geq 0$ which
matched two bulk metrics of the plus branch. We can however derive a
strong result about instability concerning bulk metrics of the minus
branch. When $\xi_L=\xi_R=-1$ we see from (\ref{extremum xis and w})
that for $1+\frac{4\alpha\Lambda}{3} \ge 0$ an extremum is not
possible. Furthermore, combining the results of Remarks
\ref{minus_minus_remark} and \ref{Force_Remark} we see that the
shell is always expanding:
\begin{proposition}\label{Repulsion}
 Let $1+\frac{4\alpha\Lambda}{3} \geq 0$
 and let $\Sigma$ be a timelike vacuum shell with wormhole orientation,
 and ${\cal V}_L$ and ${\cal V}_R$ be minus branch bulk metrics
 $(\xi_L, \xi_R) = (-1,-1)$. Then  the shell always experiences a
 repulsive force away from  $r = 0$.
\end{proposition}

So in summary, we have found some general results for the range of
parameters $1+\frac{4\alpha\Lambda}{3} \geq 0$. This range is
important because it includes the case $|\alpha\Lambda| << 1$ and
therefore applies when the Gauss-Bonnet term is a small correction.
Combining these results, we conclude that, in this range of
parameters, all timelike vacuum shells involving the minus branch
are unstable. The only vacuum shell solutions which can be static or
oscillatory are wormholes which match two regions of the exotic plus
branch.

%A similar analysis could be made for the more exotic range of
%parameters $1+\frac{4\alpha\Lambda}{3} < 0$.

%\subsection{General solution as defining a non-smooth metric}
%\label{section of one-shell non-smooth metric}

%The intrinsic and the extrinsic properties of the world-volume of
%the vacuum shell are determined by the function $a(\tau)$. As we
%explained in section 2, given this function and the orientation of
%the matching a $C^0$ spherically symmetric vacuum solution is
%defined.

%One may ask: for given (smooth) bulk metrics i.e. given $M_L,\xi_L$
%and $M_R,\xi_R$, how unique is this $C^0$ metric?

%The function $a(\tau)$ is given by the differential equation $\dot
%a^2 +V(a)=0$.

\subsection{Comment on the stability of static shells}

Dynamical equation (\ref{tonga2}) resembles the equation for a
particle moving under the influence of an effective potential
(\ref{tonga}). Nevertheless, as pointed out in section 2, this is
not strictly the case due to the presence of the vanishing energy
constraint. This is important for the case when the extremum of the
potential is at $a=a_0$ with $V(a_0)=0$, i.e. when static solutions
exist. When $V''(a_0) <0$ we conclude that the shell is unstable
with respect to the radial component of a perturbation- a slight
shift $a \to a_0 + \delta a$ will cause the shell to accelerate away
from the (unstable) equilibrium radius. When $V''(a_0)>0$ we
conclude that the shell is stable with respect to radial
perturbations. There is however a slight subtlety: as mentioned
previously the energy is unavoidably fixed instead of arising as a
constant of motion. So for a fixed potential, there is no real
solution for $a$ when $V>0$. We can consider spherically symmetric
solutions which are close-by in the space of the solutions, i.e.
with slightly different parameters $M_{L,R}$ and $w$ such that the
value of the potential at $a_e$ is slightly negative: let us say
$V(a_e)=0^{-}$. This means that such a solution oscillates between
two radii around $a_e$ at which the potential vanishes. This is
certainly a stable solution though not static, a `bounded
excursion'~\cite{VisserWiltshire}. Now if we let $a_e$ coincide with
the $a_0$ of the original static solution, this means that for
slightly different parameters than those for which $a_0$ is a static
solution, there exists an oscillating solution around $a_0$.
Therefore a static solution $a_0$ which is a minimum of the
potential gives information about when infinitesimal bounded
excursions can happen. More generally, the dynamics of the perturbed
shell can be thought of as corresponding to a perturbation of the
above equation $V(a)\to V(a)-\delta h$, provided energy $\delta h$
from an external excitation. The stable regions of the moduli space
of static solutions are plotted in figures \ref{V''} and
\ref{PlainStabilityTimelikepic_fig}. The graph will take an elegant
form in terms of the change of variables to be introduced in section
\ref{Moduli_section} (see fig \ref{ddVfinal}).\\

In the rest of this section we present some illustrative examples of
dynamical vacuum shells, first in symmetrical wormhole solutions and
then in the context of Chern-Simons gravity.

\subsection{Symmetric dynamical wormholes}

 Now let us consider the case where the masses in each
bulk region are the same, being $M_L=M_R=M$. The inequalities
(\ref{Ineq_prod}) and (\ref{Ineq_Sum}) are equivalent to:
\begin{remark}
 If  $\Sigma$ is a vacuum shell joining two bulk regions with the
 same mass $M_R = M_L$ then:
 \\
 i) The bulk solutions must have the same branch sign $\xi_L =\xi_R =
 \text{sign}(\sigma\alpha)$;
 \\
 ii) The shell must have wormhole orientation.
\end{remark}
So the spacetime is completely left-right mirror-symmetric. The
equation of motion reads
\begin{equation}
 \sigma \dot a^2+\frac{\ a^2}{4\alpha}\Big[1-\xi\frac{\ Y(a)}{2}\Big]+k = 0\
.
\end{equation}
The general solution is rather complicated. Next we proceed to
consider a simple case where both masses vanish.

The case where $M_{L,R}=0$ is an interesting special case of the
symmetric wormholes, which exists for $1+
\frac{4\alpha\Lambda}{3}>0$. The equation of motion reduces to
\begin{equation}
 \sigma \dot a^2+ \frac{\ a^2}{4\alpha}
\left(1-\xi\frac{\sqrt{1+\frac{4\alpha\Lambda}{3}}}{2}\right) +k =
0\ .
\end{equation}

\begin{remark}
Consider a timelike shell $(\sigma=+1)$, that is
$\text{sign}(\alpha)=\xi$.

Bounded motions: $\xi=+1$, $\frac{4\alpha\Lambda}{3}<3$, $k=+1$ ;
$\xi=+1$, $\frac{4\alpha\Lambda}{3}=3$, $k=0$.

Unbounded motions: $\xi=+1$, $\frac{4\alpha\Lambda}{3}=3$, $k=-1$ ;
$\xi=+1$, $\frac{4\alpha\Lambda}{3}>3$, any $k$ ; $\xi=-1$,
$\frac{4\alpha\Lambda}{3}>-1$, any $k$.

The same bounded or unbounded configurations exist in the spacelike
case provided one replaces $k$ with $-k$, for the opposite sign of
$\alpha$.
\end{remark}

The hyperbolic shell, $k=-1$, admits a stationary vacuum wormhole
solution: for $\text{sign}(\alpha)=\xi=+1$ and $4\alpha \Lambda/3 =
3$ we have that $\ddot a=0$ and $\dot a^2=1$.

When $\Lambda=0$ and $\xi=-1$ the two bulk regions have flat
Minkowski metrics. When the throat of the symmetric wormhole is a
sphere the world-volume of the shell is described by $t^2-r^2=8
\alpha/3$. When $\alpha>0$ the shell is spacelike; when $\alpha<0$
it is timelike.
% and each Minkowski bulk approaches a Rindler wedge
% as $\alpha \to 0$.

\subsection{Chern-Simons dynamical vacuum junctions}\label{CS}

When $1+\frac{4\alpha\Lambda}{3} = 0$ some very special things
happen. For this choice of coupling constants the
Einstein-Gauss-Bonnet theory (in first order formalism) is
equivalent to a Chern-Simons theory for the deSitter ($\alpha < 0$)
or Anti de Sitter ($\alpha >0$) group\footnote{The case of
Poincar\'{e} Chern-Simons theory was discussed in Ref.
\cite{Gravanis:2007ei}.}. In this case the metric function takes the
very simple form
\begin{equation}\label{}
f(r)= 1 +\frac{r^2}{4\alpha}- \mu \ ,
\end{equation}
where $\mu$ is a constant and the mass is proportional to $\mu^2
-1$\cite{BHscan}. When $\mu >0$ the bulk solution is a black hole.
When $\mu<0$ the bulk metric would have a naked singularity at the
origin.

The dynamics of vacuum shells takes a very simple form. The quantity
$a^2 \xi Y/4\alpha = -\mu$ for each bulk region is a constant and
therefore the non-harmonic part of the potential $\Delta V$ is a
constant. The equation of motion takes the form
\begin{eqnarray}\label{CS diff eq}
 \dot a^2+\frac{\sigma }{4\alpha}a^2 = {\cal E} \quad , \qquad {\cal
E}= -\sigma\left(k + \frac{3(\mu_R+ \mu_L)^2 + (\mu_R -
\mu_L)^2}{12(\mu_R+ \mu_L)} \right)\ .
\end{eqnarray}
The potential is like that of a harmonic oscillator potential (or an
upside-down  harmonic potential if $\sigma \alpha$ is negative),
although it should be remembered that the origin $r=0$ of the bulk
spacetimes is singular so the shell can not really oscillate. The
solution is constrained according to the two inequalities
(\ref{Ineq_prod}) and (\ref{Ineq_Sum}), which now read
\begin{equation}\label{CS_inequality}
 -\eta_L\eta_R (9(\mu_R +\mu_L)^2 - (\mu_R-\mu_L)^2) \ge 0\ ,
\end{equation}
\begin{equation}\label{CS_inequality2}
   -\sigma\, (\mu_R +\mu_L) >0\ .
\end{equation}
The last inequality tells that ${\cal E}>-\sigma k$. These
inequalities are generally consistent with ${\cal E}>0$ so that
solutions do indeed exist.

For instance, consider the timelike shells in this theory. Note
that, from inequality (\ref{CS_inequality2}), at least one out of
$\mu_R$ or $\mu_L$ must be negative. So it is not possible to match
two black hole spacetimes. From inequality (\ref{CS_inequality}) we
see that shells with the standard orientation must obey $\mu_R\mu_L
<0$.
\begin{remark}
 For the Chern-Simons combination $1+ \frac{4\alpha\Lambda}{3} =0$,
 timelike vacuum shells always represent either:
 \\
 i) a matching between a bulk region of a black hole spacetime with bulk region
 of a naked singularity spacetime; or
 \\
 ii) a matching, with wormhole orientation, between two bulk regions of naked singularity
 spacetimes.
\end{remark}

Now, let us analyze the de-Sitter invariant Chern-Simons gravity,
which corresponds to $\alpha \Lambda =-3/4$ with $\alpha<0$. In this
case, the potential is like an inverted harmonic oscillator centered
at the origin. There are solutions for ${\cal E}$ positive, negative
and zero.

Let us just focus on the case ${\cal E}>0$. The trajectory of a
timelike shell is then given by
\begin{equation}\label{pupo}
a(\tau)=2\sqrt{|\alpha {\cal E}|}\ \sinh\Big(\pm
\frac{\tau}{2\sqrt{|\alpha|}} + \text{const.}\Big)\ ,
\end{equation}
which is a shell either emerging from the past white hole or falling
into the future black hole, depending on the sign $\pm$ in the
argument.

For ${\cal E}<0$ the hyperbolic sine is replaced by the cosine.
${\cal E}=0$ gives an increasing and a decreasing exponential.

On the other hand, for ${\cal E}>0$ and $k=1$, one could consider
Euclideanization of the problem. Presumably, this could be relevant
in describing the decay of the exotic negative $\mu$ spacetime.
Define an angle $\chi$ by
\begin{equation}\label{}
\chi=\frac{\tau_E}{2\sqrt{|\alpha|}}
\end{equation}
up to a constant, where $\tau_E$ is the Euclidean proper time of the
shell. The metric on the Euclidean world sheet of the shell reads
\begin{equation}\label{}
ds_\Sigma^2=4|\alpha| \big(d\chi^2+{\cal E} \sin^2 \chi d\Omega^2
\big) \ .
\end{equation}
When ${\cal E}<1$ we have a deficit solid angle, and, when ${\cal
E}>1$, an excess. In both cases the space has a curvature
singularity at the poles $\chi=0$ and $\chi=\pi$. Therefore, the
smoothness of the Euclidean shell requires ${\cal E}=1$.
 This metric is spherically
symmetric in the five dimensional sense in this case, whence it
describes a 4-sphere. The 4-sphere separates a ball of Euclidean
black-hole solution with mass parameter $\mu_R$ from another
solution with $\mu_L$, obeying the relation
\begin{equation}\label{euclidean 4-sphere}
\mu_L^2+\mu_R^2+\mu_L \mu_R+6(\mu_L+\mu_R)=0\,.
 %\frac{3(\mu_R+ \mu_L)^2 + (\mu_R -
%\mu_L)^2}{24(\mu_R+ \mu_L)} =-1\, .
\end{equation}
It is interesting to note that the size of the Euclidean world sheet
depends essentially only on $\alpha$ and not on the $\mu_{L,R}$; the
latter change its shape, which is fixed to spherical by the above
relation.

The curve (\ref{euclidean 4-sphere}) is an ellipse. It is
symmetrical around the line $\mu_L=\mu_R$ and tangential with the
line $\mu_L+\mu_R=0$ at $\mu_L=\mu_R=0$. It exists completely in the
region $\mu_L+\mu_R \le 0$. In view of the inequality
(\ref{CS_inequality2}) all points of the curve are included except
$\mu_L=\mu_R=0$. Therefore the 4-sphere Euclidean world sheet does
exists for certain values of the parameters. Whether this
interesting configuration is a mere curiosity or it is related to
semiclassical transitions between the $\mu_L$ and $\mu_R$ spacetime
is an open question.

 We can also consider the Anti-de Sitter invariant
Chern-Simons theory, corresponding to $\alpha>0$. In this case, the
effective potential $V$ turns out to be a quadratic potential
centered at the singularity at the origin. The analysis is similar
to that of the dS case except that there are solutions only with
${\cal E} > 0$.

On the other hand, we can also think about the spacelike shells for
this case of Chern-Simons couplings $\Lambda \alpha =-3/4$. These
shells represent a sudden classical transition from a spacetime with
some mass parameter $\mu_{\text{in}}$ to another with a different
mass parameter $\mu_{\text{out}}$. Such transitions occur for quite
general values of $\mu_{\text{in}}$, and this is a concise
manifestation of the extreme degeneracy of the field equations of
the Chern-Simons theories.

\section{Constant solutions revisited}\label{Moduli_section}

In this section we will perform an exhaustive analysis of the space
of constant solutions, what we have called the moduli space.

\subsection{Static and instantaneous spherical vacuum
shells}\label{basic moduli section}

To begin, it will be convenient to introduce new dimensionless
parameters, defined as follows
\begin{equation}\label{define_uv}
u \equiv \sqrt{3}\sqrt{x(x+1)\Big(\frac{4}{y}+\frac{3}{x}-1\Big)}
\quad , \quad w \equiv x+1\ .
\end{equation}
On should think of $u$ and $w$ as functions of $\alpha$, $\Lambda$
and $a_0^2$, via the definitions (\ref{faithless}). Here,
\begin{equation*}
u \ge 0\ .
\end{equation*}
 The inverse transformation is given by
\begin{equation}\label{inverse}
y=\frac{12w(w-1)}{u^2+3(w^2-4w)}\quad,\quad x=w-1\ .
\end{equation}
Each point on the $(w,u)$ plane such that $u^2+3(w^2-4w)\neq 0$
uniquely determines the values of the basic dimensionless ratios $x$
and $y$ and therefore the solution\footnote{It is useful to remember
that the radius $a_0$ of the vacuum shell is given in terms of these
variables by
\begin{equation}\label{a_0 of u}
a_0^2=4\alpha\cdot \frac{12w}{u^2+3(w^2-4w)}\ .
\end{equation}}.

\begin{definition}
We will call the allowed domain on the $(u,w)$ plane as the $(u,w)$
parameter space representing the moduli space of the vacuum shell.
Similarly the allowed domain on the plane of $x$ and $y$ is the
$(x,y)$ parameter space for the vacuum shell for non-zero $\Lambda$.
The various possible pairs of parameters that uniquely represent all
possible points of the moduli space can be thought of its
coordinates.
\end{definition}

In terms of these new variables we have that the vacuum shells are
described by equations (see Proposition \ref{static instant}):
\begin{equation}\label{FF in u,w}
f_L+f_R  = \frac{2(u^2+9w^2)}{u^2+3(w^2-4w)}\ ,\qquad \sqrt{f_Lf_R}
 =\sigma \eta_R\eta_L \ \frac{u^2-9w^2}{u^2+3(w^2-4w)}\ .
\end{equation}

\begin{figure}[t]\begin{center}
  \includegraphics[width=0.6\textwidth]{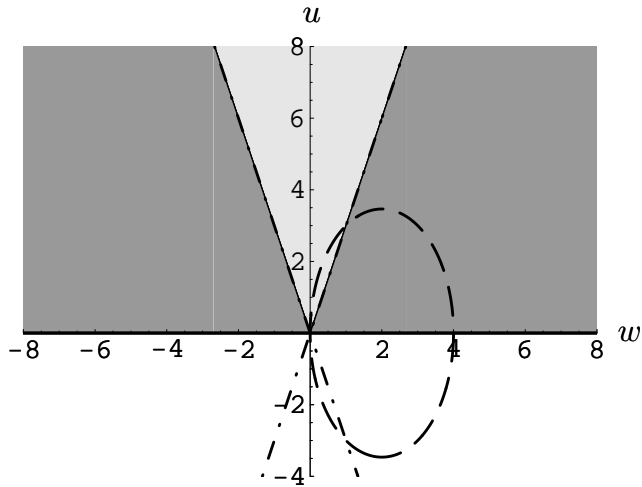}
  \caption{{
   The space of \emph{constant radius} solutions, which we have called the moduli
   space, is depicted here.
   The dimensionless variables $w$ and $u$ are defined at the beginning of section
   \ref{basic moduli section}.
   The ellipse divides solutions into spacelike (inside) and
   timelike (outside);
   The diagonal lines divide solutions into standard orientation (light grey)
   which have well-defined inner and outer region of the shell,
   and wormhole orientation (dark grey), where both regions can be thought of
   as exterior or interior
   depending on whether a non-compact or compact region is maintained.
   Solutions exist for $u\geq 0$.
\newline
   The line $w=0$, $u>0$ for finite $\alpha$ does not actually belong to the moduli space as being trivial:
   the junction condition require the metric across the shell must be continuous in this
   case.
   In terms of the couplings $\alpha$ and $\Lambda$, $w$ is
   given simply by $w \equiv 1+\frac{4\alpha\Lambda}{3}$.
   The combination of the couplings $w=0$
   corresponds to the case where Einstein-Gauss-Bonnet gravity can be written as a
   Chern-Simons theory with (A)dS gauge group. It for this combination that the smooth $C^2$ metrics
   fail to be unique~\cite{Charmousis}\cite{Z}.
   Note that the pure Gauss-Bonnet case, which formally corresponds
   to $w=0$, $\Lambda =0$ in the limit that $\alpha \to \infty$ but
   $M \alpha$ is finite, does have nontrivial solutions, which were
   considered separately in Ref. \cite{Gravanis:2007ei}.
   }}\label{egg_figure}
  \end{center}
\end{figure}

Solving this (for details  see appendix \ref{Moduli_Appendix}) we
find:

\begin{proposition}
The masses in the two bulk regions are $M_{(+)}$ and $M_{(-)}$:
\begin{equation}\label{mass formulas}
 \bar{M}_{(\pm)}=\frac{36 w^2 ((w \pm u)^2-4w)}{(u^2+3(w^2-4w))^2}\
 .
\end{equation}
The moduli space is divided into: timelike shell or spacelike shell
solutions by the inequality $u^2 + 3(w^2-4w)>0$ or $<0$
respectively; standard orientation and wormhole orientation by
$u^2-9w^2 >0$ or $<0$ respectively. (see Fig. \ref{egg_figure}).
Furthermore the branch sign of the metric in each region is given by
$\xi_{(\pm)} = \text{sign}(w\pm u)$.
\end{proposition}

The points along the curves $\pm u = 3w$ and $u^2+3(w^2-4w)$ are not
regarded as part of the moduli space. Along the curve $u^2-w^2=0$,
which we will call the \textit{branch curve}, one of the branch
signs is undetermined. The line $w=0$ is peculiar as it implies that
either $a_0=0$, or $|\alpha|=\infty$ and $\Lambda=0$. The latter is
the case of pure Gauss-Bonnet gravity, whose solutions were found in
Ref. \cite{Gravanis:2007ei}. For $|\alpha|<\infty$ the line $w=0$ is
excluded from the moduli space as it corresponds to smooth
geometries.

We further note that the signs of $\alpha$ and $\Lambda$ on the
moduli space are given by $\text{sign}(\alpha)=\text{sign}\big(w\,
(u^2+3(w^2-4w))\big)$ and
$\text{sign}(\Lambda)=\text{sign}(\alpha)\, \text{sign}(w-1)$. The
region $w \approx 1$ is where the Gauss-Bonnet coupling is small
with respect to the cosmological scale.

\begin{figure}[]
\begin{center}
{\includegraphics[width=.70\textwidth]{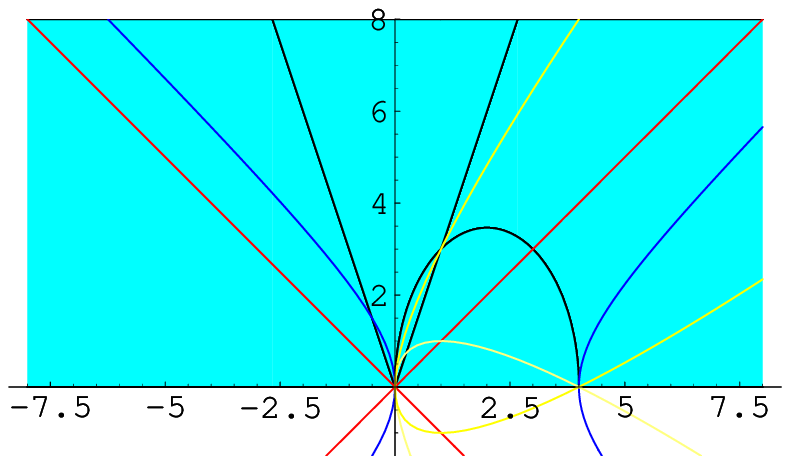}}
\end{center}
\caption{{\small {The moduli space showing the various curves listed
in section \ref{curves spectrum section}. The basic division
according to the time- or space-likeness of the world-volume of the
shell and the orientation of the matching were described in the
figure \ref{egg_figure} and are shown in black lines here: the
points along them they do not belong to the moduli space.
\newline
The diagonal (red) lines divide the space according to the branches
of the bulk metrics which we classify by the pair of signs
$(\xi_{(-)},\xi_{(+)})$ explained in section \ref{x y shells
section} and \ref{basic moduli section}: in the region on the left
the branch signs are $(+,+)$ i.e. both the bulk metrics on each side
of the shell belong to the ``exotic'' Boulware-Deser branch which
does not have a well-defined limit $\alpha \to 0$; the region in
between the diagonal lines is $(-,+)$; in the region on the left the
branches are $(-,-)$ i.e. both metrics belong to the branch with a
well-defined $\alpha \to 0$ limit (however the vacuum juncture
requires that these solutions only exist for $w \equiv
1+\frac{4\alpha\Lambda}{3}<0$, i.e. they have no asymptotics:
certain curvature singularities appear at finite
radius~\cite{GOT07}).
\newline
The hyperbola that exists on the outside of and touches the elliptic
region of spacelike solutions only at the border of the ellipse at
the points $w=0$ and $w=4$ (blue line), is what we have called the
stability curve: crossing this curve the second derivative
$V''(a_0)$ of the potential (\ref{solved-explicit}) or
(\ref{tonga2}) evaluated at the constant solutions $a=a_0$ changes
sign, which is a measure of (in)stability under perturbations. The
constants solutions for which $V''(a_0)>0$ are depicted in figure
\ref{ddVfinal}.
\newline
The remaining two lines (yellow lines) are symmetric around the
horizontal line $u=0$. Each curve corresponds to solutions such that
one of the mass parameters vanishes i.e. one of the bulk regions is
pure vacuum. Note that they exist only for $w =
1+\frac{4\alpha\Lambda}{3}>0$. These configurations are discussed in
sections \ref{Octopus section} and \ref{Black hole spectrum and
degeneracy} as an interesting example of certain non-trivial
features $C^0$ metrics acquire when Einstein gravity is supplemented
by the Gauss-Bonnet term is five dimensions.
}}}%
\label{elvis}%
\end{figure}

In what follows we further categorize the solutions according to
other physical properties. The entire information we will get is
given in the Fig.\ref{elvis}.

We first note that the formulae (\ref{mass formulas}) for $M_{(+)}$
and $M_{(-)}$ are related simply by $u \leftrightarrow -u$. Although
the true moduli space is the upper half plane $u \geq 0$, it is
useful to formally extend to $u<0$ (see Lemma \ref{mirror lemma}).
In this way we may plot inequalities for the region with mass
$M_{(-)}$ in the lower half plane and for the region corresponding
to $M_{(+)}$ in the upper half plane.

\subsection{The masses over the moduli space}

Important physical properties of the solutions have to do with what
values the masses $M_{(\pm)}$ take, w.r.t. sign, magnitude and
relative magnitude, over the moduli space. Let us comment on it
below.

\subsubsection{Equal mass solutions}

 A question with a very simple answer is where on the moduli space we
could have $M_{(+)}=M_{(-)}$. We have seen that this happens at
$u=0$. Explicitly, from (\ref{mass formulas}) we have
\begin{equation}\label{}
\bar{M}_{(+)}-\bar{M}_{(-)}=\frac{144 u
w^3}{\big(u^2+3(w^2-4w)\big)^2}\ .
\end{equation}
\begin{proposition}
$M_{(+)} = M_{(-)}$ only at the boundary $u=0$. Therefore such
solutions exist only for wormholes.
\end{proposition}
From Proposition \ref{total moduli} we have that at the points where
$M_{(+)} = M_{(-)}$ we have also that $\xi_{(+)}=\xi_{(-)}$.
\begin{lemma}
Symmetric configuration are such $M_L=M_R$ and $\xi_L=\xi_R$. They
exist only at the boundary $u=0$ of the moduli space and they can be
either time- or space-like shell wormholes.
\end{lemma}
The equal mass $\bar{M}=\bar{M}_{(\pm)}$ of the symmetric case reads
\begin{equation}\label{symmetric case mass}
\bar{M}=\frac{4w}{w-4}\ .
\end{equation}
So, symmetric configurations exist for all $w \neq 4$ and $\bar{M}$
can take all real values except 4.

\subsubsection{Zero mass solutions}\label{zero mass solutions}

The masses $\bar{M}_{(\pm)}$ change sign crossing the curves where
they vanish, and of course these curves are where $M_{(\pm)}$ vanish
too. From the formula (\ref{mass formulas}) and Proposition
\ref{total moduli} we have that $\bar{M}_{(\pm)}=0$ along the curves
\begin{equation}\label{zero mass curves}
(u \pm w)^2=4w\ ,
\end{equation}
respectively. They exist only for $w>0$. The masses cannot vanish
for $w<0$.

Using Lemma \ref{mirror lemma} it is sufficient to look only at one
of the curves e.g. $(u - w)^2=4w$, which also reads $u=\pm 2
\sqrt{w}+w$, extended over the whole plane. The other mass is
\begin{equation}\label{}
\bar{M}_{0}^{(\pm)}=\frac{9 w\sqrt{w}}{(w-3)\sqrt{w} \pm 2}\ .
\end{equation}

The curve $(u-w)^2-4w=0$ goes to negative values of $u$ for $0<w<4$.
On the $u>0$ side it appears disconnected emerging into two pieces
at $w=0$ and $w=4$, fig.~\ref{elvis}. Therefore
\begin{proposition} \label{zero mass proposition 0}
$\bar{M}_{(-)}=0$ for $u=\pm 2 \sqrt{w}+w>0$ where
$\bar{M}_{(+)}=\bar{M}^{(\pm)}_0$ respectively. $\bar{M}_{(+)}=0$
for $u=2\sqrt{w}-w>0$ where $\bar{M}_{(-)}=\bar{M}_0^{(-)}$.
\end{proposition}
Independently of whether the mass that vanishes is an $M_{(+)}$ or
an $M_{(-)}$ note also the following
\begin{remark}\label{overall zero mass}
When the zero mass is of branch $\xi$ the massive side has mass
$\bar M_0^{(-\xi)}$ and the matching happens according to $u=|w- 2
\xi \sqrt{w}|>0$. The branch of the massive side depends, as always,
on which side of the line $u=w$ we are.
\end{remark}

%We also have

%\begin{remark} For all $w>0$ there are two points
%in $\mathcal{M}$ where one mass vanishes, except: at $(1,1)$ where
%$\bar{M}_R=-9/4$, and $(4,8)$ where $\bar{M}_L=18$. This is so
%because the points $(1,3)$ and $(0,4)$ do not belong in
%$\mathcal{M}$, see remark \ref{special points}. The point $(1,1)$
%belongs to the special curve $u-w=0$. Also at $(1,1)$ $\Lambda=0$.
%Therefore we match a constant curvature spacetime,
%$f_L(r)=1+r^2/(2\alpha)$, at the curvature singularity $r_E$ of the
%indefinite branch $f_R(r)$ metric.
%\end{remark}

\subsubsection{Sign of the mass parameters}

\begin{figure}[]
\begin{center}
{\includegraphics[width=.27\textwidth]{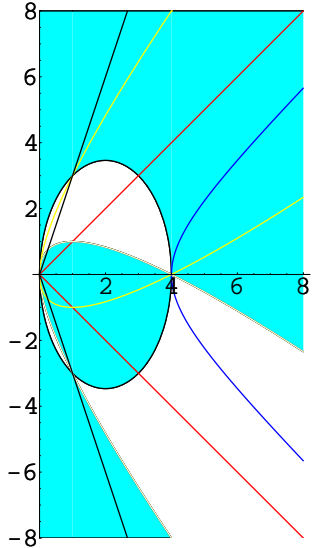}}
\end{center}
\caption{{\small {In the upper half plain, the shaded region is
where $M_{(+)}>0$. The inequality $M_{(-)} > 0$ has been plotted in
the lower half plain, making use of
 Lemma \ref{mirror lemma} (it should be remembered that in the lower
 half plane we have actually plotted
 $M_{(-)}(-u)$)).
We note that $M_{(+)}$ and $M_{(-)}$ are both negative for all
$w<0$, which is the left half of the diagram.
%and in a region within the spacelike standard shells domain for $w>0$.
}}}%
\label{M+-}%
\end{figure}

Now, let us discuss the positivity of the mass parameters.
%But, first, let us be reminded of the fact that the parameter $M$ is
%(\ref{BD_metric}) represents the mass through (\ref{m}) just for the
%case $\xi =-1$ (i.e. just for the minus branch), while for the
%exotic branch, $\xi =+1$, the mass is proportional to $-M$. This is
%easily verified by the asymptotic of (\ref{BD_metric}) considering
%the case $\alpha >0$.
The signs of $\bar{M}_{(\pm)}$ behave quite simply. From formula
(\ref{mass formulas}) we have:
\begin{proposition}\label{zero mass proposition}
$\bar{M}_{(\pm)}<0$ at the convex region defined by the curves
(\ref{zero mass curves}) i.e. where $(u\pm w)^2-4w<0$ respectively.
They have an overlap for $0\le u <2\sqrt{w}-w$, inside the spacelike
shell wormhole region. $\bar{M}_{(+)}<0$ only in this overlap.
\end{proposition}
\begin{remark} The entire curve $u-w=0$ exists
within the region where $\bar{M}_{(-)}<0$. This is also seen by the
fact that the r.h.s. of (\ref{solution=BD explicit}) vanishes there.
$\bar{M}_{(+)}>0$ along $u-w=0$.
\end{remark}
The above mean that $\bar M_{(-)}<0$ in a very large part of the
moduli space for $w>0$. Therefore the metrics $f_{(-)}$ will have
inner branch singularities, discussed in appendix \ref{BD appendix}.
The signs of $M_{(\pm)}=\alpha \bar M_{(\pm)}$ themselves are
depicted in the figure \ref{M+-} using also formula (\ref{sign of
alpha}).

\subsubsection{Mass as a function of the radius of the shell}\label{mass and
radius}

We see from (\ref{mass formulas}) that given a mass $M_{(\pm)}$, for
given $\alpha$ and $\Lambda$, the radius $a_0$ of the shell where
the matching takes place is determined by a fourth order polynomial
of $u$, namely
\begin{equation}\label{4th polynomial}
(u^2+3(w^2-4w))^2\, \bar{M}_*  -  36w^2\, (u^2+2w u+w^2-4w)=0\ ;
\end{equation}
given a $u$ we can obtain $a_0^2$ by (\ref{a_0 of u}). As discussed
in Lemma \ref{mirror lemma}, $\bar{M}_*$ is an $\bar{M}_{(+)}$ when
$u$ is non-negative and an $\bar{M}_{(-)}$ when $u$ is non-positive.

The equation above does not seem to be very enlightening. However,
we can combine it with some pieces of information we have: First we
know that $u$ takes values on the entire real line. Secondly, there
is at least one real solution $u$, since $\bar{M}_*$ is
\emph{defined} by (\ref{mass formulas}) to correspond to some real
$u$. Besides, $\bar{M}_*$ takes all real values itself as one may
verify.

So, one may ask the following: For a given $\bar{M}_*$, and a given
$w$, how many \emph{different} real solutions $u$ exist and what is
their sign? Now, the l.h.s. of (\ref{4th polynomial}) is an even
order polynomial. Then we know that there must be at least a second
$u$ producing the same $M_*$. What we a priori do not know is
whether the second $u$ is of the same sign or of the opposite.

There is one case where the second solution coincides with the
first, and therefore has the same sign. This is when the root $u$ is
also an extremum of the polynomial. It is easy to verify when this
happens. We simply differentiate the polynomial w.r.t. $u$ and use
(\ref{mass formulas}) to find the following answer
\begin{equation}\label{}
(u+3w)(u^2-w^2+4w)=0\ .
\end{equation}
The points on the orientation curve $u+3w=0$ are not included in the
moduli space. Therefore we have that there is a single $u$ when
$\bar{M}_*$ and $w$ are such that $u^2-w^2+4w=0$. We will see below
that this is the \emph{stability curve} i.e. the curve which
separates the radially stable from the unstable solutions on the
moduli space as we will see below (see also figure \ref{ddVfinal}).

A related fact is given in the following
\begin{remark}\label{stability curve and mass extrema}
For fixed $\alpha$ and $\Lambda$ we can think of the masses as
functions of the radius of the shell $a_0$$:$
$M_{(\pm)}=M_{(\pm)}(a_0)$. The function $M_{(+)}(a_0)$ has a global
minimum and the function $M_{(-)}(a_0)$ has a global maximum for
radii $a_0$ given by $u^2-w^2+4w=0$.
\end{remark}

Thus, there is simpler question one may ask: Given \emph{pair} of
masses $M_{(+)}$ and $M_{(-)}$, when can the matching happen at more
than one shell radii $a_0$?

The answer is that this can never happen:
\begin{proposition}\label{Unique_proposition}
For any $w$, any $u$ such that $(w,u)$ belongs to the moduli space
gives a pair of mass parameters $\bar{M}_{(+)}$ and $\bar{M}_{(-)}$.
Then, this is the unique $u$ that gives these mass parameters.
\end{proposition}
The proof is given in appendix \ref{Moduli_Appendix}. Therefore,
remembering that $u$ is single valued in terms of the shell radius
$a_0$, the junction conditions define a single-valued function
$a_0=a_0(M_{(-)},M_{(+)})$, in fact one-to-one on the space of the
allowed values of $M_{(\pm)}$. As we know from section
\ref{Junction_section} $a_0$ is a symmetric function of $M_L$ and
$M_R$ and it is given by $a_0=a_0(M_{(-)},M_{(+)})$, via the
correspondence implied in (\ref{swap}). Thus  given the bulk
metrics, the $a=$constant vacuum shell is \emph{unique}. So we see
that a weakened version of uniqueness of solutions does survive.
Note that for shells with standard orientation there are exactly two
inequivalent configurations corresponding to the same shell radius,
depending on whether $M_{(+)}$ is the mass of the inner or the outer
region.

\subsection{The spectrum of curves}\label{curves spectrum section}
We notice that, throughout the computations, the
quantity\begin{equation}\label{}W \equiv w^2-4w=w(w-4)\
,\end{equation}appears often. Now we comment on how it turns out to
be convenient to extract information on the moduli space. First,
notice that $W$ clearly vanishes at $w=0$ and $w=4$. We also
encounter the curves
\begin{align}\label{}u^2=\pm W,\quad u^2=\pm 3W, \quad
 \quad u=\pm w, \quad u=\pm 3w, \quad u = 0\ ;
\end{align}
which in detail correspond to
\begin{align}&
 u^2= 3(w^2-4w): \quad \text{$\bar{M}_{(+)}+\bar{M}_{(-)}=2\bar{M}$
 curve}, \nonumber\\& u^2= -3(w^2-4w): \quad \text{causality curve},
 \nonumber\\& u^2=(w^2-4w): \quad \text{stability curve},
 \nonumber\\& u^2=-(w^2-4w): \quad
 \text{$\bar{M}_{(+)}+\bar{M}_{(-)}=0$ curve},\\\label{Nice_curves}&
 u=\pm 3w: \quad \text{orientation curve}, \nonumber\\& u=\pm w:
 \quad \text{branch curve}, \nonumber\\& u=0: \quad \text{boundary
 curve (where $\bar{M}_{(\pm)}=\bar{M}$)}\ . \nonumber
\end{align}
We also found the curve where $\bar{M}_{(\pm)}=0$ to
be
\begin{eqnarray*}
 && (u-w)^2-4w=0\ , \quad  \text{i.e.}\quad  u=u^{(\pm)}=\pm 2\sqrt{w}+w : \quad
 \text{zero minus-mass curve}\ , \\&& (u+ w)^2-4w=0\ , \quad
 \text{i.e.}\quad u=-u^{(-)}=2\sqrt{w}-w : \quad \text{zero plus-mass
 curve}\ ,
\end{eqnarray*}
respectively. And notice that in terms of
$W$ this simply reads
\begin{equation}\label{}   u^{(+)}u^{(-)}=W \ .
\end{equation}

The first four curves in our list, which involve $W$, are conic
sections with symmetry axes the lines $w=2$ and $u=0$. The
orientation and branch curves on the other hand have symmetry axes
the lines $w=0$ and $u=0$. The conic sections and especially the
causality curve which is an ellipse, $u^2+3(w-2)^2=12$, break the
symmetry between positive and negative values of $w$. The image of
the causality curve around $w=0$ would be centered at $w=-2$ i.e.
$x=-3$.

The above analysis manifestly shows that the quantity $W\equiv
w^2-4w$ captures much important information about the moduli space.

\begin{figure}[]
\begin{center}{\includegraphics[width=.55\textwidth]{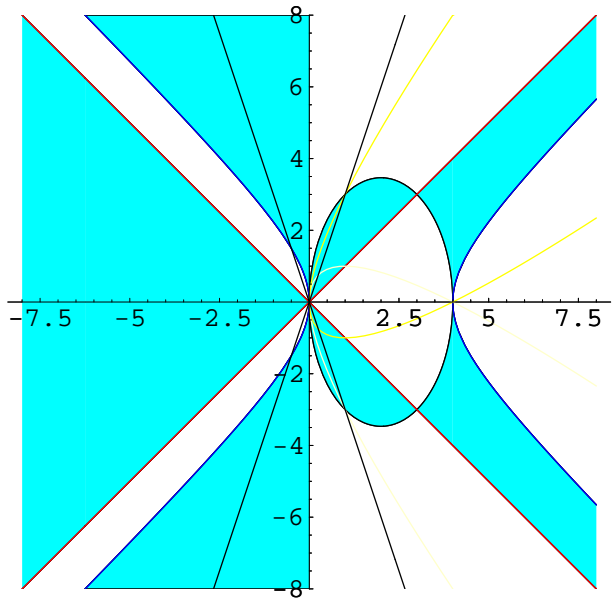}}
\end{center}\caption{{\small
{The shaded regions are where $V''(a_0)>0$. For the timelike shells
(outside of the ellipse), the unshaded regions correspond to
solutions unstable with respect to radial
perturbations.}}}\label{ddVfinal}\end{figure}

Actually, the parameterization of the space of solutions in terms of
variables ($u,w$) had shown to present advantages in order to
classify the whole set of solutions. To emphasize this, and for
completeness, let us also express the regions of radial stability
over the moduli space in terms of these variables. Such regions are
known to be characterized by the second derivative of the effective
potential, which in terms of $u$ and $w$ is seen to be
\begin{equation}\label{}V''(a_0)=-\frac{1}{a_0^2}\,\frac{w\,\big(u^2-w^2+4w\big)}{(u^2-w^2)\big(u^2+3(w^2-4w)\big)}\
.\end{equation} The regions where $V''(a_0) >0$ are shown in figure
\ref{ddVfinal}. $V''(a_0)=0$ along the curve $u^2-w^2+4w=0$ which we
have already called the stability curve, for reasons that become
clear now. According to remark \ref{stability curve and mass
extrema} this is where the mass $\bar M_{(\pm)}(a_0)$ have extrema.

\section{Nontrivial features of $C^0$ metrics}

\subsection{Topology}

In Einstein gravity in four dimensions there is a variety of smooth,
everywhere non-singular vacuum configurations in general
characterized by some non-trivial topological property, e.g. Euler
number. Any topological property they may have is an intrinsic
feature of the smooth solution.

In five dimensions and in Einstein-Gauss-Bonnet gravity similar
configurations may exist as well. The equations of motions though
are such that one can manufacture, by cut and paste along the
world-volume of vacuum shells, similar kind of solutions with the
difference that they are not smooth, i.e. not $C^1$. In this case
there is no intrinsic property in the vacuum solution, we are simply
building objects which are much simpler locally. For that reason one
may call these objects non-topological, though they certainly have
non-trivial topological features. An analogy for this would be the
difference existing between an object with an exactly given smooth
metric which has the topology, e.g. of the sphere, and tetrahedra
built out of flat pieces.

This digression leads us to recognize a great difference with
respect to four dimensions. Unlike in four dimensions, in
five-dimensional Lovelock theory, spacetimes which are defined by
some simple property locally, for example being vacuum and
spherically symmetric, are by no means well defined globally, if
smoothness is given up. For each such metric, which may itself have
non-trivial topological features, one can construct infinitely many
other spacetimes by cut and paste which locally are given by the
same simple property. That is, the theory allows for many different
topologies where one would expect it to allow only for different
coordinates.

A general analysis of the objects obtained by geometric surgery
along vacuum shells is an interesting problem and contains much of
the actual physics of five-dimensional Lovelock gravity (that is,
Einstein plus the Gauss-Bonnet term). In this work we mainly focus
on the direct implications of their existence illustrated by
appropriate examples. A systematic analysis is left for future work.
Below we analyze how a constant curvature vacuum is modified by
wormholes (and related configurations). It turns out that, the
smaller such constructions with wormholes are with respect to the
scale set by $\alpha$, the more complicated the topology can be.

\subsection{Holes in the vacuum}\label{Octopus section}

An interesting special case of a wormhole is when on one side we
have pure vacuum, as mentioned already in section 4. Starting from a
constant curvature background, by introducing the vacuum wormhole we
cut a hole in the constant curvature manifold, replacing it with an
``outgoing'' spacetime region of mass parameter $M$. Of course the
topology of the vacuum is not the same anymore; there are now
non-contractible 3-spheres. Nevertheless, it turns out that these
configurations are everywhere non-singular in the following sense:
\emph{the only singularities that exist in spacetime are
integrable}\footnote{The curvature and Lovelock tensor are singular
at the shell but only in the sense of delta functions. Local
integrals of these quantities are finite and the physical laws
defined by the field equations do not break down there. In this
sense the solutions are not singular.}.

These wormholes belong to a more general class of constructions:
Depending on whether the vacuum shell is time- or space-like and the
orientation of the matching (i.e. the different combinations of the
orientation signs $\eta_L$ and $\eta_R$), one obtains distinct types
of configurations some of which contain only integrable
singularities. Configurations we call ``instantons'' mentioned below
belong to this more general class.

%Besides, there is a second kind of feasible construction: A natural
%possibility for this construction is that of being a standard shell
%instead of a wormhole. That is, we may consider to match two metrics
%with the same orientation, with one having mass parameter equal to
%zero. In this case the topology is the same as before, up to the
%inhomogeneities introduced by the open balls which belong to a
%different spacetime. These configurations contain singularities
%since there is a singularity at the origin of each ball. This is the
%case where the inner spacetime is the massive one. A third
%conceivable type of configurations is when the inner spacetime is
%the one with zero mass. (This is actually not of ``holes in the
%vacuum'' kind exactly but it is obviously related).

%We will see below that standard shells here must be spacelike.
%$r=a_0$ spacelike shells are impossible in AdS background and one
%easily sees that in the de Sitter one the configuration must be a
%wormhole.

The analysis gets simplified and clarified if we express everything
in terms of the constant curvature. When the mass parameter is zero
the metric is defined as
\begin{equation}\label{vacuum}
f(r)=1-K r^2\ .
\end{equation}
This means that for the metric of branch $\xi$ we have:
\begin{equation}\label{K}
4\alpha K=-(1+\xi \sqrt{w})\ .
\end{equation}
These configurations exist for $w>0$.

We consult Remark \ref{overall zero mass} and also Proposition
\ref{zero mass proposition 0}. The relevant points on the moduli
space are on the curve $u=|w-2\xi \sqrt{w}|>0$. According to
Proposition \ref{pre total moduli} the points on this curve such
that $u>3w$ correspond to standard shell configurations. In detail,
standard shells are the configurations corresponding to:
$u=2\sqrt{w}-w$ for $w \in (0,1/4)$, and $u=2\sqrt{w}+w$ for $w \in
(0,1)$.

$u=|w-2\xi \sqrt{w}|$ is a continuous curve. The points with $w=0$
and $w=4$ do not belong in the moduli space. The same for the points
with $w=1/4$ and $w=1$. So in all, from (\ref{K}) we have that
$4\alpha K \ne -3,-3/2,-1, 0$.

Now from Remark \ref{overall zero mass} and Proposition \ref{zero
mass proposition 0} one finds that the mass in all cases is
\begin{equation}\label{mass on vacuum}
M=9\alpha\, \frac{(4\alpha K+1)^3}{(4\alpha K)^2(4\alpha K+3)}\ .
\end{equation}
Also
\begin{equation*}
\Lambda=6 K+ 12 \alpha K^3.
\end{equation*}

We have that this construction is possible when $w>0$. Therefore,
from Remark \ref{alpha>0}, we have that the sign of $\alpha$ depends
solely on the causal character of the vacuum shell. Namely, it is
$\alpha>0$ when the shell is timelike, and $\alpha<0$ when the shell
is spacelike. We have the following
\begin{remark}
 All standard orientation shell configurations with zero mass in one of the bulk regions
 are spacelike.
\end{remark}
\begin{remark}
The variable $4 \alpha K$ takes values on the entire real line with
the exception of the points $-3, -3/2,-1,0$. With these exceptions
in mind we have:

Spacelike shells  i.e. $\alpha<0$$:$ $4 \alpha K \in (-3,0)$. In the
interval $(-3/2,0)$ exist all the standard shell configurations.

Timelike shells i.e. $\alpha>0$$:$ $4\alpha K \in (- \infty, -3)
\cup (0, \infty).$
\end{remark}

The mass $M$ has poles at the boundary of the spacelike shell
region. One may note that thought of as a function of $\alpha$ both
poles are of first order.

From the formula $u=|w-2\xi \sqrt{w}|>0$ we find for the radius of
the shell in all cases is
\begin{equation}\label{}
a_0^2=K^{-1}\left(1+\frac{4\alpha K}{3}\right)^{\!-1}.
\end{equation}

\begin{figure}[t]
\begin{center}
{\includegraphics[width=0.6\textwidth,angle=270]{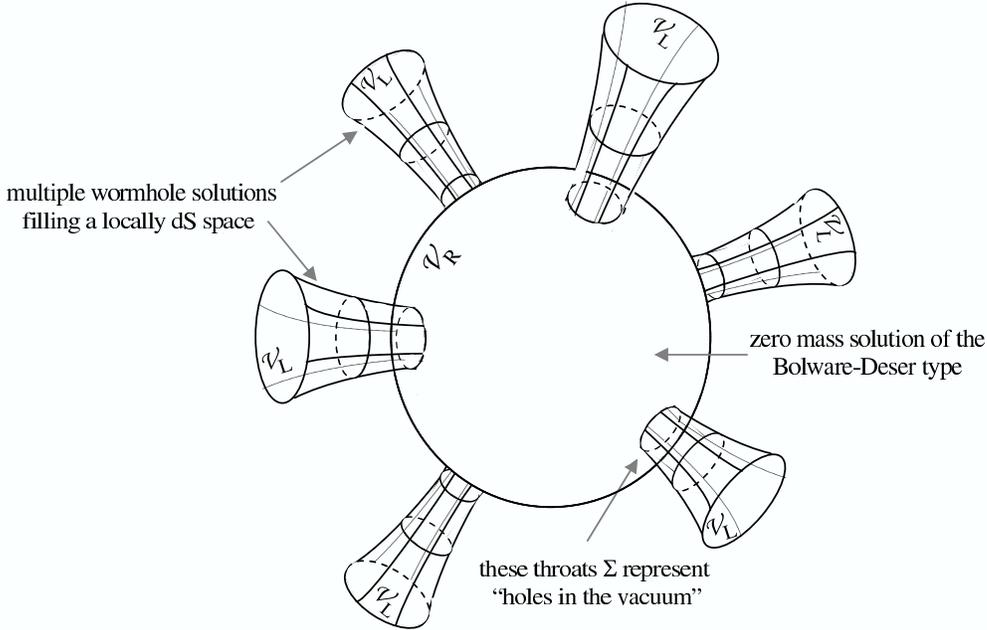}}
\end{center}
\caption{{\small{$K>0$. When $\alpha>0$ the solution is a
multi-soliton, with multiple asymptotic massive regions joined to a
de Sitter space. A spatial slice of a multi-soliton is sketched
above. When $\alpha<0$ and $0< K<3/8|\alpha|$ the solutions are
multi-instantons. The radii of the instantons have a lower bound:
$a_0^2>16|\alpha|/3$. }}}%
\label{octopus}%
\end{figure}

The vacuum of constant curvature $K$ is a locally homogeneous
spacetime and in particular is locally spatially homogeneous. Having
placed one vacuum shell around some arbitrarily chosen origin, we
have seen that outside of the shell the homogeneity is everywhere
maintained. As long as it does not cross the first, we may place a
second vacuum shell and in fact an arbitrary number of them
modifying the manifold in a way depicted in Fig. \ref{octopus}.

Let $K>0$ and $\alpha>0$. It is useful to rewrite this as
\begin{equation}\label{}
\alpha K=\frac{3}{8}\left\{\sqrt{\frac{16}{3
}\frac{\alpha}{a_0^2}+1}-1\right\}\ .
\end{equation}
It is clear from both the last formula that in units of $\alpha$ the
radius of the shell is an increasing function of the radius of
universe $1/\sqrt{K}$. When the shell is microscopic, i.e. small in
units of $\alpha$, we have that $K a_0^2 \ll 1$. When the shell is
macroscopic we have that $K a_0^2 \simeq 1$. A microscopic universe
could fit roughly
\begin{equation}\label{}
(K a_0^2)^{-2} =\frac{1}{4}\left(\sqrt{\frac{16}{3
}\frac{\alpha}{a_0^2}+1}+1\right)^{\!\! 2}
\end{equation}
vacuum shells of radius $a_0$. So the more microscopic the universe
the more complicated its topology can be.

\subsection{Black hole spectrum and degeneracy}\label{Black hole spectrum and degeneracy}

Reversing in a sense our viewpoint from the previous section, we may
think of the matching of a massive metric with one of constant
curvature along a vacuum shell, as a way to eliminate the
singularity at the origin, or better to replace it with an
integrable singularity along the vacuum shell.

Consider for example configurations along the curve $u=2 \sqrt{w}+w$
and $\alpha>0$ i.e. $w>1$ (and the vacuum shell is static). Then
$K>0$, that is $M>0$ and $\Lambda>0$, and the massive branch is an
exotic branch ($\xi=+1$). This metric alone has a naked singularity
at the origin. By constructing the vacuum wormhole we have managed
to replace a region around the origin with a region of a de Sitter
spacetime which contains the horizon. That is, the spacetime which
asymptotically looks like an exotic branch, massive, Boulware-Deser
spacetime is actually everywhere non-singular and has
%the causal
%diagram of a black hole. The black hole in fact does not conceal a
%singularity.
horizons. We might reasonably expect that thermal effects of the
horizons will be felt in this would-be singular spacetime.

The mass parameter $M$ in the massive region is determined by the
curvature $K$ of the de Sitter region. So then, the de Sitter space
mimics a particle, or some fairly localized mass, as viewed from
sufficiently far away. The entropy $S$ related to the existence of
the de Sitter horizon depends on $K$ and therefore on $M$. We expect
$\partial S/\partial K>0$. We know that $\partial M/\partial K<0$.
Therefore that entropy will decrease with $M$. This is not
surprising since a positive mass in the exotic branch behaves
effectively like a negative gravitational mass.

The previous example shows that the spectrum of black holes in
Einstein gravity modified by the Gauss-Bonnet term is not the same
when $C^0$ metrics are allowed, compared to the smooth
Boulware-Deser metrics. A space which by an asymptotic observer who
thinks in terms of smooth metrics would not be recognized as a black
hole might actually be one. Conversely, a spacetime which
asymptotically would be a recognized as a Boulware-Deser black hole,
could actually be a spacetime with a naked singularity, or a black
hole different to the one expected.

Consider, for example, the case $\Lambda=0$ and $\alpha<0$. From the
analysis in appendix \ref{BD appendix} we see that this spacetime is
a black hole for $M>|\alpha|$ (and $\xi=-1$). The horizon hides an
inner branch singularity. Our analysis in section
\ref{Moduli_section} shows that we can cut this spacetime and match
it with wormhole orientation along a spacelike shell, i.e. within
the horizon, with another spacetime of exotic branch metric. In that
spacetime $r$ is everywhere a timelike variable. Thus although
outside the horizon spacetime looks like a specific Boulware-Deser
black hole spacetime it can actually be a different one. The two
different states have the same energy as measured at spatial
infinity and horizons with the same properties: as black holes they
must be degenerate. Whether the usual entropy calculations take into
account the effects of this degeneracy in the number of states is
not clear to us. It is amusing to think that the modifications to
the usual Bekenstein-Hawking formula in the presence of the
Gauss-Bonnet term, see e.g.~\cite{Cai:2001dz}, are essentially due
to such degeneracies.

\subsection{Other types of shells}

The analysis  has focused on the spherically symmetric case ($k=1$).
This can readily be extended to the case of $k= -1$ (where the bulk
metrics are taken to be topological black holes, with horizons some
compactified hyperbolic space, or the corresponding naked
singularity spacetime). Similar features are expected to occur
(wormholes and shells of standard orientation exist, typically
involving the exotic plus branch.) Also the case of $k=0$ for
toroidal black holes or naked singularity spacetimes, can be
investigated.

We have seen that spacelike shells exist, representing a sudden
transition from one solution to another. These present problems in
terms of the predictability of the field equations. It would be
useful to know whether the shells are generic or if they only occur
for a certain range of the coupling constants and mass parameters.

The Euclidean version of the $C^0$ wormholes may be important for
estimating the transition rate between the (unstable) plus branch
and the (stable) minus branch solutions.

These are left for future work.

\subsection{On uniqueness and staticity of
solutions}\label{uniqueness}

In this work we construct and analyze solutions of
Einstein-Gauss-Bonnet gravity whose metric is class $C^0$, piecewise
analytic in the coordinates. The solutions are made by joining
together two spherically symmetric pieces. Since the shell itself
admits $SO(4)$ isometry group, the resulting global spacetime is
spherically symmetric. To put things into this context and discuss
the special implications of low differentiability we start by
reviewing the existing relevant theorems in Einstein and Lovelock
gravity.

We start with a uniqueness and staticity theorem, applying to
Lovelock gravity in general, which imposes the stronger conditions
on differentiability.
\begin{theorem}[Ref.~\cite{Z}] For generic values of the
couplings (including the cosmological constant), class $C^2$
solutions of the Lovelock gravity field equations with spherical,
planar or hyperbolic symmetry are isometric to the corresponding
static solutions.

In particular, in Einstein-Gauss-Bonnet gravity in five dimensions
$C^2$ solutions with spherical symmetry are isometric to the
Boulware-Deser solutions when $\Lambda \ne -3/4\alpha$.
\end{theorem}
%\begin{corollary}
%$ The spherically symmetric vacuum $C^2$ solution of five-dimensional
%Einstein-Gauss-Bonnet gravity with cosmological constant, is a
%unique (up the branch
%$\xi$) one-parameter family of metrics, when the couplings $\alpha$
%and $\Lambda$
%satisfy $\Lambda \ne -3/4\alpha$.
%\end{corollary}

When we let the metric become merely continuous at hypersurfaces, we
have seen already that one can construct many different
time-independent solutions of the vacuum field equations: for
example, when $\Lambda=0$ with $\alpha>0$, one can construct
multiple concentric vacuum discontinuities separating Boulware-Deser
solutions. So uniqueness of black hole solutions does not hold for
$C^0$ metrics in Lovelock gravity. In fact neither does staticity.
We return to discuss this below, after we revisit the corresponding
theorems in Einstein gravity.

\begin{theorem}[Ref.~\cite{PapapetrouTreder}\cite{Bergmann}]
A differentiability class $C^0$ and spherically symmetric vacuum
solution of Einstein gravity is: $i$) static, $ii$) equivalent to
the Schwarzschild solution.
\end{theorem}
That a spherically symmetric vacuum solution is static can be shown
by finding a timelike Killing vector, which also happens to be
hypersurface orthogonal, even when the solution is given in forms
that don't look very much like the usual Schwarzschild metric and
which assume lower differentiability \cite{Petrov}, see
\cite{Bergmann}.

\begin{theorem}[Ref.~\cite{PapapetrouTreder}]
A $C^0$ solution of the Einstein gravity field equations is well
defined as the limit of a sequence of (at least) $C^2$ solutions.
The metric is assumed to become $C^0$ only at smooth hypersurfaces.
\end{theorem}
Fields of low differentiability, e.g. with a discontinuous first
derivative, can be understood as solutions of field equations in the
weak sense, as limits of sequences of smoother fields. The fact that
this limit is well defined makes the junction conditions of Israel
well defined (the above work appeared earlier than Israel's famous
work). Now based on the junction conditions one may conclude: any
hypersurfaces where the metric is not smooth must be a null
hypersurfaces (we may call them shock waves). Then one may show that
there are no spherically symmetric shock waves in Einstein gravity,
see e.g. \cite{Bergmann}.

The result regarding limits of smooth metrics holds in
Einstein-Gauss-Bonnet and in fact in Lovelock gravity in general
(see the appendix of Ref. \cite{Gravanis:2007ei}).
\begin{theorem}\label{Limit_Theorem}
A $C^0$ solution of Lovelock gravity field equations is well defined
as the limit of a sequence of (at least) $C^2$ solutions. The metric
is assumed to become $C^0$ only at smooth hypersurfaces and their
intersections.
\end{theorem}
 So considerations
related to uniqueness or non-uniqueness similar to the above are
meaningful in Lovelock gravity as well. In this paper we have
demonstrated:
\begin{theorem}\label{Non_unique}
 There exist spherically symmetric $C^0$ solutions of Einstein-Gauss-Bonnet
 gravity in five dimensions which are not
 given by the Boulware-Deser metric, but rather they are piecewise
 of the Boulware-Deser form.
 There exist solutions which are not static.
\end{theorem}
In section \ref{Moduli_section} we found that for any value of the
couplings $\alpha$ and $\Lambda$ such that\footnote{According to
that section, $y (w-1) w=y x (x+1)>0$ for timelike i.e. static
shells and $y x (x+1)<0$ for spacelike i.e. instantaneous shells.
Via the simple relations of $x$ and $y$ to the couplings these read
for non-zero $\Lambda$: $3/4\alpha+\Lambda>0$ and
$3/4\alpha+\Lambda<0$ respectively. As can be seen from the results
of section \ref{Moduli_section} they actually hold for $\Lambda=0$
as well.}~$\Lambda> -3/4\alpha$, there exist static
(time-independent) vacuum shells: spherically symmetric $C^0$ vacuum
metrics are not unique for a wide range of couplings $\alpha$ and
$\Lambda$ in Einstein-Gauss-Bonnet gravity in five dimensions.
%Also there is
%evidence of this for generic values of $\alpha$ and $\Lambda$.
One can in fact construct arbitrarily complicated spherically
symmetric configurations by having an infinity of concentric
discontinuities. The exotic branch ($\xi=+1$) is typically involved.
Though the radius of a static vacuum shell is uniquely fixed by the
metrics in the bulk, $C^0$ static metrics are to a high degree
non-unique as one does not a priori know how many vacuum shells
there may be in spacetime.

Now recall section \ref{dynamical general section}. The
time-dependent solutions, i.e.~non-static ones, exist \emph{always}:
For any non-zero value of $\alpha$ and any value of $\Lambda$ there
exists\footnote{In fact, it exists for a wide range of the bulk
metric masses $M_L$ and $M_R$, possibly for all values of the masses
for which the metrics are real. What is more important, for given
values of the couplings $\alpha$ and $\Lambda$, for any given
Boulware-Deser metric one can construct a time-dependent vacuum
shell for some other Boulware-Deser metric on the other side.} a
time-dependent vacuum shell solution $a(\tau)$. The shell radius
function $a(\tau)$ and the orientation signs $\eta_L$ and $\eta_R$,
completely define the world-volume of the shell intrinsically as
well as its embedding in spacetime (section \ref{the sting}). That
is, they define a $C^0$ metric in spacetime. Therefore a non-static
$C^0$ metric which respects everywhere spherical symmetry can always
be constructed in Einstein-Gauss-Bonnet gravity with cosmological
constant (which can be also zero). However in section \ref{dynamical
general section} we have obtained a general result concerning shells
with bulk metrics which have a well defined General Relativistic
limit as $\alpha \to 0$ (the $\xi=-1$ branch). In the range of
parameters $1+\frac{4\alpha \Lambda}{3}>0$ all shell solutions
involving the minus branch are unstable in the sense that they can
not be in a stable static state, neither can they perform bounded
oscillations.

%Looking back at the propositions and remarks of the previous
%sections one may come up with a conjecture.
%\begin{conjecture}
%Consider a subspace of the space of $C^0$ solutions of
%Einstein-Gauss-Bonnet gravity such that, $i$) they are piecewise
%smooth, such that all smooth regions have a well-defined Einstein
%gravity limit ($\alpha \to 0$), $ii$) they do not contain naked
%singularities. Then, vacuum wormhole or standard shell solutions are
%not possible and uniqueness and staticity theorems essentially hold.
%\end{conjecture}

Theorem \ref{Non_unique} shows that uniqueness does not apply to
$C^0$ metrics. How is this to be interpreted? One could simply
reject non-smooth metrics as unphysical. However, according to
Theorem \ref{Limit_Theorem} these $C^0$ solutions are well defined
as the limit of a family of smooth geometries. As such, they
approximate arbitrarily closely to some smooth solution of the
theory. Now suppose $g_{\mu\nu}^{(n)}$ is a family of smooth metrics
which converge to a spherically symmetric vacuum shell solution as
$n \to \infty$. For finite $n$, $g^{(n)}_{\mu\nu}$ can not be a
spherically symmetric vacuum solution, because the uniqueness
theorem holds for smooth metrics. So it must either deviate slightly
from spherical symmetry or have some small amount of matter as
source. Assuming that suitable $g_{\mu\nu}^{(n)}$ can be constructed
which obey the energy conditions, our results can be taken as
evidence for the generic existence of such exotic features as smooth
wormholes in this theory.

\section{Conclusions}

In this paper we have presented a method for generating new exact
vacuum solutions of five-dimensional Lovelock theory of gravity. The
solutions we obtained are spherically symmetric spacetimes whose
metrics are $C^0$ functions, and are composed by patches of
different five-dimensional Boulware-Deser spacetimes.

The proof of the Birkhoff's theorem  for this theory (see e.g. Refs.
\cite{DF}, \cite{Wheeler}) involves an assumption of
differentiability \cite{Z}. We have seen that if this assumption is
relaxed then there are $C^0$ metrics (which satisfy the field
equations in the distributional sense). The Birkhoff's theorem still
holds, but merely in a (weaker) piece-wise form. Uniqueness and
staticity of the metric turns out to be valid only locally in
regions where the metric is differentiable.

We have used a geometric surgery procedure, employing the junction
conditions of Einstein-Gauss-Bonnet theory to join spherically
symmetric pieces of spacetime. This lead us to find different
geometries with quite interesting global structure. In particular,
we have shown that vacuum wormholes do exist in this theory. The
wormholes connect two different asymptotically (Anti)de-Sitter
spaces and, in certain sense, they represent gravitational solitons
in five-dimensions. Although their metrics are not $C^1$ functions,
they are globally static vacuum solutions of gravity equations of
motion and have finite mass. These metrics, being non differentiable
where the wormhole throat is located, still represent exact
solutions defined everywhere, provided the junction conditions are
obeyed. This is ultimately due to cancelations among different terms
in the junction conditions.

We have analyzed both static and dynamical solutions and, related to
this, we have pointed out a new type of classical instability that
arises in Einstein-Gauss-Bonnet gravity for certain range of the
Gauss-Bonnet coupling. This concerns fundamental aspects such as
predictability and uniqueness.

\[
\]

\textbf{Acknowledgements:} C.G. and G.G. thank J. Oliva and R.
Troncoso, for useful discussions. They are grateful to the Centro
de Estudios Cient\'{\i}ficos CECS for the hospitality during their
stays, where part of this work was done. Also, they specially
thank M. Leston for helpful discussions. G.G. also thanks C.
Bunster, A. Gurzinov, G. Gabadadze and M. Kleban for conversations; and
thanks the financial support of Fulbright Commission, Universidad
de Buenos Aires, CONICET and ANPCyT through grants UBACyT X816,
PIP6160 and PICT34557. C.G. is doctoral fellow of CONICET,
Argentina. S.W. wishes to thank A. Giacomini, H. Maeda, J. Oliva
and R. Troncoso for many useful discussions and IAFE and
Universidad de Buenos Aires for warm hospitality. S.W. gratefully
acknowledges funding through FONDECYT grant $N^{o}$ 3060016 and
support to CECS from Empresas CMPC, the Millennium Science
Initiative, Fundaci\'{o}n Andes, the Tinker Foundation.

\section*{Appendix}

\appendix

\numberwithin{equation}{section}

\section{Spherically symmetric solutions in Einstein-Gauss-Bonnet gravity}

The spherically symmetric static solution of Einstein-Gauss-Bonnet
theory of gravity was obtained by Boulware and Deser in 1985, [Ref].
In five dimensions, and in terms of a suitable Schwarzschild-like
ansatz (\ref{ansatz}), the metric is given by (\ref{BD_metric}).
%\begin{equation}
%ds^{2}=-f_{\pm }(r)dt^{2}+\frac{1}{f_{\pm }(r)}dr^{2}+r^{2}d\Omega
%_{3}^{2} \label{intervalo}
%\end{equation}%
%where $d\Omega _{3}^{2}$ is the metric of a unitary $3$-sphere;
%namely
%\begin{equation*}
%d\Omega _{3}^{2}=d\theta ^{2}+\sin ^{2}\theta d\chi ^{2}+\sin
%^{2}\theta \sin ^{2}\chi d\varphi ^{2},
%\end{equation*}%
%and where the function $f_{\pm }(r)$ is%
%\begin{equation}
%f_{\pm }(r)=1+\frac{r^{2}}{4\alpha }\pm \frac{r^{2}}{4\alpha }\sqrt{1+\frac{%
%16\alpha M}{r^{4}}+\frac{4\alpha \Lambda }{3 }}. \label{sol}
%\end{equation}%
%The existence of two different solutions, $f_{+}(r)$ and $f_{-}(r),$
%is due to the fact that the equations of motion of
%Einstein-Gauss-Bonnet theory are quadratic in the derivatives of the
%metric. These two solutions are called the plus branch and the minus
%branch respectively. The parameter $M$ arises as an integration
%constant, and it is related to the mass of the solution $m$
%through the relation $m=\frac{6\pi ^{2}}{\kappa ^{2}}M$. Metric (\ref%
%{intervalo}) represents a black hole solution with horizon located at $%
%r=r_{H}>0$, with $f_{\pm }(r_{H})=0$.
On the other hand, if the
higher dimensional theory with the dimensionally extended quadratic
Gauss-Bonnet term is considered, then the black hole solutions take
a similar form, namely
\begin{equation}
f_{\pm }(r)=k+\frac{r^{2}}{4\alpha }\pm \frac{r^{2}}{4\alpha }\sqrt{1+\frac{%
\alpha M_{D}}{r^{D-1}}+\alpha \Lambda _{D}}  \label{solnueva2t}
\end{equation}%
where $M_{D}$ is an integration constant, and $\Lambda _{D}$ is a
numerical factor proportional to $\Lambda $ and which only depends
on the dimension $D$. The ambiguity in expressing the sign $\pm $
reflects the existence of two branches, and corresponds to the
parameter $\xi $ introduced in section 2. The parameter $k$ takes
the value $k=1$ in the case where the base manifold corresponds to
the unitary $(D-2)$-sphere $d\Omega _{D-2}^{2}$; besides, $k$ may
take the values $-1$ or $0,$ if the base manifold is of negative or
vanishing constant curvature, respectively.

Now, let us analyse the large distance limit of the solution (\ref{BD_metric}%
). Asymptotically, this solution tends to the five-dimensional
Schwarzschild solution when $\alpha \rightarrow 0$, as it is
naturally expected. Namely
\begin{equation}
f_{-}(r)\simeq 1-\frac{2M}{r^{2}}-\frac{\Lambda }{6}r^{2},
\label{sol1}
\end{equation}%
which represents a (A)dS-Schwarzschild black hole in five
dimensions. Notice that the large $r^{2}/\alpha $ limit of the
solution $f_{+}(r)$ acquires a large additional  cosmological
constant term $\sim \frac{r^{2}}{2\alpha }$. In particular, this
implies that (A)dS space-time is a solution of the theory even for
the case $\Lambda =0$.

It is also worth noting that in the case of non-vanishing
cosmological constant, besides the leading term in the expansion
(\ref{sol1}), we find finite-$\alpha $ corrections to the black hole
parameters \cite{AFG}. Namely
\begin{equation}
f(r)=1-\frac{2m_{d}}{\pi r^{2}}-\frac{\Lambda _{d}}{6}r^{2}+\mathcal{O}%
(\alpha r^{-6})  \label{jjj45}
\end{equation}%
where the \textit{dressed} parameters $m_{d}$ and $\Lambda _{d}$ are
given by
\begin{equation*}
\Lambda _{d}=\Lambda \left( 1+\sum_{n=2}^{\infty }c_{n} \ x
^{n-1}\right) = 1-\sqrt{1+x} ,\quad m_{d}=m\left(
1+\sum_{n=2}^{\infty }n\ c_{n} \ (-x) ^{n-1}\right) ,
\end{equation*}%
with
\begin{equation*}
c_{n}=\frac{(2n-3)!!}{2^{n-1}n!}\ ,\ \ \ x := \frac{4}{3}\Lambda
\alpha .
\end{equation*}%
It is important to emphasize the difference existing between
(\ref{sol1}) and (\ref{jjj45}): While the first corresponds to the
actual limit $\alpha \rightarrow 0$, the second represents the large
$r^{2}/\alpha $ regime which
takes into account finite-$\alpha $ contributions. For instance, the finite-$%
\alpha $ corrections to the mass are found by simply collecting the
coefficients of the Newtonian term $\sim r^{2}$. The parameter $x $
controls the \textit{dressing} of the whole set of black hole
parameters. The above power expansion converges for values such that
$x <1$. On the other hand, for $x > 1$ we find a different
expansion, leading to the following \textit{dressed} parameters in
the large $r$ regime
\begin{equation*}
m_{d} = \frac {m}{\sqrt{| x | }} \left( 1+\sum _{n=2}^{\infty} n \
c_n \ (-x) ^{1-n} \right)
\end{equation*}
Thus, we note that the Newtonian term $\sim m_d r^{-2}$ vanishes in
the limit $| \Lambda \alpha | \to \infty$. The particular case $x =
1$ is discussed below. Moreover, it is possible to see that, if one
considers the
case $\alpha\Lambda > 0$, the effective cosmological constant in the large $%
x $ limit turns out to be
\begin{equation*}
\Lambda _d = \sqrt{\frac{3\Lambda}{\alpha}} -\frac {3}{2\alpha }+
\mathcal{O} (1/\sqrt {| x |}) \ .
\end{equation*}

One of the relevant differences existing between the black hole
solutions in Einstein theory and in Einstein-Gauss-Bonnet theory is
the fact that, in the latter, the metric does not diverge at the
origin of Schwarzschild
coordinates, $r=0,$ though its curvature is still singular. From (\ref{BD_metric}%
), we easily observe%
\begin{equation*}
f_{\pm }(r=0)=1\pm \sqrt{\frac{M}{\alpha }}.
\end{equation*}%
In particular, this implies that the metric presents a angular
deficit around the origin, and, also, that massive objects with no
even horizon exist; thus, these correspond to naked singularities.

Another interesting feature of the presence of the Gauss-Bonnet term
is
that, for the particular choice of the parameters $\alpha \Lambda =-\frac{3}{%
4}$, the solution takes the form%
\begin{equation}
f_{\pm }(r)=\frac{r^{2}}{4\alpha }-\mathcal{M}  \label{gaston}
\end{equation}%
where we have considered $\Lambda <0$ and $\alpha >0$, and where $\mathcal{M}%
+1=\sqrt{\frac{M}{\alpha }}$. This solution resembles the Ba\~{n}%
ados-Teitelboim-Zanelli black hole \cite{BHTZ,BTZ}. Actually, the solution (\ref%
{gaston}) shares several properties with the three-dimensional black
hole
geometry, as it is the case of its thermodynamics properties. Parameter $%
\mathcal{M}$ in Eq. (\ref{gaston}) plays the role of the mass $M$ in
the BTZ solution. For instance, just like $AdS_{3}$ space-time is
obtained as a
particular case of the BTZ geometry by setting the negative mass $%
M=-(8G)^{-1}$, the five-dimensional Anti-de Sitter space corresponds
to setting $\mathcal{M}=-1$ in Eq. (\ref{gaston}). Moreover, notice
that in the
large $\mathcal{M}^{-1}$ limit the solution becomes the metric to which $%
AdS_{5}$ tends in the near boundary limit. Similarly, the massless
BTZ corresponds to the boundary of $AdS_{3}$. Besides, as it was
already mentioned, a conical singularity is found in the range
$0<M<\alpha $ (corresponding to $-1<\mathcal{M}<0$), and this
completes the parallelism with the three-dimensional\ black hole.

\section{Properties of the Boulware-Deser metric}
\label{BD appendix}

The Boulware-Deser(-Cai) metric is given by (\ref{ansatz}) with
metric function (\ref{BD_metric}).\footnote{Only the the spherically
symmetric case $k=1$ was discussed by Boulware and Deser. The cases
$k=0,-1$ were analyzed later by Cai in \cite{Cai:2001dz}. As we are
mainly interested in the spherically symmetric case we will call
this metric Boulware-Deser.} The metric has two branches for given
cosmological constant $\Lambda$ and energy $M$: $\xi=\pm 1$. [These
we call as the plus- and minus- branch respectively; also, more
descriptively, as the ``exotic'' and the ``good'' branch.] Therefore
solving the vacuum field equations for spherically symmetric metrics
we obtain \emph{two} solutions. Asymptotically they read
\begin{equation}\label{BDasympt}
 f = \frac{1+\xi \sqrt{w}}{4\alpha} \  r^2+1+ \frac{2 \xi M}{\sqrt{w}
 \ r^2}+{\cal O}(r^{-4})\ ,
\end{equation}
where we use our variable $w \equiv 1+\frac{4\alpha\Lambda}{3}>0$.

For $\Lambda=0$, the $\xi=+1$ branch depends on $\alpha$
asymptotically, while the asymptotically flat branch $\xi=-1$ does
not. Also, the sign of the Schwarzschild type of term depends on the
branch: the two branches view the energy $M$ differently, i.e. the
exotic metric of the Boulware-Deser solution does not reduce to
Einstein solution in the ``infrared'' limit.

The sign $\xi$ is in some sense a charge which determines how a
certain energy $M$ enters a metric and thus if the field will be
attractive or repulsive. As noted in \cite{BD}, the graviton is a
ghost on the asymptotic $\xi=+1$ branch, because the linear Einstein
tensor appears to have the opposite overall sign (that is, this
metric is classically unstable). This wrong sign is reflected in the
inverted sign of the Schwarzschild term.

An interesting issue about the Boulware-Deser solutions is that it
contains a square root, whose reality imposes constraints. From
(\ref{BD_metric}) we see that: when $w<0$ there is a \emph{maximum}
radius; when $M/\alpha<0$ there is a \emph{minimum} radius in
spacetime. At those finite radii there exists curvature
singularities, known as branch singularities~\cite{GOT07}. We call
them outer and inner branch singularities, respectively to the cases
above. These unusual spacetimes can also have horizons behind which
the singularities are hidden.

We turn now to discuss the horizon structure of the Boulware-Deser
spacetimes. The following does not intend to be an exhaustive
analysis, it is rather a list of general formulas in our notation
useful for our purposes. We will use the dimensionless parameters
$w$ and $\bar{M}$. Recall the Boulware-Deser metric function $f(r)$
given in (\ref{BD_metric}) and define $r_H$ by $f(r_H)=0$. One finds
that if $w \ne 1$
\begin{equation}\label{r_H}
r_{H\pm}^2=4\alpha\,\frac{1\pm \sqrt{\bar{M}(w)}}{w-1}\ .
\end{equation}
We have defined the useful quantity
\begin{equation}
\bar{M}(w)=w+(1-w)\bar{M}\ ,
\end{equation}
which looks an interpolation between $\bar{M}$ and $1$.

From the definition of $r_{H+}$ we see that $r_{H+}>0$ if:
\begin{equation}\label{rH+>0}
\frac{3}{\Lambda}=\frac{4\alpha}{w-1}>0\ .
\end{equation}
That is $\Lambda>0$. Also $r_{H-}>0$ one finds that it is equivalent
to $M>\alpha$. Therefore we have:
\begin{remark}\label{elementary horizons}
Elementary conditions for the existence of $r_{H+}$ is $\Lambda>0$
and for the existence of $r_{H-}$ the condition $M>\alpha$.
\end{remark}

 When $0<|\alpha|<\infty$, ~$w=1 \Leftrightarrow \Lambda=0$. So the previous formula holds for
non-zero $\Lambda$. When $\Lambda=0$, the correct result can be
obtained as the limit $w\to 1$ of the previous formula for $r_{H-}$.
It reads
\begin{equation}\label{}
r_{H-}^2=2\alpha\, (\bar{M}-1)\ .
\end{equation}

We must substitute (\ref{r_H}) back to $f(r_H)=0$ to solve for the
signs. We have:
\begin{equation}\label{horizon 1}
-\xi=\text{sign}\left(\frac{w\pm\sqrt{\bar{M}(w)}}{1\pm\sqrt{\bar{M}(w)}}\right)\
,
\end{equation}
for $r_{H\pm}$ respectively. Again the case $w=1$ i.e. $\Lambda=0$
can be correctly obtained from the limit $w\to 1$ for $r_{H-}$.
Explicitly it reads
\begin{equation}\label{horizon 1 w=1}
-\xi=\text{sign}\left(\frac{\bar{M}+1}{\bar{M}-1}\right)\ .
\end{equation}
We have used the sign function defined by $\text{sign}(x)=x/|x|$.
When $x=0$ it is ambiguous.

Before continuing note the following. One implicit inequality that
should be respected for horizons to exist is
\begin{equation}\label{horizon_2}
\bar{M}(w) \ge 0 \ .
\end{equation}
This is related to the reality of the square root of the
Boulware-Deser metric function (\ref{BD_metric}). $\bar{M}$ and $w$
cannot be both negative. That is, if $w\, \bar{M} \ge 0$ then it
must be $w+\bar{M} \ge 0$. This is precisely what is guarantied by
(\ref{horizon_2}).
\\

From remark \ref{elementary horizons} we have
\begin{remark}  \label{rH>0 remark}
$r_{H+}>0$ is equivalent to ${\rm sign}(\alpha)={\rm sign}(w-1)$.
$r_{H-}>0$ is equivalent to ${\rm sign}(\alpha)={\rm sign}(\bar
M-1)$.
\end{remark}
Note that, as we will solve the problem of existence for the real
numbers $r^2_{H\pm}/4\alpha$ the positivity conditions above
essentially restrict the sign of $\alpha$.

 Now
\begin{equation}\label{rh+-rh-}
r_{H+}^2-r_{H-}^2=\frac{8\alpha}{w-1} \sqrt{\bar M(w)}\ ,
\end{equation}
and (\ref{rH+>0}) tell us that
\begin{remark}\label{rh+-rh- remark}
If $r_{H+}$ exists then $r_{H+} \ge r_{H-}$.
\end{remark}

A solution $r_H$ corresponds to a horizon if $r_H>0$ and there exist
$r_1<r_H<r_2$ such that $f(r_1)f(r_2)<0$.
\\

Recall (\ref{horizon 1}). We have some \textit{Special cases}:

$i$). $w\pm\sqrt{\bar{M}(w)}=0 \Leftrightarrow \bar{M}=-w$.
 (Note
again that the correct result for $w=1$ is obtained as the limit).
Then one of the two solutions $r_{H\pm}$ coincides with a
\emph{branch singularity}. I.e. in this case the branch singularity
is \emph{null}. [This is possible for $\alpha<0$ otherwise this
solution doesn't exist.]

There is a single horizon solution given by
%\begin{equation}\label{rH not rE}
$r_H^2=4\alpha\, \frac{w+1}{w-1}$. %\ ,
%\end{equation}
It is a horizon of the branch $\xi$ according to
%\begin{equation}\label{}
$- \xi=\text{sign}(w\, (w+1))$. %\ .
%\end{equation}
[Of course the case $w=0=\bar M$ does not have two branches.] The
case $w=-1$ i.e. $\bar M=1$ does not have a horizon as $r_H$
vanishes (if $\xi=-\text{sign}(\alpha)$), or the metric function
$f$, which reads
\begin{equation}\label{}
f=1+\frac{r^2}{4\alpha}+\xi \text{sign}(\alpha)
\sqrt{1-\left(\frac{r^2}{4\alpha}\right)^{\!\!2}}\ ,
\end{equation}
can vanish only at the branch singularity when $\alpha<0$.

Finally one should bear in mind that $r_H=r_{H-}$ when $\bar M=-w>0$
and $r_H=r_{H+}$ when $\bar M=-w<0$.

$ii$). $1-\sqrt{\bar{M}(w)}=0 \Leftrightarrow \bar{M}(w)=\bar{M}=1$.
(Note again that the correct result for $w=1$ is obtained as the
limit). We just learned that when we also we have $w=-1$ there are
no horizons. So we assume that $w \ne -1$. We observe that
$r_{H-}=0$. This actually happens if $\xi=-\text{sign}(\alpha)$
otherwise this solution doesn't exist.

The single horizon solution is
%\begin{equation}\label{}
$r_{H+}^2=4\alpha\, \frac{2}{w-1}=\frac{6}{\Lambda}$.  %\ .
%\end{equation}
It is a horizon of the branch $\xi$ according to
%\begin{equation}\label{}
$- \xi=\text{sign}(w+1)$. %\ .
%\end{equation}

 $iii$).
$\bar{M}(w)=0$. This is the saturated case where the two radii
coincide: $r^2_{H\pm}=4\alpha/(w-1)=3/\Lambda$. Condition
(\ref{horizon 1}) works well in this case: $\xi=-\text{sign}(w)$.
Also from $r_H^2>0$ we have $\text{sign}(\alpha)=\text{sign}(w-1)$.

In this case $r_H$ is not a horizon radius. It is the (single) zero
of $f$ which has the same sign everywhere else. There are three
non-trivial cases. $w<0$: Then there is an outer branch singularity
and $f(r) \ge 0$. $0<w<1$: Then there is an inner branch singularity
and $f(r) \le 0$. $w>1$: Then $0<r<\infty$ and $f(r) \le 0$. $\Box$
\\

Recall (\ref{horizon 1}).
\begin{proposition}\label{horizon exist prop}
With the exception of cases covered in $i)$ and $ii)$ we have: The
radius $r_{H+}$ is a horizon of the branch $\xi$ if
\begin{equation}  \label{plus r sigma}
 -\xi={\rm sign}\!\left(w+\sqrt{\bar{M}(w)}\right)\ ;
\end{equation}
the radius $r_{H-}$ is a horizon of the branch $\xi$ if
\begin{equation}  \label{minus r sigma}
 -\xi={\rm sign}((\bar M+w)(\bar M-1))   \,   {\rm sign}\!\left(w+\sqrt{\bar{M}(w)}\right)\ .
\end{equation}
\end{proposition}

The type of the horizon, i.e. whether it is black hole, inner or
cosmological horizon, can be determined by the sign of the first
derivative of $f(r)$ (combined with Remark \ref{rh+-rh- remark}). We
have
\begin{equation}\label{horizon type}
r_{H\pm} f'(r_{H\pm})=\mp 2\ \sqrt{\bar{M}(w)}\ \cdot
\frac{1\pm\sqrt{\bar{M}(w)}}{w\pm\sqrt{\bar{M}(w)}}\ .
\end{equation}

Therefore for $\bar{M}(w)>0$, when $r_{H-}$ or $r_{H+}$ does
correspond to a horizon, the type is determined by
\begin{equation}\label{horizon type simple}
\text{sign}(f'(r_{H\pm}))=\pm\xi\ .
\end{equation}

Remarks \ref{rH>0 remark} and \ref{rh+-rh- remark}, Proposition
\ref{horizon exist prop}, and formula (\ref{horizon type simple})
provide criteria for the existence and the type of horizons for each
branch $\xi$ of the Boulware-Deser metric.

For the exotic branch $(\xi=+1)$ a black hole horizon must be an
$r_{H+}$. This is not possible by (\ref{plus r sigma}). Thus there
no black holes in the exotic branch. For the good branch $(\xi=-1)$
a black hole horizon must be an $r_{H-}$. From (\ref{elementary
horizons}), this is possible only for $M> \alpha$.

\section{ The Junction conditions}

 For our purposes, a singular shell $\Sigma$ is a
submanifold of codimension one at which the metric is continuous but
the extrinsic curvature has a finite discontinuity. The field
equations of Einstein-Gauss-Bonnet theory are given by
(\ref{The_field_equations}). Integrating the field equations across
$\Sigma$, one obtains the junction conditions
\begin{equation*}
\label{explicit junction ij}   ({\frak Q}_R)_{a}^{b}-({\frak
Q}_L)_{a}^{b} =- \kappa^{2} S_{a}^{b}\ ,
\end{equation*}
with ${\frak Q}_{a}^{b}$ given by
 \footnote{The notation of Ref. \cite{Gravanis:2007ei} has been
used. However in that reference there was an unconventional sign
convention used (in equation A3) for the definition of extrinsic
curvature. Although none of the results of that paper were affected
by this, unfortunately the formulae B13-B17 for the
Einstein-Gauss-Bonnet in the appendix were a mixture of inconsistent
sign conventions. Here we correct this sign error by choosing the
standard sign convention as in Refs. \cite{I} and
\cite{Davis:2002gn}. The
 developed expression is:
\begin{equation}
\Big[ \sigma (K_{b}^{a} -\delta _{b}^{a}K) + 2\alpha \left(
3J_{b}^{a}-\delta _{b}^{a}J - 2\varsigma P_{\ \
bd}^{ac}K_{c}^{d}\right)\Big]^+_- = -\kappa^2 S_{b}^{a}\, ,
\end{equation}
where $\sigma$ is $\pm1$ for a timelike/spacelike shell, $J_{ab}:=
(2KK_{ac}K^c_b + K_{cd}K^{d}K_{ab} - 2K_{ac}K^{cd}K_{db}
-K^2K_{ab})/3$ and $P_{abcd}:= R_{abcd} +2R_{b[c}g_{d]a} -
2R_{a[c}g_{d]b} +Rg_{a[c}g_{d]b}$ is the trace-free part of the
intrinsic curvature. In the case of a timelike shell ($\sigma =
+1$), this expression agrees with that given in Ref.
\cite{Davis:2002gn,Gravanis:2002wy}.} (\ref{explicit_Q}). Lower case
Roman letters from the beginning of the alphabet $a$, $b$ etc.
represent four-dimensional tensor indices on the tangent space of
the world-volume of the shell. The $R^{ab}_{\ \ cd}$ appearing in
the junction condition is the four-dimensional intrinsic curvature.
The antisymmetrized Kronecker delta is defined as $\delta^{a_1 \dots
a_p}_{b_1 \dots b_p} \equiv p!\, \delta^{a_1}_{[b_1} \cdots \delta^{
a_p}_{b_p]}$.

Now we calculate the intrinsic curvature of the world-volume of a
spherical shell of radius $a(\tau)$ and the extrinsic curvature
(which takes a diagonal form). There are two cases: For the timelike
case the components are
\begin{align*}
R_{\ \ \tau\phi}^{\tau\phi}  & =\frac{\overset{..}{a}}{a}\, ,\qquad
R_{\ \ \phi\theta}^{\phi\theta} = R_{\ \
\theta\chi}^{\theta\chi}=R_{\ \
\chi\phi}^{\chi\phi}=\frac{(k+\overset{.}{a}^{2})}{a^{2}}\, ,
\end{align*}
\begin{gather*}
K_{\tau}^{\tau}   = \eta\frac{\ddot{a}+\frac{1}{2}f^{\prime}}%
{\sqrt{\dot{a}^{2}+f}}\, , \qquad
K_{\theta}^{\theta}   =K_{\phi}^{\phi}=K_{\chi}^{\chi}= \frac{\eta}{a}%
\sqrt{\dot{a}^{2}+f} \, ;
\end{gather*}
while for the spacelike case these are
\begin{align*}
R_{\ \ \tau\phi}^{\tau\phi}  & =-\frac{\overset{..}{a}}{a}\,
,\qquad R_{\ \ \phi\theta}^{\phi\theta}
=\frac{(k-\overset{.}{a}^{2})}{a^{2}}\, ,
\end{align*}
\begin{gather*}
K_{\tau}^{\tau}   = \eta\frac{\ddot{a}-\frac{1}{2}f^{\prime}}%
{\sqrt{\dot{a}^{2}-f}}\, , \qquad
K_{\theta}^{\theta}   =K_{\phi}^{\phi}=K_{\chi}^{\chi}= \frac{\eta}{a}%
\sqrt{\dot{a}^{2}-f} \, .
\end{gather*}

In this paper we are interested in pure vacuum shells, i.e. when
$S^a_b = 0$. It is clear that in this case one can pull out a
factor of $\Delta K^d_c \equiv (K_+ - K_-)^d_c$, which is the jump
in the extrinsic curvature across the shell.
\begin{gather}
 \Delta K^d_c \left(\cdots \right) =S^a_b   = 0\, .
\end{gather}
In the case of interest in this paper, the extrinsic curvature is
diagonal. Thus, one expects each component of the junction
conditions to factorize conveniently.

Using the above formulae, we derive ${\frak Q}^\tau_\tau$ given in
(\ref{Qtt}). The angular components are, for the timelike case:
\begin{gather}
 {\frak Q}_\theta^\theta=-2!\ a^{-2}  \left\{ \eta\
\frac{\frac{1}{2}f'\{a^2+4\alpha(k-f)\}}{\sqrt{\dot a^2+f}} + \eta\
2a \sqrt{\dot a^2+f}+ \eta\ 4 \alpha\ \frac{\ddot a}{\sqrt{\dot
a^2+f}}\ \Big(k+f+2\dot a^2+\frac{a^2}{4\alpha}\Big) \right\}.
\end{gather}

\section{The derivatives of the potential}\label{Potential_Appendix}

As before, let us denote the derivative with respect to $a$ by a
prime. In analysing dynamical shells and the stability of static
shells it is useful to calculate the derivatives of $V(a)$ with
respect to $a$, $V^{\prime}$, $V^{\prime\prime}$ etc. First we
recall the definition of $Y(a)$; namely
\begin{gather}\label{Def_Y}
f(a) \equiv k + \frac{a^2}{4\alpha}\left(1 + \xi Y(a)\right)\,,
\qquad Y:= \sqrt{ w+ \frac{16M \alpha}{a^4} }\, .
\end{gather}
Note that $Y$ obeys the simple differential equation:
\begin{gather}\label{Y_diff_equation}
 (Ya^2)^\prime = \frac{2w a}{Y},
\end{gather}
where we recall that $w:= 1+\frac{4\alpha\Lambda}{3}$.

In terms of $Y_R$ and $Y_L$, the effective potential defined in
(\ref{solved-explicit}) takes the form:
\begin{gather}
\sigma V=  \left(k+\frac{a^2}{4\alpha}\right)
 - \frac{a^2}{12\alpha}
 \left( \xi_R  Y_R + \xi_L Y_L  - \frac{ \xi_R\xi_L Y_R Y_L}
 {\xi_R Y_R + \xi_L Y_L}\right)  .
\end{gather}
This can be also written as
\begin{equation}
 V(a)=  \sigma\left(k+\frac{a^2}{4\alpha}\right)
 - \frac{\sigma a^2}{4\alpha}
 \left(  \frac{3(\xi_R  Y_R + \xi_L Y_L )^2 + (\xi_R  Y_R - \xi_L Y_L )^2 }
 {12(\xi_R Y_R + \xi_L Y_L)}\right) \, .
\end{equation}

By repeated application of the
differential equation (\ref{Y_diff_equation}) we obtain:
\begin{align}\label{Vprime}
 \sigma V^\prime & =  \frac{a}{2\alpha} \left(
 1 -  \frac{ w}{\xi_R  Y_R + \xi_L Y_L}\right)\, ,
\\
 \sigma V^{\prime\prime} & =  \frac{1}{2\alpha}\left( 1 - \frac{3 w}
 {\xi_R Y_R + \xi_L Y_L} +
 \frac{ 2 w^2}{\xi_R \xi_L Y_R Y_L(\xi_RY_R
 + \xi_LY_L)}\right)\, ,
\end{align}
Note that the second derivative of $V$ depends on $a$ only
implicitly through $Y(a)$.

Let $a_e$ be the radius at which $V$ is an extremum,
$V^\prime(a_e) =0$. From (\ref{Vprime}) we have
\begin{gather}
 \xi_R Y_R(a_e)   + \xi_L Y_L(a_e) = w\, .
\end{gather}
It is of interest to know whether the extremum is minimum or
maximum. The second derivative evaluated at the extremum is:
\begin{gather}\label{Min_Max}
  V^{\prime\prime}(a_e) =  \frac{\sigma}{\alpha}\left(
 \frac{  w}{\xi_R \xi_L Y_R(a_e) Y_L(a_e) }  -1\right)\, ,
\end{gather}
If the right hand side of (\ref{Min_Max}) is positive, the extremum
is a minimum.

Let us look for a solution where the minimum of the potential
coincides with $V =0$. Imposing at some radius $a_0$ that $V(a_0) =
V^{\prime}(a_0) =0$ implies:
\begin{align}
\xi_R Y_R + \xi_L Y_L &= w\, ,\label{JunctionwY1} \\
\xi_R \xi_L Y_R Y_L &= w^2 -  \left( 3+\frac{12k\alpha}{a_0^2}
\right)w\, .
\end{align}

One can verify as a consistency check that the static and
instantaneous shell solutions of section \ref{Static_Section} are
recovered. In terms of the metric functions $f$ the above two
equations are:
\begin{equation*}
f_R + f_L = \left( \frac{3}{4\alpha} + \frac{\Lambda}{3}\right)a_0^2
+ 2k\, , \qquad f_Rf_L = \left( \frac{\Lambda a_0^2}{3}-k\right)^2\,
,
\end{equation*}
c.f. the junction conditions for static and instantaneous shells in
proposition \ref{static instant}. Upon imposing the inequalities
(\ref{timed ineq}-\ref{spacelike real roots}) we recover exactly the
solutions of that section.

It is important in analyzing the stability of the static ($\sigma =
+1$) vacuum  shells to know the sign of $V^{\prime\prime}$ evaluated
at the static radius $a_0$.
\begin{gather}\label{Min_Max_static}
  V^{\prime\prime}(a_0) =  \frac{1}{\alpha}\left(
 \frac{  w}{ w^2 -  \left( 3+\frac{12k\alpha}{a_0^2}\right)}  -1\right)\, ,
\end{gather}
Note that this can also be written
\begin{equation*}
 V^{\prime \prime }(a_0) = -\frac{1}{\alpha}\left( 1 +
 \frac{ \frac{k a_0^2}{4\alpha}  }{3 + (2-\frac{4\alpha
 \Lambda}{3})\frac{k a_0^2}{4\alpha}}\right) = -\frac{1}{\alpha
}\frac{xy-3kx-3y}{xy-3kx-2y}
\end{equation*}
in terms of the original variables and of the variables of section
\ref{Static_Section} respectively.

\section{Some details of the space of constant solutions} \label{Moduli_Appendix}

In terms of the variables $w$ and $u$ introduced in section
\ref{Moduli_section}, the junction conditions for static or
instantaneous shells are given by equation (\ref{FF in u,w})
 with $f_{L,R}>0$ for the timelike vacuum shell (which
corresponds to $\sigma=+1$), and $f_{L,R}<0$ for the spacelike
vacuum shell ($\sigma=-1$). After squaring the equations above, we
obtain
\begin{equation}\label{fpm in u and w}
f_{(\pm)}=\frac{(3w\pm u)^2}{u^2+3(w^2-4w)}\ .
\end{equation}
which turns out to be always real. The solution for $f_{L,R}$ is
given by (\ref{swap}) as discussed there. Then from (\ref{FF in
u,w}) we first have
\begin{proposition}\label{pre total moduli}
Let the total moduli space be described in the $(w,u)$ parameter
space. Then it necessarily is a subset of the upper half plane $u
\ge 0$ from which the points on the curves $\pm u= 3w$ and
$u^2+3(w^2-4w)=0$ are excluded.
 The four disconnected
regions are divided according to the type of the matching by
combinations of the following. Timelike: $u^2+3(w^2-4w)>0$.
Spacelike: $u^2+3(w^2-4w)<0$. Same orientation i.e. $\eta_L\eta_R>0$
$:$ $u^2-9w^2>0$. Opposite orientation, i.e. $\eta_L\eta_R<0$ $:$
$u^2-9w^2<0$.
\end{proposition}
It is good to remember
\begin{remark}\label{special points} The points $(0,0)$, $(1,3)$ and $(0,4)$ in the
$(w,u)$ plain do not belong to the moduli space. The point $(1,3)$
corresponds to the line $x=0$ and $y \ne 0$.
\end{remark}
We have already used the fact that $f_{L,R}(r)$ are the
Boulware-Deser metric functions. In order to completely solve our
problem we must substitute for $f_{L,R}$ using the Boulware-Deser
expression evaluated at $r=a_{0}$ given in equation (\ref{BD}),
\begin{equation}
\label{solution=BD}f_L=f_L(a_{0}) \quad, \quad f_R=f_R(a_{0}) \ .
\end{equation}
Recall (\ref{swap}). Similarly to equations
(\ref{General_Solution_mess_1}) and (\ref{General_Solution_mess_2})
we have that, within the space of Proposition \ref{pre total
moduli}, (\ref{solution=BD}) amount to
\begin{equation}\label{solution=BD explicit}
w\, (w\pm u)=2w\, \xi_{(\pm)}\,
\sqrt{w+\frac{\big(u^2+3(w^2-4w)\big)^2}{144 w^2}\bar{M}_{(\pm)}}\ .
\end{equation}

The solution $w=0$ is possible only if $\bar{M}_{L,R}=0$. Then for
$|\alpha|<\infty$ we have that $M_{L,R}=0$ and the bulk metrics are
simply $f_{L,R}(r)=1+r^2/(4\alpha)$. We have
\begin{remark}
The line $w=0$, which lies in the ``cone'' $u^2-9w^2>0$ and entirely
within the timelike standard shell region, is excluded from the
moduli space as it merely corresponds to smooth geometries.
\end{remark}
Therefore we work with non-zero $w$. Squaring the previous relation
we find the mass parameters of $f_{(\pm)}(r)$ which are consistent
with the vacuum shell solution; they are given by equation
(\ref{mass formulas}).

Substituting back into (\ref{solution=BD explicit}) we have the
condition
\begin{equation}\label{solution=BD simple}
 \xi_{(\pm)}\, |w\pm u|=w \pm u\ .
\end{equation}
The sign of $\xi_{(+)}$ is completely determined over the moduli
space if $u+w \ne 0$ by $\xi_{(+)}(w+u)>0$. Similarly, the sign of
$\xi_{(-)}$ is determined if $w-u \ne 0$ by $\xi_{(-)}(w-u)>0$. Now
for $w+u=0$ we find that $\xi_{(-)}|w|=w=-u<0$. Similarly for
$w-u=0$ we find that $\xi_{(+)}>0$, and this happens for $w>0$.

We see that the signs $\xi_{\pm}$ are specified for each point on
the moduli space, i.e. a solution of the vacuum shell. We will say
that this is a solution of the vacuum shell of \textit{type}
$(\xi_{(-)},\xi_{(+)})$. The exception is along the \textit{branch
curve} $u^2-w^2=0$ where one of the signs is undetermined. We can
summarize
\begin{proposition}\label{total moduli}
The moduli space consists of the regions of the parameter space
$(w,u)$ given in Proposition \ref{pre total moduli} such that: i$)$
the line $w=0$ is excluded, ii$)$ according to the branch signs
$(\xi_{(-)},\xi_{(+)})$ of the bulk regions the parameter space is
divided as follows: $(+,+)$ for $ u < w$; $(-,+)$ for $-u<w<u$,
$(-,-)$ for $w < -u$.

The points along the branch curve $u^2-w^2=0$ satisfy: if $w>0$ then
$\xi_{(+)}>0$ and $\xi_{(-)}$ arbitrary, if $w<0$ then $\xi_{(-)}<0$
and $\xi_{(+)}$ arbitrary. The mass parameters $M_{(\pm)}$ are well
defined and given over the moduli space by formula (\ref{mass
formulas}).
\end{proposition}
Propositions \ref{pre total moduli} and \ref{total moduli}
categorize the allowed spherically symmetric vacuum shell solution
at constant $r$ in terms of spacelike/timelike and branch signs.
This is plotted in figure \ref{egg_figure}.

 Note also the following: Formula (\ref{mass formulas})
says that we can define a function
\begin{equation}\label{single mass formula}
\bar{M}_{*}(w,u):=\frac{36 w^2 ((w + u)^2-4w)}{(u^2+3(w^2-4w))^2}\ ,
\end{equation}
defined on the whole of the $(w,u)$ plain (minus the curve
$u^2+3(w^2-4w)=0$) and not only on the upper half. Then for $u>0$,
$\bar{M}_{(+)}=\bar{M}_{*}(w,u)$ and
$\bar{M}_{(-)}=\bar{M}_{*}(w,-u)$. More generally, recalling also
equations (\ref{fpm in u and w}), and (\ref{solution=BD simple}),
one may extend also $f(a_0)$ and $\xi$, regarded as functions of $w$
and $u$, over the whole of the $(w,u)$ plane.
\begin{lemma}\label{mirror lemma} Let $X$ denote any of the
quantities $\bar{M}$, $f$, $\xi$, or combinations of them. One may
define a function $X_{*}(w,u)$ such that $X_{(+)}=X_{*}(w,u)$ for $u
\geq 0$. Then, $X_{(-)}=X_{*}(w,-u)$. At $u=0$ we have
$X_{(+)}=X_{(-)}$ i.e. $X_R=X_L$.

The parameter space can be extended over the whole of the $(u,w)$
plane. The mirror transformation $u \to -u$ has the effect of
sending $(+) \leftrightarrow (-)$. So one may specify one type of
quantities $\xi_{(+)}$, $\bar{M}_{(+)}$ on the whole plane and
mirror image the results to obtain the values of $\xi_{(-)}$ and
$\bar{M}_{(-)}$.
\end{lemma}

Now, we will also return to discuss in a more detailed manner the
two most basic distinct types of constructions here (recall
Definition \ref{the definition}): matching with the same
orientation, i.e. standard shell solutions, and matching with
opposite orientation, which we call collectively wormholes. Though
the following definition has been already in use in our work, it is
useful to formalize the following
\begin{definition}\label{plus minus metrics}
A plus-metric, corresponding to the metric function $f_{(+)}(r)$, is
one whose mass parameters is given by $\bar{M}(w,u)=M_{(+)}$ and
branch by $(w+u)/|w+u|=\xi_{(+)}$ over the moduli space. A
minus-metric, corresponding to the metric function $f_{(-)}(r)$), is
one whose mass parameters is given by $\bar{M}(w,-u)=\bar{M}_{(-)}$
and branch by $(w-u)/|w-u|=\xi_{(-)}$ over the moduli space.
%The
%$\bar{M}_{(\pm)}$ will be called plus- and minus-mass, and
%$\xi_{(\pm)}$ will be called plus- and minus-sign respectively.
\end{definition}

Now, let us make a remark on the sign of $\alpha $. As $a_0^2>0$ it
can be determined by the sign of $y/x$ and is given by
\begin{equation}\label{sign of alpha}
\text{sign}(\alpha)=\text{sign}\big(w\, (u^2+3(w^2-4w))\big)\ .
\end{equation}
Therefore we have
\begin{remark}\label{alpha>0}
$\alpha>0$ only for timelike vacuum shells and in the region $w>0$
(for standard or wormhole orientation). Inside the ellipse of the
spacelike vacuum shells (see fig. \ref{egg_figure}), or for $w<0$,
we have$:$ $\alpha<0$.
\end{remark}

From the definition of $w$, the sign of the cosmological constant
$\Lambda$ is determined according to
\begin{equation}
\text{sign}(\Lambda)=\text{sign}(\alpha)\, \text{sign}(w-1).
\end{equation}
When $\Lambda=0$ i.e. $w=1>0$, the sign of $\alpha$ depends on the
whether the shell is time- or space-like as we mentioned just above.
\\

\textbf{Proof of Proposition \ref{Unique_proposition}: } For a given
$w$, let $u_0$ be such that the corresponding point ($u_0, w$)
belongs to the moduli space. We have
\begin{equation}\label{a0 uniqueness}
 \bar{M}_{(+)}=36w^2\, \frac{(u_0+w)^2-4w}{(u_0^2+3(w^2-4w))^2}\quad
,\quad \bar{M}_{(-)}=36w^2\,
\frac{(u_0-w)^2-4w}{(u_0^2+3(w^2-4w))^2}.
\end{equation}
Of course $u_0 \ge 0$.

There are two special cases to deal with before proceeding. First,
consider $\bar{M}_{(+)}=\bar{M}_{(-)}$. We know that this is
possible if and only if $u_0=0$. So for non-unique solutions we may
restrict ourselves to $u_0>0$. The second case is when
$\bar{M}_{(+)}+\bar{M}_{(-)}=0$. This happens in the moduli space
along the circle: $u_0^2+w^2-4w=0$. Clearly there is a unique
positive $u_0$ solving this equation.

Therefore it is adequate to consider $u_0>0$ and masses such that
$\bar{M}_{(+)}\pm\bar{M}_{(-)} \ne 0$. The proof is by
contradiction. Let us suppose that $u_0$ is not unique in the sense
that there exists some $u_1>0$ in the moduli space such that $u_1
\ne u_0$ and which gives the same masses
\begin{equation}\label{a0 uniqueness 2}
\bar{M}_{(+)}=36w^2\, \frac{(u_1+w)^2-4w}{(u_1^2+3(w^2-4w))^2} \quad
,\quad \bar{M}_{(-)}=36w^2\,
\frac{(u_1-w)^2-4w}{(u_1^2+3(w^2-4w))^2} \ .
\end{equation}
With a little rearranging subtracting the respective equations we
have
\begin{eqnarray*}\label{}
&&
(u_1-u_0)\Big\{(u_1+u_0)\big(u_1^2+u_0^2+6(w^2-4w)\big)\bar{M}_{(+)}-36w^2\,
(u_1+u_0+2w)\Big\}=0\ , \\
&&
(u_1-u_0)\Big\{(u_1+u_0)\big(u_1^2+u_0^2+6(w^2-4w)\big)\bar{M}_{(-)}-36w^2\,
(u_1+u_0-2w)\Big\}=0\ .
\end{eqnarray*}
$u_1 \ne u_0$ so the quantities in the big brackets vanish. Adding
and subtracting them we obtain the equations
\begin{equation}\label{u1}
u_1^2+u_0^2+6(w^2-4w)=\frac{72w^2}{\bar{M}_{(+)}+\bar{M}_{(-)}}
\quad , \quad u_1+u_0=2w\,
\frac{\bar{M}_{(+)}+\bar{M}_{(-)}}{\bar{M}_{(+)}-\bar{M}_{(+)}}\ .
\end{equation}
Via (\ref{a0 uniqueness}) these equations express $u_1$ in terms of
$u_0$ and $w$. The second of these tells us that
\begin{equation}\label{}
u_1=-\frac{3w^2}{u_0}<0\ .
\end{equation}
So we conclude that $u_1$ is negative, contradicting the assumption.
$\Box$

\clearpage

\section{Diagrams of the moduli space}\label{Plots_Appendix}

Here we collect the diagrams referred to in section
\ref{Static_Section}.

\begin{figure}[h]
\begin{center}
  \includegraphics[height=.9\textwidth,angle=270]{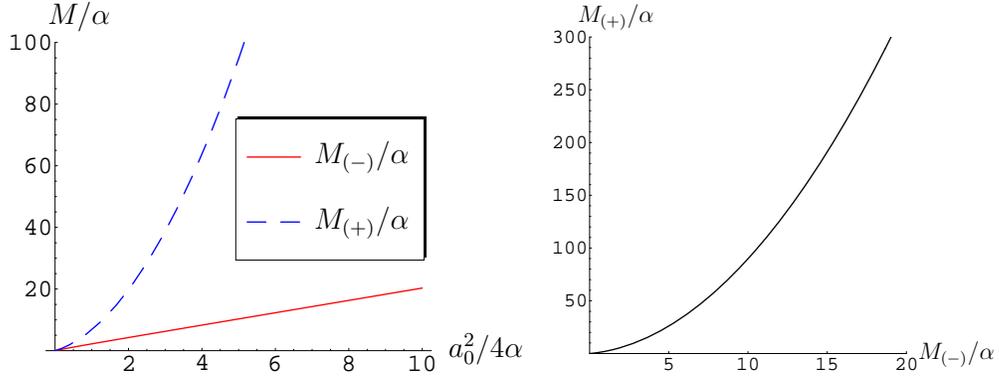}\\
  \caption{ For $\Lambda =0$, spherically symmetric shells exist only with standard
  orientation and for $\alpha>0$.
   Masses $M_{(-)}$, $M_{(+)}$ and shell radius $a_0$ are measured in units of
   the Gauss-Bonnet coupling, $\alpha$.}\label{ZeroLmassPics}
  \end{center}
\end{figure}

\begin{figure}[h]
\parbox{.5\textwidth}{ \includegraphics[width=.47\textwidth]{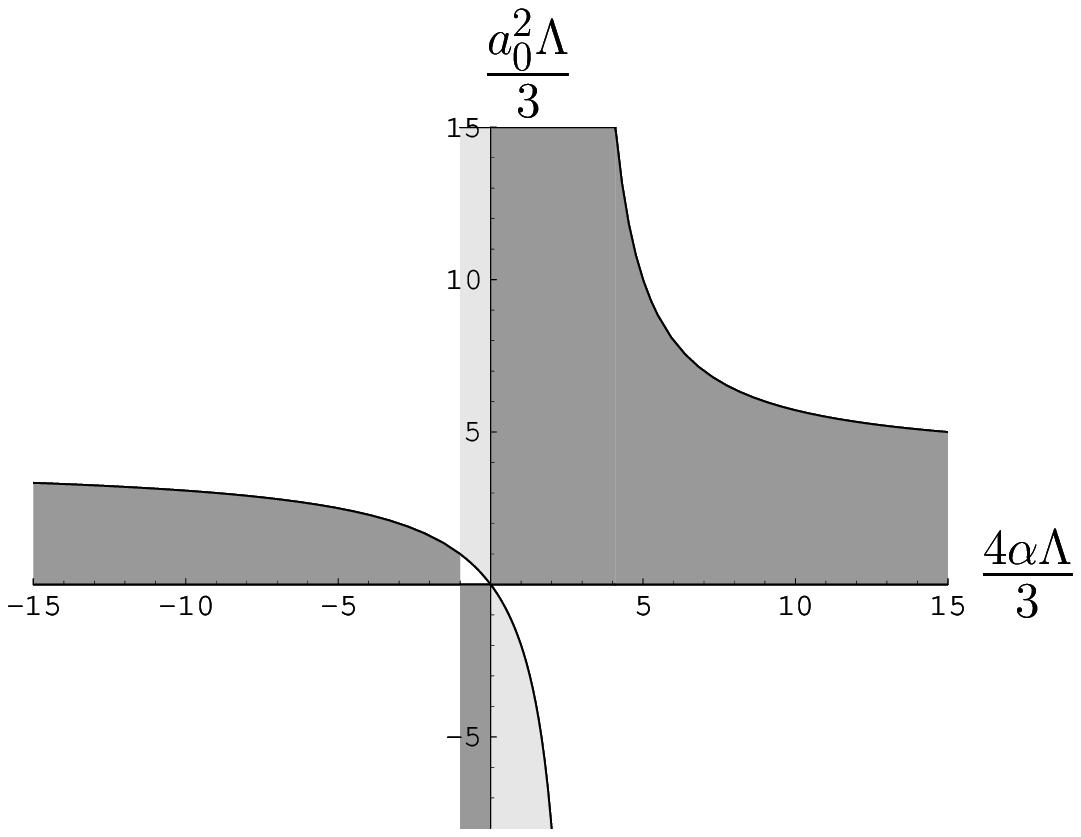}\\
  \caption{Static vacuum shells exist in the dark grey region.
  Instantaneous vacuum shells with $a =a_0$ exist in the light grey region. }\label{SpacelikeAndTimelike_fig}
 } \;
\parbox{.5\textwidth}{\includegraphics[width=.47\textwidth]{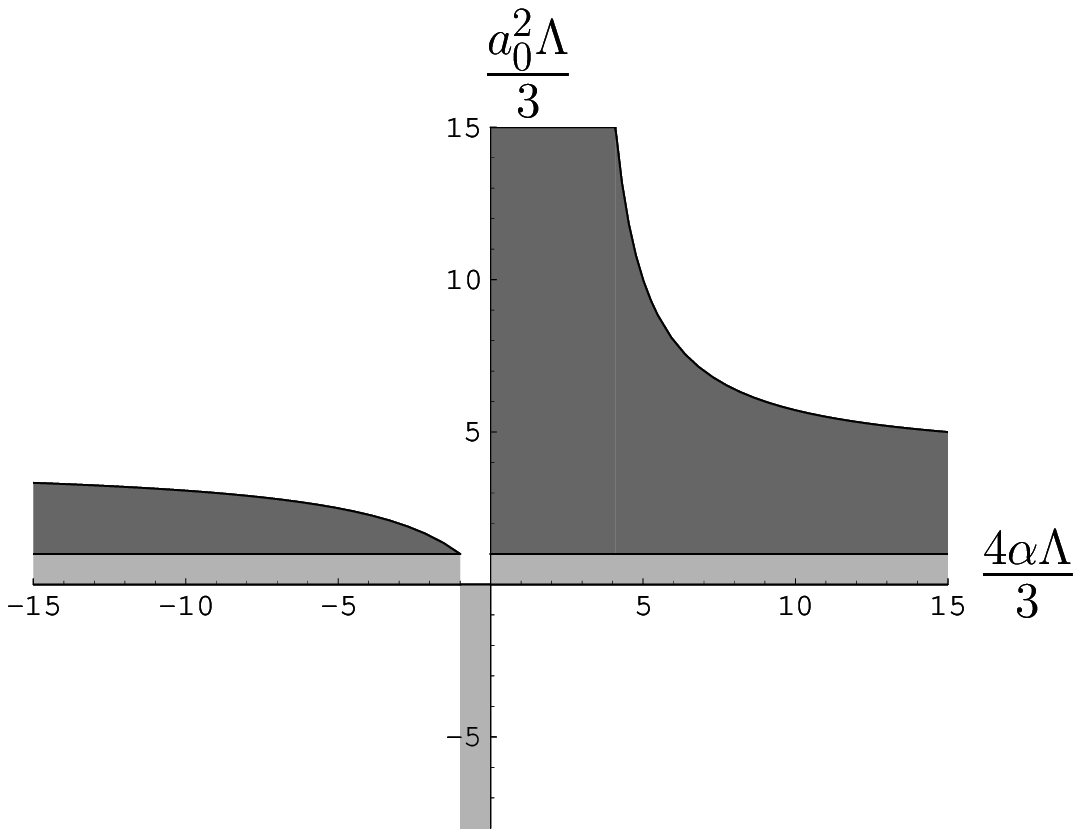}\\\\\\
  \caption{The static vacuum shells can have the standard orientation $\eta _L \eta _R >0$ (light grey)
  or wormhole orientation $\eta _L \eta _R <0$ (dark grey). }\label{WormholeAndPlain_fig}
    }
\end{figure}

\begin{figure}[b]
\parbox{.5\textwidth}{
\includegraphics[width=.47\textwidth]{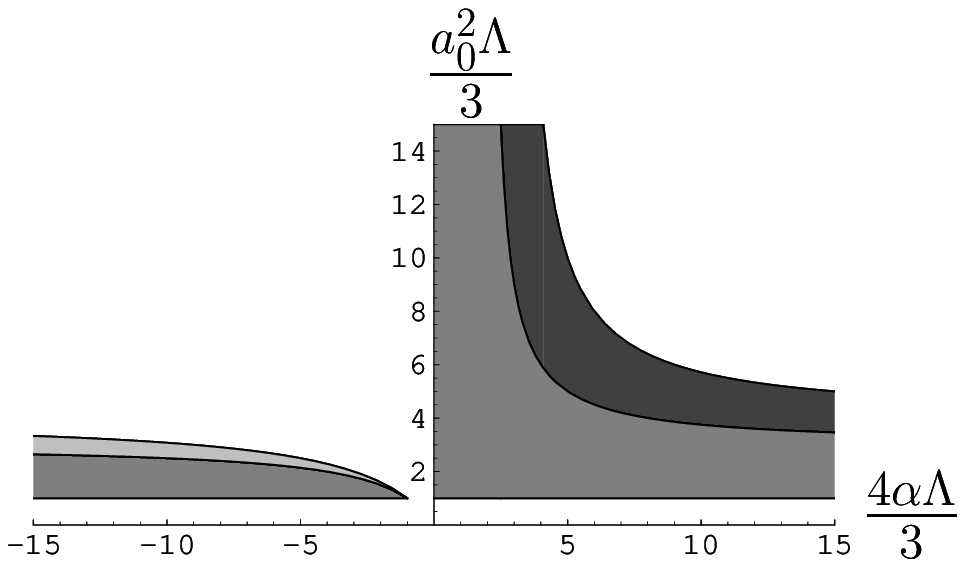}\\\\
  \caption{There are three types of static wormholes according to
  the branch signs $(\xi_L, \xi_R)$ in each bulk region:
  $(-,-)$ lightest grey; $(-,+)$ medium grey; $(+,+)$ dark grey. }\label{BranchesTimelikeWormhol_fig}
 } \;
\parbox{.5\textwidth}{  \includegraphics[width=.47\textwidth]{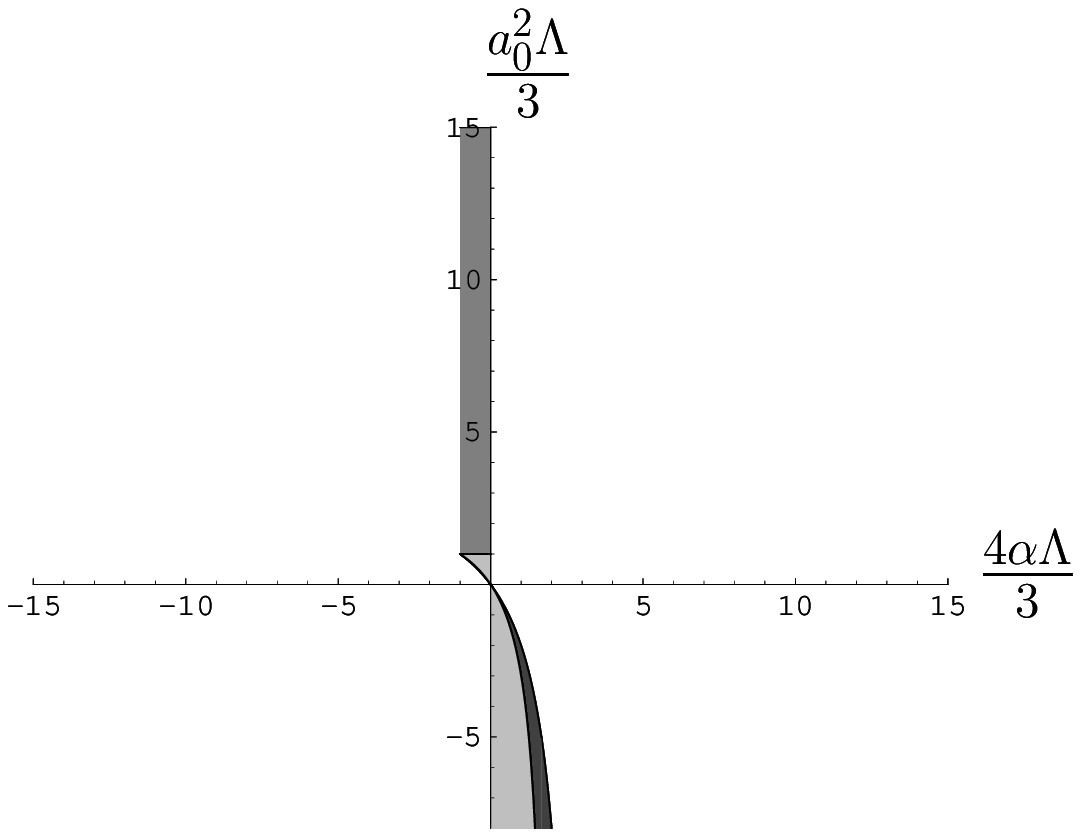}
  \caption{The different types of constant $a$ instantaneous shells are:
  $(-,+)$ branch standard orientation (light grey);
  $(-,+)$ branch wormhole orientation (medium grey);
  $(+,+)$ branchstandard orientation (dark grey).  }\label{SpacelikeRegions_fig}
 }
\end{figure}

\begin{figure}[thb]
\parbox{.5\textwidth}{
\includegraphics[width=0.5\textwidth]{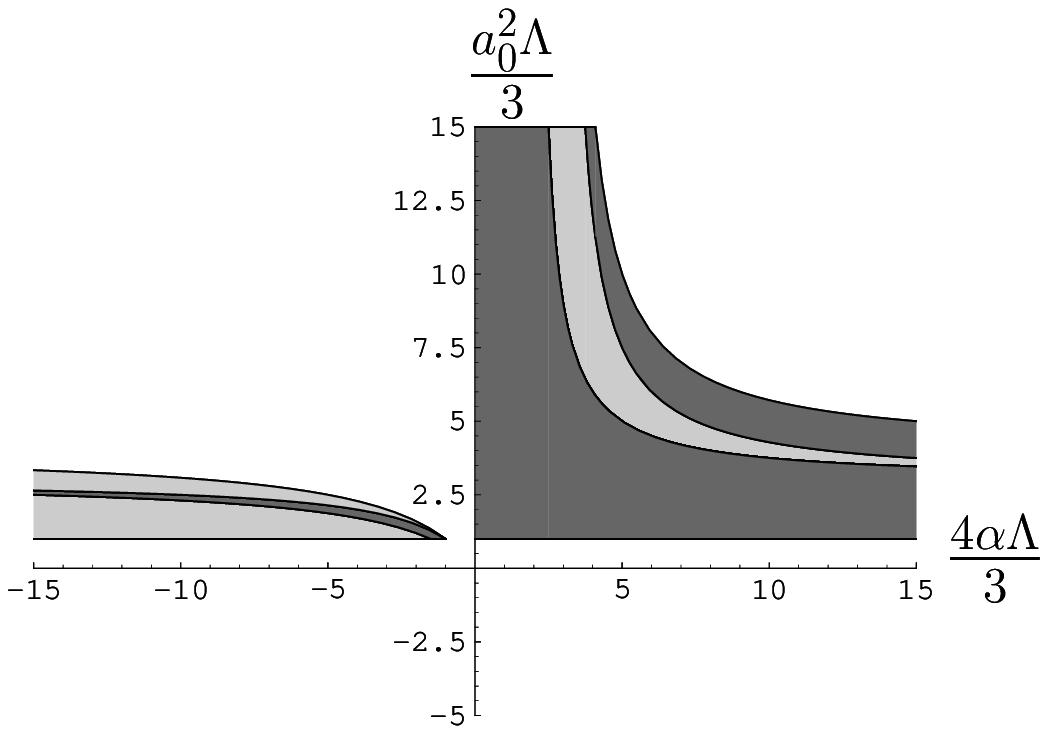}
\caption{{ {The stable region $V^{\prime%
\prime}(a_{0})>0$ for wormholes is shown in light grey. For positive
$\alpha$ this is a region of the (+,+) branch soluions. For negative
$\alpha$ it includes all except a small region of the $(-,+)$
branch.}} }\label{V''} } \;
\parbox{.5\textwidth}{
\includegraphics[width=0.5\textwidth]{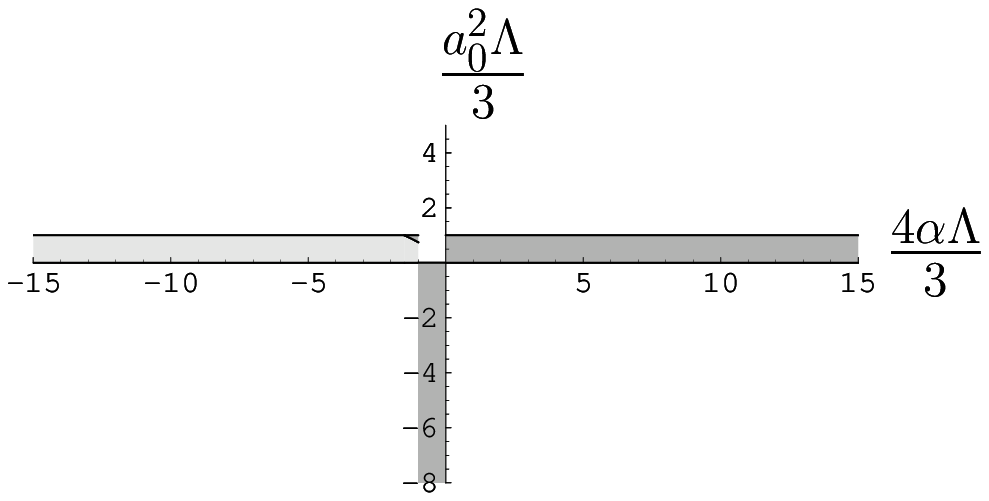}\\\\\\
  \caption{The stability of the standard shells.
  The stable regions are shown in light grey and the unstable regions in dark grey.
  All standard shells are $(-,+)$.}
  \label{PlainStabilityTimelikepic_fig}
 }
\end{figure}


\clearpage

\begin{thebibliography}{99}


\bibitem{Lovelock} D. Lovelock, J. Math. Phys. \textbf{12}, 498 (1971).
%%CITATION = JMAPA,12,498;%%


\bibitem{Zwiebach} B.~Zwiebach,
%``Curvature Squared Terms And String Theories,''
Phys.\ Lett.\ B \textbf{156} (1985) 315. %%CITATION = PHLTA,B156,315;%%



\bibitem{Zumino} B.~Zumino,
%``Gravity Theories In More Than Four-Dimensions,''
Phys.\ Rept.\ \textbf{137} (1986) 109. %%CITATION = PRPLC,137,109;%%

\bibitem{GS} D. Gross and J. Sloan, Nucl. Phys. \textbf{B 291}, 41 (1987).

\bibitem{mas} E. Fradkin and A. Tseytlin, Phys. Lett. {\bf B158},
316 (1986); Nucl. Phys. {\bf B261}, 1 (1985).

\bibitem{GW} D. Gross and E. Witten, Nucl. Phys. \textbf{B 277}, 1 (1986).

\bibitem{Ferrara:1996hh}
  S.~Ferrara, R.~R.~Khuri and R.~Minasian,
  %``M-Theory on a Calabi-Yau Manifold,''
  Phys.\ Lett.\  B {\bf 375}, 81 (1996)
  [arXiv:hep-th/9602102];
  %%CITATION = PHLTA,B375,81;%%
%\cite{Antoniadis:1997eg}
  I.~Antoniadis, S.~Ferrara, R.~Minasian and K.~S.~Narain,
  %``R**4 couplings in M- and type II theories on Calabi-Yau spaces,''
  Nucl.\ Phys.\  B {\bf 507}, 571 (1997)
  [arXiv:hep-th/9707013].
  %%CITATION = NUPHA,B507,571;%%


\bibitem{Teitelboim-87}
C.~Teitelboim and J.~Zanelli, Class. Quant. Grav. {\bf 4}, L125
(1987).

\bibitem{Choquet-Bruhat-88}
Y.~Choquet-Bruhat, J. Math. Phys. {\bf 29}, 1891 (1988).

\bibitem{Davis:2002gn}  S.~C.~Davis,
%``Generalised Israel junction conditions for a Gauss-Bonnet brane world,''
Phys.\ Rev.\ D \textbf{67}, 024030 (2003) [arXiv:hep-th/0208205].
%%CITATION = PHRVA,D67,024030;%%
%\cite{Gravanis:2002wy}


\bibitem{Gravanis:2002wy}
E.~Gravanis and S.~Willison,
%``Israel conditions for the Gauss-Bonnet theory and the Friedmann equation on
%the brane universe,''
Phys.\ Lett.\ B \textbf{562}, 118 (2003) [arXiv:hep-th/0209076].
%%CITATION = PHLTA,B562,118;%%



\bibitem{Meissner}
  K.~A.~Meissner and M.~Olechowski,
  Phys.\ Rev.\ Lett.\  {\bf 86}, 3708 (2001)
  [arXiv:hep-th/0009122];
  %%CITATION = HEP-TH 0009122;%%
 A.~Iglesias and Z.~Kakushadze,
  Int.\ J.\ Mod.\ Phys.\ A {\bf 16}, 3603 (2001)
  [arXiv:hep-th/0011111];
  %%CITATION = HEP-TH 0011111;%%
    J.~E.~Kim, B.~Kyae and H.~M.~Lee,
  Phys.\ Rev.\ D {\bf 64}, 065011 (2001)
  [arXiv:hep-th/0104150];
  %%CITATION = HEP-TH 0104150;%%
  M.~Hassaine, R.~Troncoso and J.~Zanelli,
  %``Eleven-dimensional supergravity as a gauge theory for the M-algebra,''
  Phys.\ Lett.\  B {\bf 596}, 132 (2004)
  [arXiv:hep-th/0306258].
  %%CITATION = PHLTA,B596,132;%%


\bibitem{Gravanis:2007ei}
  E.~Gravanis and S.~Willison,
  %```Mass without mass' from thin shells in Gauss-Bonnet gravity,''
  Phys.\ Rev.\  D {\bf 75}, 084025 (2007)
  [arXiv:gr-qc/0701152].
  %%CITATION = PHRVA,D75,084025;%%


\bibitem{Boulware:1985wk}  D.~G.~Boulware and S.~Deser,
%``String Generated Gravity Models,''
Phys.\ Rev.\ Lett.\ \textbf{55}, 2656 (1985).
%%CITATION = PRLTA,55,2656;%%

\bibitem{mar} %\cite{Myers:1987qx}
%\bibitem{Myers:1987qx}
  R.~C.~Myers,
  %``SUPERSTRING GRAVITY AND BLACK HOLES,''
  Nucl.\ Phys.\  B {\bf 289}, 701 (1987).
  %%CITATION = NUPHA,B289,701;%%


\bibitem{Cai:2001dz}  R.~G.~Cai,
%``Gauss-Bonnet black holes in AdS spaces,''
Phys.\ Rev.\ D \textbf{65}, 084014 (2002)  [arXiv:hep-th/0109133].
%%CITATION = PHRVA,D65,084014;%%

\bibitem{Aros:2000ij}
  R.~Aros, R.~Troncoso and J.~Zanelli,
  %``Black holes with topologically nontrivial AdS asymptotics,''
  Phys.\ Rev.\  D {\bf 63}, 084015 (2001)
  [arXiv:hep-th/0011097].
  %%CITATION = PHRVA,D63,084015;%%

\bibitem{BHscan}
  J.~Crisostomo, R.~Troncoso and J.~Zanelli,
  %``Black hole scan,''
  Phys.\ Rev.\  D {\bf 62}, 084013 (2000)
  [arXiv:hep-th/0003271].
  %%CITATION = PHRVA,D62,084013;%%




\bibitem{Torii-05}
  T.~Torii and H.~Maeda,
  %``Spacetime structure of static solutions in Gauss-Bonnet gravity:  Neutral
  %case,''
  Phys.\ Rev.\  D {\bf 71}, 124002 (2005)
  [arXiv:hep-th/0504127].
  %%CITATION = PHRVA,D71,124002;%%

\bibitem{Wiltshire} D. Wiltshire, Phys. Rev. D38 (1988) 2445.

\bibitem{Wiltshire2} D. Wiltshire, Phys. Lett. B169 (1986) 36.

\bibitem{Wheeler} J. T. Wheeler, Nucl. Phys. {\bf B268}, 737 (1986); Nucl. Phys. {\bf B273}, 732 (1986).



\bibitem{Charmousis}
  C.~Charmousis and J.~F.~Dufaux,
  %``General Gauss-Bonnet brane cosmology,''
  Class.\ Quant.\ Grav.\  {\bf 19}, 4671 (2002)
  [arXiv:hep-th/0202107].
  %%CITATION = HEP-TH 0202107;%%



\bibitem{Aliev:2007dp}
  A.~N.~Aliev, H.~Cebeci and T.~Dereli,
  %``Exact Solutions in Five-Dimensional Axi-dilaton Gravity with
  %Euler-Poincare Term,''
  Class.\ Quant.\ Grav.\  {\bf 24}, 3425 (2007)
  [arXiv:gr-qc/0703011].
  %%CITATION = CQGRD,24,3425;%%


\bibitem{Z}
 R.~Zegers,
  %``Birkhoff's theorem in Lovelock gravity,''
  J.\ Math.\ Phys.\  {\bf 46}, 072502 (2005)
  [arXiv:gr-qc/0505016].
  %%CITATION = JMAPA,46,072502;%%


\bibitem{GOT07}
  G.~Dotti, J.~Oliva and R.~Troncoso,
  %``Exact solutions for the Einstein-Gauss-Bonnet theory in five dimensions:
  %Black holes, wormholes and spacetime horns,''
  Phys.\ Rev.\  D {\bf 76}, 064038 (2007)
  [arXiv:0706.1830 [hep-th]].
  %%CITATION = PHRVA,D76,064038;%%



\bibitem{Deruelle-03}
N.~Deruelle and J.~Madore, [arXiv: gr-qc/0305004].


\bibitem{I} W. Israel, Nuovo Cim. \textbf{B44S10} (1966) 1 [Erratum, Nuovo
Cimento \textbf{B48} (1967) 463].


\bibitem{PoissonVisser}
  E.~Poisson and M.~Visser,
  %``Thin shell wormholes: Linearization stability,''
  Phys.\ Rev.\  D {\bf 52}, 7318 (1995)
  [arXiv:gr-qc/9506083].
  %%CITATION = PHRVA,D52,7318;%%


\bibitem{Poisson} P. Brady, J. Louko and E. Poisson, Phys. Rev. D44 (1991) 1891-1894.

\bibitem{VisserWiltshire}
 M.~Visser and D.~L.~Wiltshire,
  %``Stable gravastars - an alternative to black holes?,''
  Class.\ Quant.\ Grav.\  {\bf 21}, 1135 (2004)
  [arXiv:gr-qc/0310107].
  %%CITATION = CQGRD,21,1135;%%


\bibitem{BD} D. Boulware and S. Deser, Phys. Lett. \textbf{B175} (1986) 409.

\bibitem{Vacuum bubbles}
V. A. Berezin, V. A. Kuzmin and I. I. Tkachev, Phys. Lett. {120B},
(1983) 91; K. Maeda, Gen. Rel. Grav. {\bf 18}, (1986) 931;  H. Sato,
Prog. Theor. Phys. {\bf 76} (1986) 1250; S.~T.~Blau,
E.~I.~Guendelman and A.~Guth, Phys. Rev. {\bf D 35}, 1747 (1987);
A.~Aguirre and M.~C.~Johnson, Phys.\ Rev.\  D {\bf 72}, 103525
(2005) [arXiv:gr-qc/0508093];
%%CITATION = PHRVA,D72,103525;%%
  S.~V.~Chernov and V.~I.~Dokuchaev,
  arXiv:0709.0616.
  %%CITATION = ARXIV:0709.0616;%%


%\bibitem{T} C. Teitelboim, Phys. Lett. \textbf{B158 }(1985) 293.


\bibitem{Ghoroku:1992tz}
  K.~Ghoroku and T.~Soma,
  %``Lorentzian wormholes in higher derivative gravity and the weak energy
  %condition,''
  Phys.\ Rev.\ D {\bf 46}, 1507 (1992);
  %%CITATION = PHRVA,D46,1507;%%
B.~Bhawal and S.~Kar, Phys. Rev. D {\bf 46}, 2464 (1992).

\bibitem{wormjulin}
  G.~Dotti, J.~Oliva and R.~Troncoso,
  %``Static wormhole solution for higher-dimensional gravity in vacuum,''
  Phys.\ Rev.\  D {\bf 75}, 024002 (2007)
  [arXiv:hep-th/0607062].
  %%CITATION = PHRVA,D75,024002;%%



\bibitem{DF} S. Deser and J. Franklin, Class. Quant. Grav. \textbf{22}, L103
(2005).

\bibitem{AFG}
  M.~Aiello, R.~Ferraro and G.~Giribet,
  %``Exact solutions of Lovelock-Born-Infeld black holes,''
  Phys.\ Rev.\  D {\bf 70}, 104014 (2004)
  [arXiv:gr-qc/0408078].
  %%CITATION = PHRVA,D70,104014;%%



\bibitem{PapapetrouTreder} A. Papapetrou and A. Treder, Math. Nach.
23, 371 (1962).

%\bibitem{Papapetrou} A. Papapetrou, Compt. Rend. {\bf 258}, (1964)
%6085.

\bibitem{Petrov} A. Z. Petrov, JETP {\bf 17}, (1963) 1026.

\bibitem{Bergmann} P. G. Bergmann, M. Cahen and A. B. Komar, J.
Math.  Phys. {\bf 6}, (1965), 1.

\bibitem{BTZ} M. Ba\~nados, C. Teitelboim and J. Zanelli, Phys.Rev.Lett. {\bf 69} (1992) 1849.

\bibitem{BHTZ} M. Ba\~nados, M. Henneaux, C. Teitelboim and J.
Zanelli, Phys.Rev. {\bf D48} (1993) 1506.


\end{thebibliography}
\end{document}